\documentclass{aa}
\usepackage{graphicx}
\usepackage{amssymb}
\usepackage{natbib}
\def\ssim{\setbox0=\hbox{$\propto$}%
\setbox1=\hbox{$<$}\dimen0=\ht1%
\advance\dimen0by-1.2pt\,\lower.6\dimen0%
\copy0\kern-\wd0\raise.4\dimen0\copy1 \,}

\def\gsim{\setbox0=\hbox{$\propto$}%
\setbox1=\hbox{$>$}\dimen0=\ht1%
\advance\dimen0by-1.2pt\,\lower.6\dimen0%
\copy0\kern-\wd0\raise.4\dimen0\copy1\,}

\def\lambdab{\lambda\mkern-9mu\lower1.2pt\hbox{$\mathchar'26$}}%

\begin{document}
   \title{Stellar evolution with rotation XII: Pre--supernova models}
   \titlerunning{Rotation: pre--supernova evolution}


 \author{R. Hirschi \inst{1,2} \and G. Meynet \inst{1} \and A. Maeder \inst{1}}
 \authorrunning{R. Hirschi et al}
 \institute{Geneva Observatory CH--1290 Sauverny, Switzerland \and
              Universiti Malaya, 50603 Kuala Lumpur, Malaysia}
\offprints{R. Hirschi \email{Raphael.Hirschi@obs.unige.ch}}

   \date{Received  / Accepted }

\abstract{We describe the latest developments of the
Geneva stellar evolution code in order to model the pre--supernova
evolution of rotating massive stars. Rotating and non--rotating 
stellar models at solar
metallicity with masses equal
to 12, 15, 20, 25, 40 and 60 $M_{\sun}$ were computed from the ZAMS 
until the end of the core silicon burning phase. 
We took into account meridional circulation, secular shear
instabilities, horizontal turbulence and dynamical shear 
instabilities.
We find that dynamical shear 
instabilities mainly smoothen
the sharp angular velocity gradients but do  
not transport 
angular momentum or chemical species over long distances.

Most of the differences between the pre--supernova structures obtained 
from rotating and non--rotating stellar models have their origin in the 
effects of rotation during
the core hydrogen and helium burning phases. The advanced stellar evolutionary 
stages
appear too short in time to allow the rotational instabilities 
considered in this work to have
a significant impact during the late stages. In particular the internal 
angular momentum does not change significantly
during the advanced stages of the evolution. We can therefore have a good
estimate of the final angular momentum at the end of the core helium burning phase.

The effects of rotation on pre--supernova models are significant between
15 and 30 $M_{\sun}$. Indeed,
rotation increases the core sizes (and the yields) by a factor $\sim 1.5$. 
Above 20 $M_{\sun}$, rotation may change the radius or colour of the 
supernova progenitors (blue instead of red supergiant) and
the supernova type (Ibc instead of II).
Rotation affects the lower mass limits for radiative
core carbon burning, for iron core
collapse and for black hole formation.
For Wolf-Rayet stars ($M \gtrsim 30\,M_{\sun}$), 
the pre--supernova structures are mostly affected by the intensities
of the stellar winds and less by rotational mixing.
 }

\maketitle
%

\section{Introduction}
Over the last years, the development of the Geneva evolutionary 
code has allowed
the study of rotating star evolution from the ZAMS until the end of the
core carbon burning
phase. 
Various checks of the validity of the rotating stellar models have been made.
In particular, it has been shown that rotating models well reproduce the
observed surface enrichments \citep{HL002,ROTV}, the ratios of blue to red supergiants
in the Small Magellanic Cloud \citep{ROTVII},
and the variations of the Wolf--Rayet (WR hereinafter) star populations 
as a function of the metallicity \citep{ROTX}. For all these features
non--rotating models cannot reproduce observations.
The goal of
this paper is to follow the evolution of these models, which well
reproduce the above observed features,
during the pre--supernova evolution. 
Section 2 describes the modifications done in
order to model the advanced stages. In Sect. 3 we present 
the stellar evolution in the Hertzsprung--Russell diagram and the
lifetimes of the different burning stages. In Sect. 4 and 5 we discuss 
the evolution of rotation and internal structure respectively. 
Section 6 describes the structure of the
pre--supernova models. Finally,
in Sect. 7, we compare our results with the literature.
\section{Model physical ingredients}
The computer model used here is the same as the one described in
 \citet{ROTX} except for the wind anisotropy which here is
not taken into account. The model therefore includes secular
shear and meridional circulation. Convective stability is determined by the 
Schwarzschild criterion. Overshooting is only considered for 
H-- and He--burning cores with an overshooting parameter, 
$\alpha_{\rm{over}}$, of 0.1 H$_{\rm{P}}$. 
The modifications made in order to 
follow the advanced
stages of the evolution are described below.
\subsection{Internal structure equations}
The internal structure equations used are described in \citet{ROTI}.
These equations have been discretised according to Sugimoto's
prescription \citep[see][]{SU70} in order to damp instabilities
which develop during the advanced stages of stellar evolution.
We note that the equations are still hydrostatic (no acceleration term) 
as in the pre--supernova models of \citet{LC03}.

\subsection{Nuclear reaction network}
The choice of the nuclear reaction network is a compromise between the
number of chemical elements one wants to follow and the computational cost
(CPU and memory). The network used for hydrogen (H) and helium (He) burnings is
the same as in \citet{ROTX}. For carbon (C), neon (Ne), oxygen (O) and
silicon (Si) burnings, we chose to minimise the computational
cost without losing accuracy for the energy production and
the evolution of the abundance of the main elements. 
For this purpose, the chemical species followed during the
advanced stages are $\alpha$, $\element[][12]{C}$, $^{16}$O,
$^{20}$Ne, $^{24}$Mg, $^{28}$Si, $^{32}$S, $^{36}$Ar, $^{40}$Ca,
$^{44}$Ti, $^{48}$Cr, $^{52}$Fe and $^{56}$Ni. This network 
 is usually called an $\alpha$--chain network. Note that
\citet{THW00} and \citet{HT98} show that even a network of seven
elements is sufficient for this purpose. 

The system of equations describing the changes of the abundances
by the nuclear reactions is resolved by the method of \citet{AT69}.
This method has been chosen because it is very stable and rapid. 
It is therefore suitable to be included in an evolutionary code. 
We ensured that we used small enough
time steps to keep it very accurate. The use of small time steps
ensures on top of it a good treatment of the
interplay between nuclear burning and diffusion since 
these two phenomena
are treated separately (in a serial way) although they occur 
simultaneously. 

The reactions rates are taken from the
NACRE \citep{NACRE} compilation for the experimental reaction rates
and from the NACRE website (http://pntpm.ulb.ac.be/nacre.htm) for the
theoretical ones. 

The nuclear energy production rates 
 are derived from the individual reaction rates for C, Ne and O--burning
 stages. 
 During Si--burning, two quasi--equilibrium groups form around 
$^{28}$Si and $^{56}$Ni respectively. 
\citet{HT98} therefore only follow explicitly 
the reactions between $^{44}$Ti and $^{48}$Cr and assume nuclear
statistical equilibrium between the other elements heavier than 
$^{28}$Si. They choose the reaction between $^{44}$Ti and $^{48}$Cr 
because it is the
bottleneck between the two quasi--equilibrium groups. 
We followed explicitly the abundance evolution of 
the 13
elements cited above. However, for the energy production, we followed
the method of \citet{HT98} during Si--burning.
We therefore only considered the reaction
rate between $^{44}$Ti and $^{48}$Cr and multiply them by the energy
produced by the transformation of $^{28}$Si into $^{56}$Ni.
\subsection{Dynamical shear}
The criterion for stability against dynamical shear instability is the
Richardson criterion:

\begin{equation}
Ri=\frac{N^2}{(\partial U/\partial z)^2} > \frac{1}{4}=Ri_c,
\end{equation} where
$U$ is the horizontal velocity, $z$ the vertical coordinate and $N^2$
the Brunt-V\"ais\"al\"a frequency:
\begin{equation}
N^2=\frac{g \delta}{H_P} [\nabla_{ad}-\nabla + \frac{\varphi}{\delta} \nabla_{\mu}]
\end{equation} where
g is the gravity, $\delta=-\mathrm{\partial\,ln}\rho/\mathrm{\partial\,ln}T)_{\mu,P}$,
 H$_{\rm{P}}$ is the pressure scale height,
$\nabla_{ad}=\mathrm{d\,ln}T/\mathrm{d\,ln}P)_{ad}$,
$\nabla=\mathrm{d\,ln}T/\mathrm{d\,ln}P$,
$\nabla_{\mu}=\mathrm{d\,ln}\mu/\mathrm{d\,ln}P$ and
$\varphi=\mathrm{\partial\,ln}\rho/\mathrm{\partial\,ln}\mu)_{T,P}$.\\

The critical value, $Ri_c=1/4$, corresponds to the situation where the excess
kinetic energy contained in the differentially rotating layers
is equal to the work done against
the restoring force of the density gradient (also called buoyancy force). It is
therefore used by most authors as the limit for the occurrence of the dynamical shear.
However, recent studies by \citet{CA02} show that turbulence may occur as
long as $Ri \lesssim Ri_c \sim 1$. This critical value is consistent with numerical
simulations done by \citet{BH01} where they find shear
mixing for values of $Ri$ greater than $1/4$ (up to about $1.5$).\\

Different dynamical shear diffusion coefficients, $D$, can be found 
in the literature. 
\citet{HLW00} use:
\begin{equation}
D=[\min \{d_{inst},H_P\}(1-\max \{\frac{Ri}{Ri_c},0\})]^2/\tau_{\mathrm{dyn}}
\end{equation} where
$\tau_{\mathrm{dyn}}=\sqrt{r^3/(Gm_r)}$ is the dynamical timescale and $d_{inst}$
the spatial extent of the unstable region, which is limited to one 
H$_{\rm{P}}$.
\\

\citet{BH01} use another formula and
they do numerical simulations to study the dependence of $D$ on $Ri$.
They find the following result:

\begin{equation}
D=\frac{0.6\ 10^{10}}{Ri}
\end{equation}
\subsubsection{The recipe}\label{rcp}
The following dynamical shear coefficient is used, as suggested
 by J.--P. Zahn \citetext{priv.\ comm.}:

\begin{equation}
D=\frac{1}{3}vl
=\frac{1}{3}\ \frac{v}{l}\ l^2
=\frac{1}{3}\ r\frac{\mathrm{d}\,\Omega}{\mathrm{d}\,r} \ \Delta r^2
=\frac{1}{3}\ r\Delta\Omega\ \Delta r
\label{formds}
\end{equation}where $r$ is the mean radius of the zone where the 
instability occurs,
$\Delta\Omega$ is the variation of $\Omega$ over this zone and
$\Delta r$ is the extent of the zone. The zone is the reunion of consecutive shells
where $Ri< Ri_c$. 
This is valid if $P_e>1$, where $P_e$, 
 the Peclet Number, is the ratio of cooling to dynamical timescale of a
 turbulent eddy.
We calculated three $\upsilon_{\rm{ini}}=300$\,km\,s$^{-1}$ 15 $M_{\sun}$
models to see the impact of dynamical shear and the importance of the
value of $Ri_c$ \citep{HMM02}: 
one without dynamical shear, one with $Ri_c=1/4$ and the last one
with $Ri_c=1$. See Sect. \ref{ds} for a discussion of the results.
In the present grid of pre--supernova models, 
the dynamical shear is included with $Ri_c=1/4$.

\subsubsection{Solberg--H\o iland instability}
Solberg--H\o iland stability criterion corresponds to the inclusion of 
the effect of rotation (variation of centrifugal force) in the convective 
stability criterion. It is a
combination of the Ledoux (or possibly Schwarzschild) and the Rayleigh 
criteria \citep{MM00,HLW00}. Both the dynamical shear and 
Solberg--H\o iland instabilities occur in the case of a very large angular
velocity decrease outwards (usual situation in stars, see Fig. 
\ref{omds}). Note that if there is a large increase
outwards, dynamical shear
instability occurs but not the Solberg--H\o iland instability.

Both instabilities, shear instability and Solberg--H\o iland stability, occur on the dynamical
timescale. We therefore expect them to have similar effects. The question is
which instability sets in first?
By comparing the stability criteria of the dynamical shear
and of the Solberg--H\o iland instability:
$$1/4\,(\rm{d}\Omega /\rm{d} r)^2r^2<N^2\ \ \ {\rm dynamical\ shear}$$
$$-2\,\Omega [2\,\Omega + (\rm{d}\Omega /\rm{d} r)\,r] <N^2\ \ \ {\rm Solberg-H\o
iland,}$$ 
\noindent where
$\Omega$ is the angular velocity,
$r$ the radius and
$N^2$ the Brunt-V\"ais\"al\"a frequency, it can be demonstrated that whenever a zone is
unstable towards the Solberg--H\o iland instability, it is also unstable 
towards 
the dynamical shear instability. 
Indeed:
$$1/4\,(\rm{d}\Omega /\rm{d} r)^2r^2 >
-2\,\Omega [2\,\Omega + (\rm{d}\Omega /\rm{d} r)\,r]$$ because
$$1/4\,(\rm{d}\Omega /\rm{d} r)^2r^2 +2\,\Omega [2\,\Omega +
(\rm{d}\Omega /\rm{d} r)\,r]=$$ 
$$1/4[ (\rm{d}\Omega /\rm{d} r)\,r +4\,\Omega]^2>0$$ 
This means that the treatment of the dynamical
shear instability alone is sufficient (since the timescales are similar). 
We therefore did not include explicitly the
Solberg--H\o iland instability in our model.
\subsection{Convection}
Convective diffusion replaces instantaneous convection from oxygen 
burning onwards 
because the mixing timescale becomes longer than the evolution timescale
 at that
point. The numerical method used for this purpose is the method
used for rotational diffusive mixing \citep{dm03}. The 
mixing length theory is used to derive the corresponding diffusion
coefficient. Note that multi--dimensional studies have been started on this
subject \citep{BA98}.

\section{Hertzsprung--Russell (HR)  diagram and lifetimes}
\begin{table*}
\caption{Initial properties and lifetimes of central burning stages of
solar metallicity models. Also given are the total mass and the
different core masses at the end of central silicon burning as
well as at the last time step of our calculations. These last
models correspond approximatively to the end of the first shell
silicon burning for non--rotating models and slightly later then
central silicon burning for rotating ones. All masses are in
solar mass units. Lifetimes are in years with exponent in brackets
(2.14 (-2)=$2.14\ 10^{-2}$). Velocities are in km\,s$^{-1}$.}
\begin{tabular}{l r r r r r r r r r r}
\hline \hline
\multicolumn{11}{c}{Initial model properties} \\
\hline
$M_{\rm{ZAMS}}$    &   15        &   15        &   20        &   20        &  25         &   25        &   40        &   40        &   60        &  60         \\
$\upsilon_{\rm{ZAMS}}$    &   0         &   300       &   0         &   300       &   0         &   300       &   0         &   300       &   0         &   300       \\
\hline \\
\multicolumn{11}{c}{Lifetime of burning stages} \\
\hline
$\tau_{\rm{H}}$    &   1.13 (7) &   1.43 (7) &   7.95 (6) &   1.01 (7) &   6.55 (6) &   7.97 (6) &   4.56 (6) &   5.53 (6) &   3.62 (6) &   4.30 (6) \\
$\tau_{\rm{He}}$   &   1.34 (6) &   1.13 (6) &   8.75 (5) &   7.98 (5) &   6.85 (5) &   6.20 (5) &   4.83 (5) &   4.24 (5) &   3.85 (5) &   3.71 (5) \\
$\tau_{\rm{C}}$    &   3.92 (3) &   1.56 (3) &   9.56 (2) &   2.82 (2) &   3.17 (2) &   1.73 (2) &   4.17 (1) &   8.53 (1) &   5.19 (1) &   5.32 (1) \\
$\tau_{\rm{Ne}}$   &   3.08     &   0.359     &  0.193      &  8.81 (-2)  &  0.882      &  0.441      &  4.45 (-2)  &  6.74 (-2)  &  4.04 (-2)  &  4.15 (-2)  \\
$\tau_{\rm{O}}$    &   2.43     &   0.957     &  0.476      &  0.132  &  0.318      &  0.244      &  5.98 (-2)  &  0.176  &  5.71 (-2)  &  7.74 (-2)  \\
$\tau_{\rm{Si}}$   &   2.14 (-2) &   8.74 (-3) &   9.52 (-3) &   2.73 (-3) &   3.34 (-3) &   2.15 (-3) &   1.93 (-3) &   2.08 (-3) &   1.95 (-3) &   2.42 (-3) \\
\hline \\
\multicolumn{11}{c}{End of central silicon burning} \\
\hline
$M_{\rm{total}}$   &   13.232    &  10.316     &  15.694     &   8.763     &  16.002     &  10.042     &  13.967     &  12.646     &  14.524     &  14.574     \\
$M_{\alpha}^{75}$  &    4.211    &   5.677     &   6.265     &   8.654     &   8.498     &  10.042     &  13.967     &  12.646     &  14.524     &  14.574     \\
$M_{\rm{CO}}^{\rm{int}}$&    2.441    &   3.756     &   4.134     &   6.590     &   6.272     &   8.630     &  12.699     &  11.989     &  13.891     &  13.955     \\
$M_{\rm{CO}}^{01}$ &    2.302    &   3.325     &   3.840     &   5.864     &   5.834     &   7.339     &  10.763     &   9.453     &  11.411     &  11.506     \\
$M_{\rm{Si}}^{50}$ &    1.561    &   2.036     &   1.622     &   2.245     &   1.986     &   2.345     &   2.594     &   2.212     &   2.580     &   2.448     \\
$M_{\rm{Fe}}^{50}$ &    1.105    &   1.290     &   1.110     &   1.266     &   1.271     &   1.407     &   1.464     &   1.284     &   1.458     &   1.409     \\
\hline \\
\multicolumn{11}{c}{Last model} \\
\hline
$M_{\rm{Si}}^{50}$ &    1.842    &   2.050     &   2.002     &   2.244     &   2.577     &   2.894     &   2.595     &   2.868     &   2.580     &   2.448     \\
$M_{\rm{Fe}}^{50}$ &    1.514    &   1.300     &   1.752     &   1.260     &   1.985     &   1.405     &   2.586     &   1.286     &   2.440     &   1.409     \\
\hline
\end{tabular}
\label{table1}
\end{table*}
Stellar models of
12, 15, 20, 25, 40 and 60 $M_{\sun}$ at solar metallicity, with initial rotational velocities 
of 0 and 300 km\,s$^{-1}$ respectively have been computed. The value of the initial velocity
corresponds to an average velocity of about 220\,km\,s$^{-1}$ on the Main
Sequence (MS) which is
very close to the observed average value \citep[see for instance][]{FU82}. 
The calculations start at
the ZAMS for the 12, 15, 20 and 25 $M_{\sun}$ models and at the end of central He--burning
for the 40 and 60 $M_{\sun}$ models \citep[for these models, we take over
the calculations done by][]{ROTX}.
The calculations reach the end of central Si--burning 
with models of rotating stars and the end of shell
Si--burning with models of non--rotating stars. 
For the non--rotating 12 $M_{\sun}$ star,
Ne--burning starts at
a fraction of a solar mass from the centre but does not reach the centre and the 
calculations stop there. For the rotating 12 $M_{\sun}$ star, the model stops after
O--burning.  

The major characteristics of the models are summarised in
Tables \ref{table1} and \ref{table1b}. In order to calculate lifetimes of the central burning stages, 
we take the start of a burning
stage when $0.003$ in mass fraction of the main burning fuel is burnt. 
We consider that a burning
stage is finished when the main fuel mass fraction drops below
$10^{-5}$. The results would be the same if we had chosen $10^{-4}$ or $10^{-6}$. 
Neon burning is an exception because neon abundance does not drop
significantly before the end of oxygen burning. We therefore consider the end of Ne--burning
when its abundance drops below $10^{-3}$. 
Therfore the lifetimes for Ne--burning 
are to be considered
as estimates. 
Other authors use the duration of the convective core as the lifetime. 
We note that using the
duration of convective cores as central burning lifetimes instead of
threshold values of the central abundance of the main fuel would yield 
results very
similar to those given in Table \ref{table1}
for H, He, O and Si--burning stages. 
The core sizes are given at the end of central silicon--burning 
and at the last model calculated (which corresponds to a different evolutionary
stage in the non--rotating and the rotating models as seen above).
The inner limit of each core is the star centre. The
outer limit is the point in mass where the sum of the mass 
fraction of the main burning 
products  
(helium for $M_{\alpha}$, carbon and oxygen for $M_{\rm{CO}}$,
 $^{28}$Si--$^{44}$Ti for $M_{\rm{Si}}$ and 
$^{48}$Cr--$^{56}$Ni for $M_{\rm{Fe}}$) becomes less than $0.75$ (superscript 75)
or $0.50$ (superscript 50). 
Another possibility to define the outer limit of a core is to consider
the lagrangian mass where the mass fraction of the main fuel (helium for
the CO cores) drops below $10^{-2}$. The CO cores thus obtained are
given in Tables 1 and 2 (superscript 01).
These limits are suitable for most masses 
(see Fig. \ref{a20neosi}). However, for very massive stars (see
Fig. \ref{a60s3endsi}), shell He--burning transforms most helium 
into carbon and oxygen and one could also consider that $M_{\rm{CO}}$ includes 
the whole star. In that case we suggest another
definition of $M_{\rm{CO}}$, which we name $M_{\rm{CO}}^{\rm{int}}$, defined by 
$M_{\rm{CO}}^{\rm{int}}= M_{\rm{CO}}^{01} + \int_{M_{\rm{CO}}^{01}}^{M_{\rm{\alpha}}^{01}}
X_{\rm{CO}}\ dm$,
 where $X_{\rm{CO}}$ is the sum of $^{12}$C and $^{16}$O mass fractions.
This definition gives an intermediate value between $M_{\rm{CO}}^{01}$ and the total
actual mass of the star.
\begin{table}
\centering
\caption{Same as Table \ref{table1} for the 12 $M_{\sun}$ models.
The non--rotating model starts Ne--burning off--centre and
the burning never reaches the centre. 
The unburnt Ne--O core, is given by $M_{\rm{Ne-O}}$. 
}
\begin{tabular}{l r r }
\hline \hline
\multicolumn{3}{c}{Initial model properties} \\
\hline 
$M_{\rm{ZAMS}}$    &   12        &   12    \\
$\upsilon_{\rm{ZAMS}}$    &   0         &   300    \\
\hline 
\multicolumn{3}{c}{Lifetime of burning stages} \\
\hline 
$\tau_{\rm{H}}$    &   1.56 (7) &   2.01 (7) \\
$\tau_{\rm{He}}$   &   2.08 (6) &   1.58 (6) \\
$\tau_{\rm{C}}$    &   6.47 (3) &   6.09 (3) \\
$\tau_{\rm{Ne}}$   &    -       &   1.138    \\
$\tau_{\rm{O}}$    &    -       &   4.346    \\
\hline 
\multicolumn{3}{c}{End of calculation} \\
\hline 
$M_{\rm{total}}$   &   11.524      &  10.199 \\
$M_{\alpha}^{75}$  &    3.141      &   3.877 \\
$M_{\rm{CO}}^{\rm{int}}$&    1.803      &   2.258 \\
$M_{\rm{CO}}^{01}$ &    1.723      &   2.077 \\
$M_{\rm{Si}}^{50}$ &    0.805      &   1.340 \\
$M_{\rm{Ne-O}}$    &    0.096      &   -     \\
\hline
\end{tabular}
\label{table1b}
\end{table}

\subsection{Hertzsprung--Russell (HR) diagram}
\begin{figure}[!tbp]
\centering
   \includegraphics[width=8.5cm]{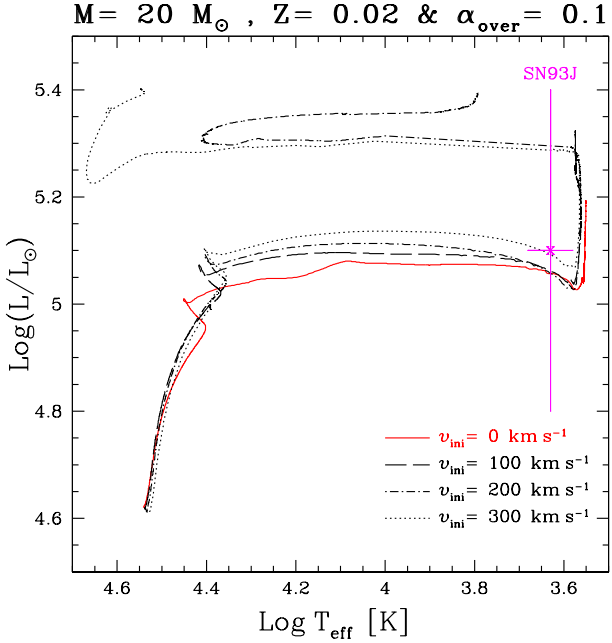}
\caption{HR--diagram for 20 $M_{\sun}$ models: solid, dashed, dotted-dashed and dotted lines correspond 
respectively to
$v_{\rm{ini}}$= 0, 100, 200 and 300 km\,s$^{-1}$.
We also indicate the position of the progenitor of SN1993J.}
\label{hr20}
\end{figure}
\begin{figure}[!tbp]
\centering
   \includegraphics[width=8.8cm]{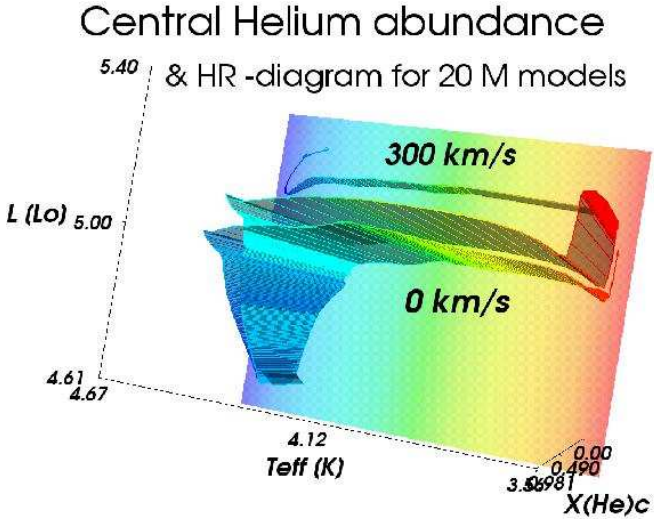}
\caption{3D HR diagram with central helium mass fraction as the third
dimension for non--rotating and rotating 20 $M_{\sun}$ models.}
\label{hr3d}
\end{figure}
\begin{figure}[!tbp]
\centering
   \includegraphics[width=8.8cm]{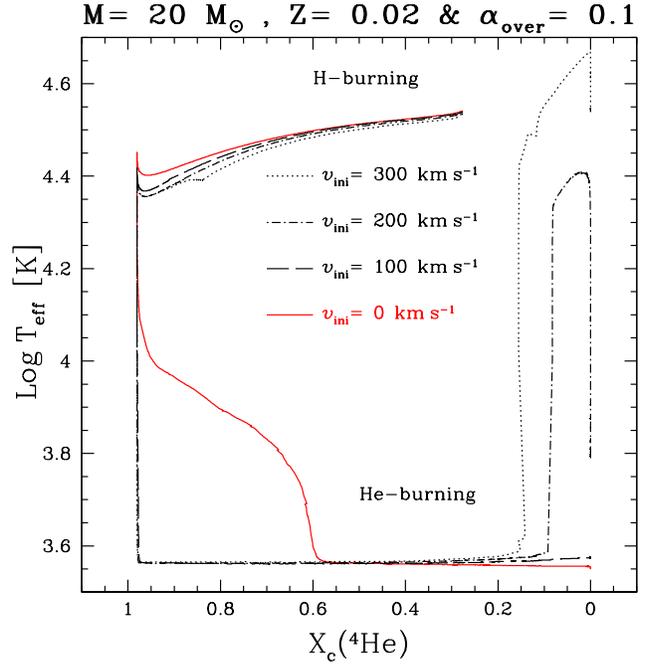}
\caption{$T_{\rm{eff}}$ vs central helium mass fraction for 20 $M_{\sun}$ 
models: solid, dashed, dotted-dashed and dotted lines correspond 
respectively to $v_{\rm{ini}}$= 0, 100, 200 and 300 km\,s$^{-1}$.}
\label{teffhem20}
\end{figure}
The models calculated in the present work follow the same 
tracks as the models from
\citet{ROTX}. This is expected since
the only difference between the two sets of models is the 
inclusion of dynamical shear in the present models. 
Here we concentrate on the 20 $M_{\sun}$ 
models. For that purpose, we also calculated 20 $M_{\sun}$ models with initial rotation 
velocities of 100 and 200 km\,s$^{-1}$ \citep{HMMG03}. The tracks of 
the 20 $M_{\sun}$ models are
presented in Figs. \ref{hr20}, \ref{hr3d} and \ref{teffhem20}. 
Figure  
\ref{hr20}  shows the evolutionary tracks of the four
different   20\,$M_{\odot}$   stars  in  the  HR--diagram.  The
non--rotating model ends up as a red supergiant (RSG) like the model of 
other groups \citep[see][]{HL002,LSC00}. However, the
rotating  models  show  very  interesting features. Although the
100\,km\,s$^{-1}$  model remains  a  RSG,  the  200\,km\,s$^{-1}$  
model undergoes  a  blue  loop to finish as a yellow--red supergiant
whereas  the  300\,km\,s$^{-1}$  model  ends  up as a blue supergiant 
(BSG).  Thus rotation may have a strong impact on the nature
of the supernova progenitor (red, blue supergiant or even Wolf--Rayet star)
and thus on some observed characteristics of the supernova explosion.
For instance the shock wave travel time
through  the  envelope is proportional to the radius of the star. 
Since RSG radii are about hundred times BSG ones, this travel time may
differ by two orders of magnitude depending on the initial rotational velocity.

When does the star
evolve back to the blue after a RSG phase and why ?
Figure \ref{hr3d} is a 3D plot of the HR--diagram 
(in the plane) and the extra dimension represents the central helium 
mass fraction, $X_{\rm{c}}(^4\rm{He})$. The extra
dimension allows us to follow H, He and post He--burnings within a
same diagram. Indeed, $X_{\rm{c}}(^4\rm{He})$
increases during the main sequence, then decreases during
He--burning and finally is equal to zero during the post He--burning
evolution. We can see that:
\begin{itemize}
\item For the non--rotating model, He--burning 
starts when the star crosses the HR--diagram (Log\,$T_{\rm{eff}} \sim 4$) and the
star only reaches the RSG stage halfway through He--burning. Finally, 
the star luminosity rises during shell He--burning.

\item For the $\upsilon_{\rm{ini}}=300$\,km\,s$^{-1}$ model, the star 
is more luminous and
becomes a RSG before He--burning ignition. 
These two factors favour higher mass loss rates and
the star loses most of its
hydrogen envelope before
He--burning is finished. Thus the star
evolves towards the zone of the HR diagram where homogeneous helium stars are found,
{\it i.e.} in the blue part of the HR diagram.
We can see that the star track still evolves 
during shell He--burning. 
\end{itemize} 

Figure \ref{teffhem20} is a projection of Fig. \ref{hr3d} in the
Log\,$T_{\rm{eff}}$ versus $X_{\rm{c}}(^4\rm{He})$ plane. Although less intuitive
than the 3D plot, it is more quantitative and still allows us to follow
the various burning stages described above. Figure \ref{teffhem20}
shows that all the rotating models become RSG before the beginning of the He--burning phase. The  
100\,km\,s$^{-1}$ model luminosity is lower than for the
300\,km\,s$^{-1}$ model and therefore less mass is lost during
He--burning and the burning ends before the hydrogen envelope is
removed. The star therefore remains a RSG. The 200\,km\,s$^{-1}$ model
evolution is similar to the 300\,km\,s$^{-1}$ model but the extent of
its blue loop is smaller. At the end of He--burning for the 200\,km\,s$^{-1}$ model, 
Log\,$T_{\rm{eff}}=4.28$ and
the star becomes redder before C--burning starts.

Although the models discussed here are for solar metallicity, one can
note that the behaviours of the models with
$\upsilon_{\rm{ini}}$  between 200 and 300\,km\,s$^{-1}$ are reminiscent
of  the  evolution  of  the progenitor of SN1987A. Let us recall
that this supernova had a blue progenitor which evolved from a RSG stage
\citep[see e.g. the review by][]{Ar89}. In Fig. \ref{hr20}, we also
indicate the position of the progenitor of SN 1993J. SN 1993J probably 
belongs to a binary system \citep{P93J}. 
Nevertheless it has
common points with our 
$\upsilon_{\rm{ini}}=$200\,km\,s$^{-1}$ 20 $M_{\sun}$ model: the star model
and the progenitor of SN 1993J have
approximately the same metallicity, 
they have a similar position 
in the HR--diagram taking into account the
uncertainties 
and they both have a small hydrogen rich envelope,
making possible a change from type II to type Ib some time after the
explosion.
\subsection{Lifetimes}\label{ltau}
\begin{figure}[!tbp]
\centering
\includegraphics[width=8.8cm]{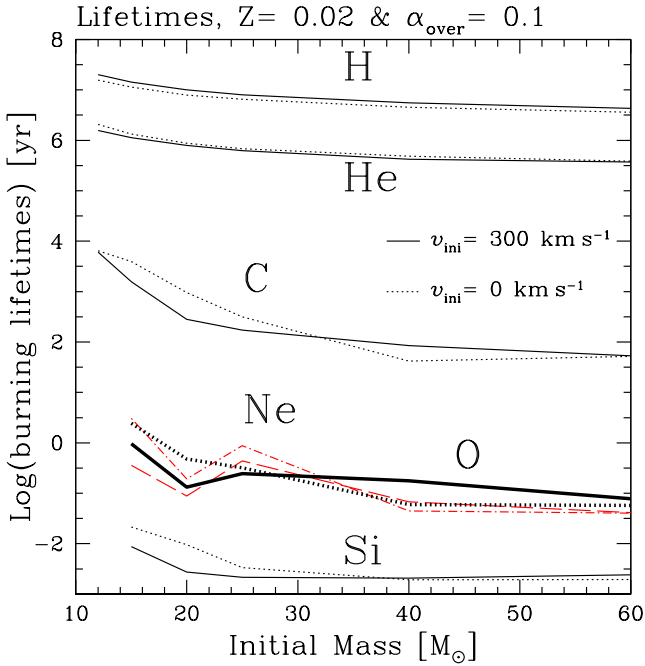}
\caption{Burning lifetimes as a function of the initial mass and 
velocity. Solid and dotted lines correspond respectively to rotating and
non--rotating models. Long--dashed and dotted--dashed lines are used for
rotating and non--rotating Ne--burning lifetimes to point out that they
are to be considered as estimates (see text).}
\label{tau}
\end{figure}
The lifetimes are presented in Tables \ref{table1} and \ref{table1b} 
and plotted in Fig. \ref{tau}. 
We focus here our discussion on the effects of rotation on the lifetimes of the
advanced burning phases. A discussion of the earlier stellar evolutionary phases
can be found in previous papers \citep{ROTX,HLW00}.
For C--burning onwards, we have two patterns:
\begin{description}
\item[$M \lesssim 30 M_{\sun}$:] Since the He--burning temperature
is higher in rotating stars, the $\frac{\rm{C}}{\rm{O}}$ ratio is 
smaller at the end of He--burning and therefore the C--burning lifetimes 
are shorter. If C--burning is less important, less neon is produced 
and neon burning is also
shorter. The trends for O-- and Si--burnings are similar. 
\item[$M \gtrsim 30 M_{\sun}$:]  The rotating stars become more 
rapidly WR stars and are more eroded by winds. 
The central temperatures for rotating models are 
therefore equal
or even smaller than for non-rotating models. This leads to higher
$\frac{\rm{C}}{\rm{O}}$ ratios, longer C-- and Ne--burnings phases. 
\end{description}
These two groups correspond to mass ranges where rotational mixing 
($M \lesssim 30 M_{\sun}$) or
mass loss ($M \gtrsim 30 M_{\sun}$) dominates the other process. 

\section{Rotation evolution}
\subsection{Dynamical shear}\label{ds}
\begin{figure}[!tbp]
\centering
\includegraphics[width=8.8cm]{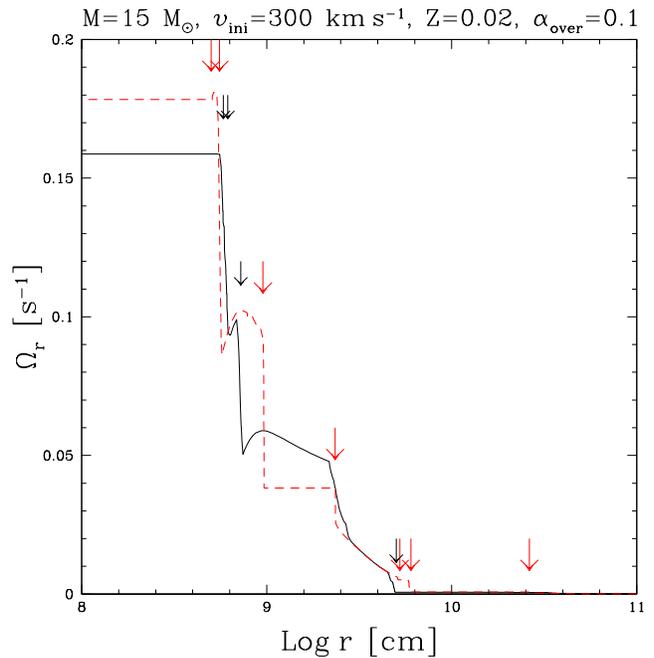}
\caption{$\Omega$ variations as a function of the radius inside 15 
$M_{\sun}$ models: the dashed line is a profile from a model without 
dynamical shear and 
the solid line from a model with dynamical shear and $Ri_c=1$ during core O--burning. 
The long and short
arrows indicate the zones where $Ri<1$ in the model without and with 
dynamical shear respectively. Note that 
the profiles do not differ significantly.}
\label{omds}
\end{figure}
As said in Sect. \ref{rcp}, we calculated three 
 $\upsilon_{\rm{ini}}=300$\,km\,s$^{-1}$ 15 $M_{\sun}$
models to see the impact of dynamical shear and the importance of the
value of $Ri_c$: 
one model without dynamical shear, one with $Ri_c=1/4$ and the last one
with $Ri_c=1$.
In Fig.~\ref{omds}, the variation of the angular velocity, $\Omega$, 
as a function of the radius is shown
inside 15 $M_\odot$ stellar models in the core O--burning phase. Arrows indicate the zones
which are unstable against dynamical shear instability. These zones remain unstable during the whole
post core He--burning phase.
Our simulations show that the characteristic timescale of the dynamical 
shear ($\propto R^2/D$) is always very short
when using
 Eq. (\ref{formds}) for the dynamical shear diffusion coefficient. 
Indeed, we obtain diffusion coefficients between $10^{12}$
and $10^{14}$ cm$^2$ sec$^{-1}$. This is in general one or two orders of 
magnitude larger than using
the expressions given by  
\citet{BH01} or \citet{HLW00}. 
However, the extent of the unstable zones is very small, a few thousandths 
of $M_{\sun}$. Therefore the shear mainly smoothens
the sharp $\Omega$-gradients as can be seen in Fig. \ref{omds} but does 
not transport 
angular momentum or chemical species over long distances. 
The general structure 
and the convective
zones are similar between the model without dynamical shear and the one 
with dynamical shear.

Concerning the Richardson criterion, there is no significant difference 
between the models using $Ri_c=1/4$ and $Ri_c=1$. 
Except for the 15 $M_\odot$ model discussed in this subsection, all the other
models were computed with $Ri_c=1/4$.

\subsection{Angular velocity, $\Omega$, and momentum
evolution}\label{omjevo}
\begin{figure}[!tbp]
\centering
   \includegraphics[width=8.8cm]{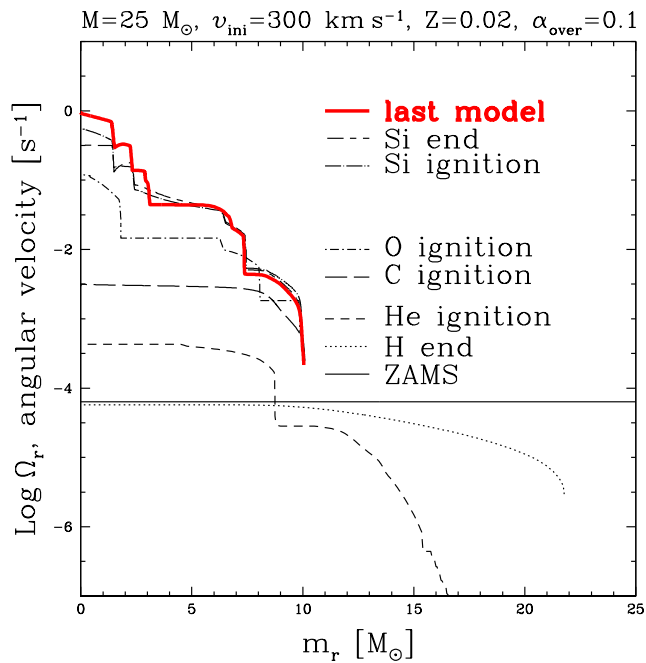}
\caption{Angular velocity as a function of the lagrangian mass
coordinate, $m_r$ inside the 25 $M_{\sun}$ model ($v_{\rm{ini}}$=
300 km\,s$^{-1}$) at various evolutionary stages.
}
\label{omevo}
\end{figure}
\begin{figure}[!tbp]
\centering
   \includegraphics[width=8.8cm]{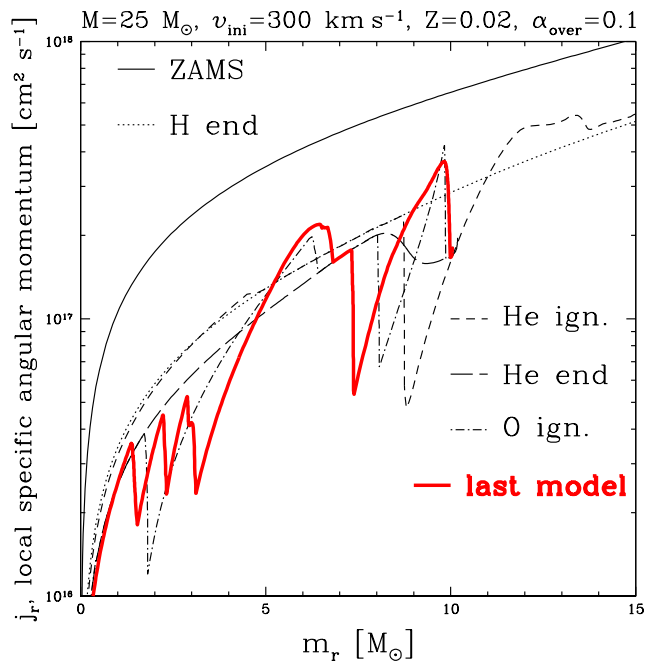}
\caption{Local specific angular momentum profiles for the 25 $M_{\sun}$ model ($v_{\rm{ini}}$=
300 km\,s$^{-1}$) at different evolutionary stages.
}
\label{jevo}
\end{figure}

Figure \ref{omevo} shows the evolution of $\Omega$ inside the
25 $M_\odot$ model from the ZAMS until the end of the core Si--burning phase.
The evolution of $\Omega$ results from many different processes:
convection enforces solid body rotation,
contraction and expansion respectively increases and 
decreases $\Omega$ in order to conserve angular momentum,
shear (dynamical and secular) erodes
$\Omega$--gradients while meridional circulation may erode or build 
them up and finally
mass loss may remove angular momentum from the surface.
If during the core H--burning phase, all these processes may be important,
from the end of the MS phase onwards, the evolution
of $\Omega$ is mainly determined by convection, the local conservation of the angular momentum
and, for the most massive stars only during the core He--burning phase, by mass loss.

During the MS phase, $\Omega$ decreases in the whole star. 
When the star becomes a red supergiant (RSG), $\Omega$ at the surface
decreases significantly due to the expansion of the outer layers. Note that the envelope
is gradually lost by winds in the 25 $M_{\sun}$ model. 
In the centre, $\Omega$ significantly increases when the core contracts
and then the $\Omega$ profile flattens due to convection. $\Omega$ reaches values of the 
order of $1\,s^{-1}$ at the end of Si--burning. It never
reaches the local break--up angular velocity limit, $\Omega_c$, although, when
local conservation holds, $\Omega_r/\Omega_c \propto r^{-1/2}$.

 Figure \ref{jevo} shows the evolution of the specific angular momentum,
$j_r=2/3\, \Omega_r  r^2$, in the central region of a 25 $M_\odot$ stellar model. 
The specific angular momentum remains constant under the effect
of pure contraction or expansion, but varies when transport mechanisms are active.
One sees that the transport processes remove angular momentum from the
central regions. Most of the removal occurs during
the core H--burning phase.
Still some decrease occurs during the core He--burning phase, then the evolution is mostly
governed by convection, which 
transports the angular momentum from the inner part of 
a convective zone to
the outer part of the same convective zone. 
This produces the teeth seen in Fig.~\ref{jevo}.
The 
angular momentum of the star at the end of Si--burning is 
essentially the same as at the end of He--burning (by end of
He--burning, we mean the time when the central helium mass fraction
becomes less than $10^{-5}$). 
This result is very similar to the conclusions of \citet{HLW00} on 
this issue. They
find that the angular momentum profile does not vary substantially 
after C--burning ignition (see Sect. \ref{compam} for a comparison).
It means that 
we can estimate the pre--supernova angular momentum by 
looking at its value  at the end of He--burning. 
We calculated, for the 25 $M_{\sun}$ model, the angular
momentum of its remnant (fixing the remnant mass to 3 $M_{\sun}$). 
We obtained 
${\cal L}_{\rm rem}=2.15\,10^{50}$g\,cm$^2$\,s$^{-1}$
at the end of He--burning and 
${\cal L}_{\rm rem}=1.63\,10^{50}$g\,cm$^2$\,s$^{-1}$
at the end of Si--burning. This corresponds to a loss of only 24\%. 
In comparison, the angular momentum
is decreased by a factor $\sim$5 between the ZAMS and the end of
He--burning. This shows the importance of correctly treating the 
transport of angular momentum during the Main Sequence phase.

\section{Internal structure evolution}
\subsection{Central evolution}
\begin{figure*}[!tbp]
\centering
\includegraphics[width=8.8cm]{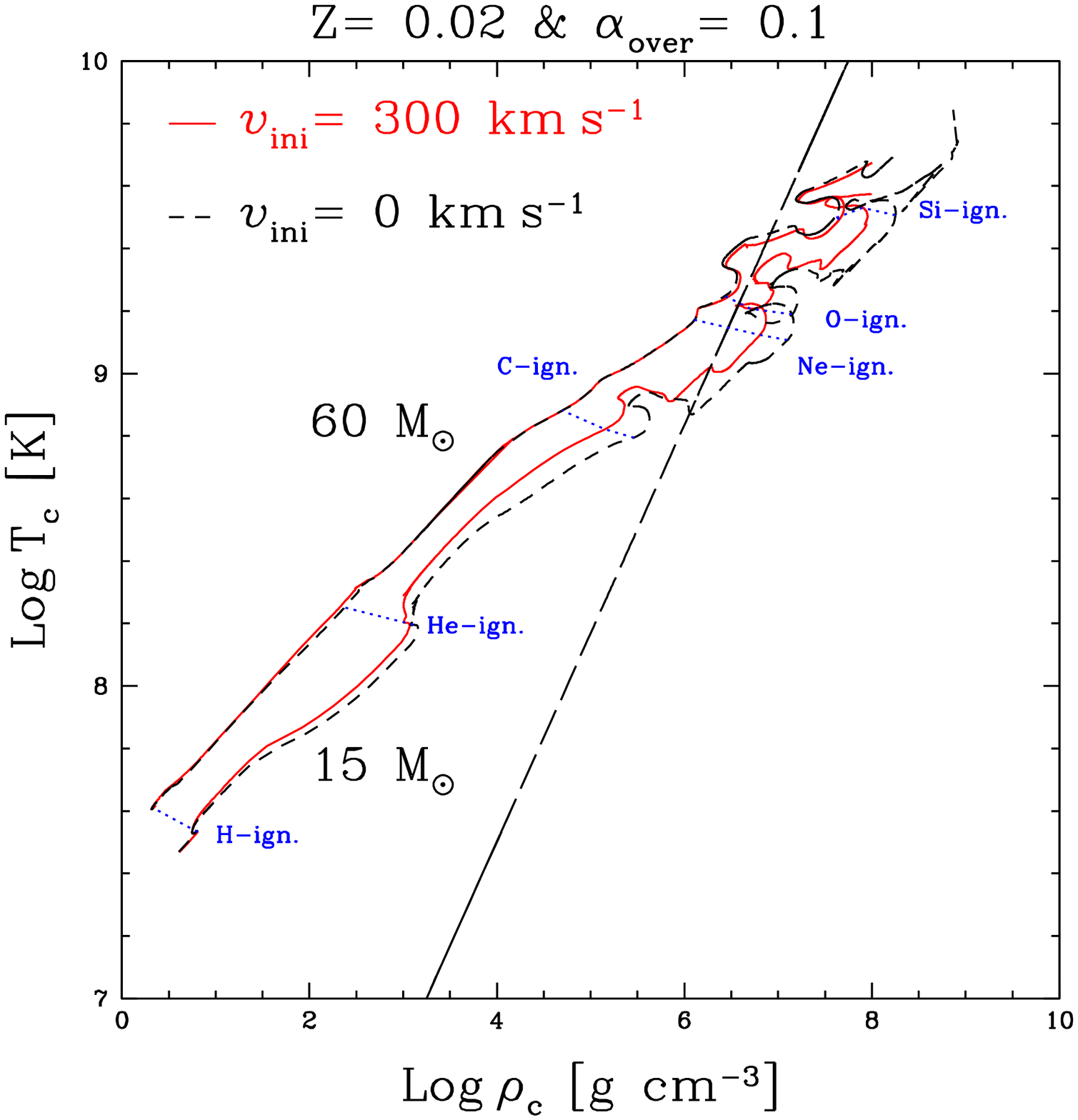}\includegraphics[width=8.8cm]{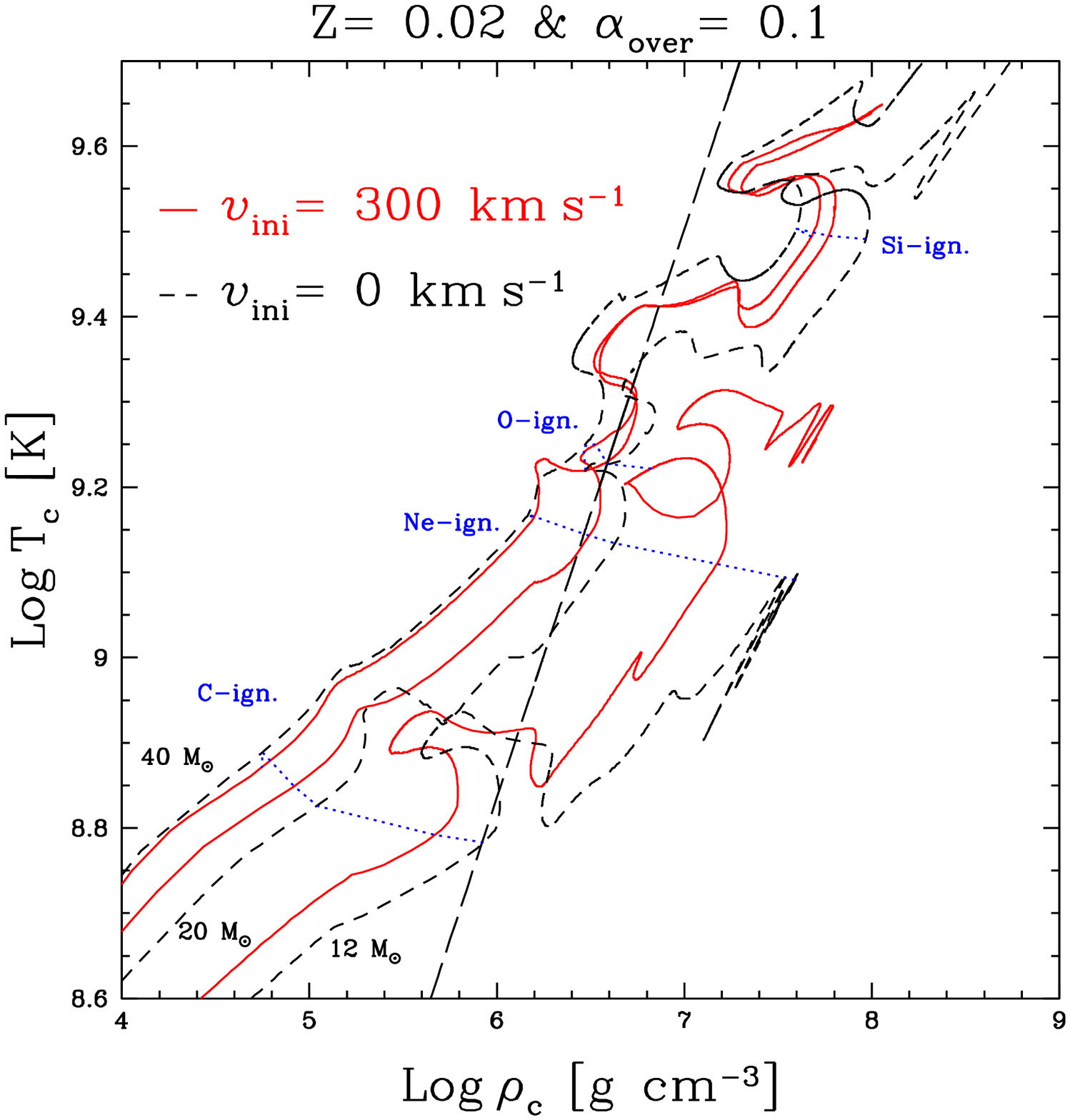}
\caption{Log\,$T_{\rm{c}}$  vs Log\,$\rho_{\rm{c}}$ diagrams: 
Left: evolutionary tracks for the 15 and 60 $M_{\sun}$ models.
Right: evolutionary tracks zoomed in the advanced stages for the 12, 20 and 40 $M_{\sun}$ models.
Solid lines are rotating models and dashed lines are non--rotating models. 
The ignition points of every burning stage are connected with dotted lines.
The additional long dashed line
corresponds  to the limit between non--degenerate and degenerate
electron
gas ($P^{\rm{el}}_{\rm{perfect\,gas}}=P^{\rm{el}}_{\rm{degenerate\,gas}}$).
}
\label{trsl}
\end{figure*}

\begin{figure*}[!tbp]
\centering
   \includegraphics[width=8.8cm]{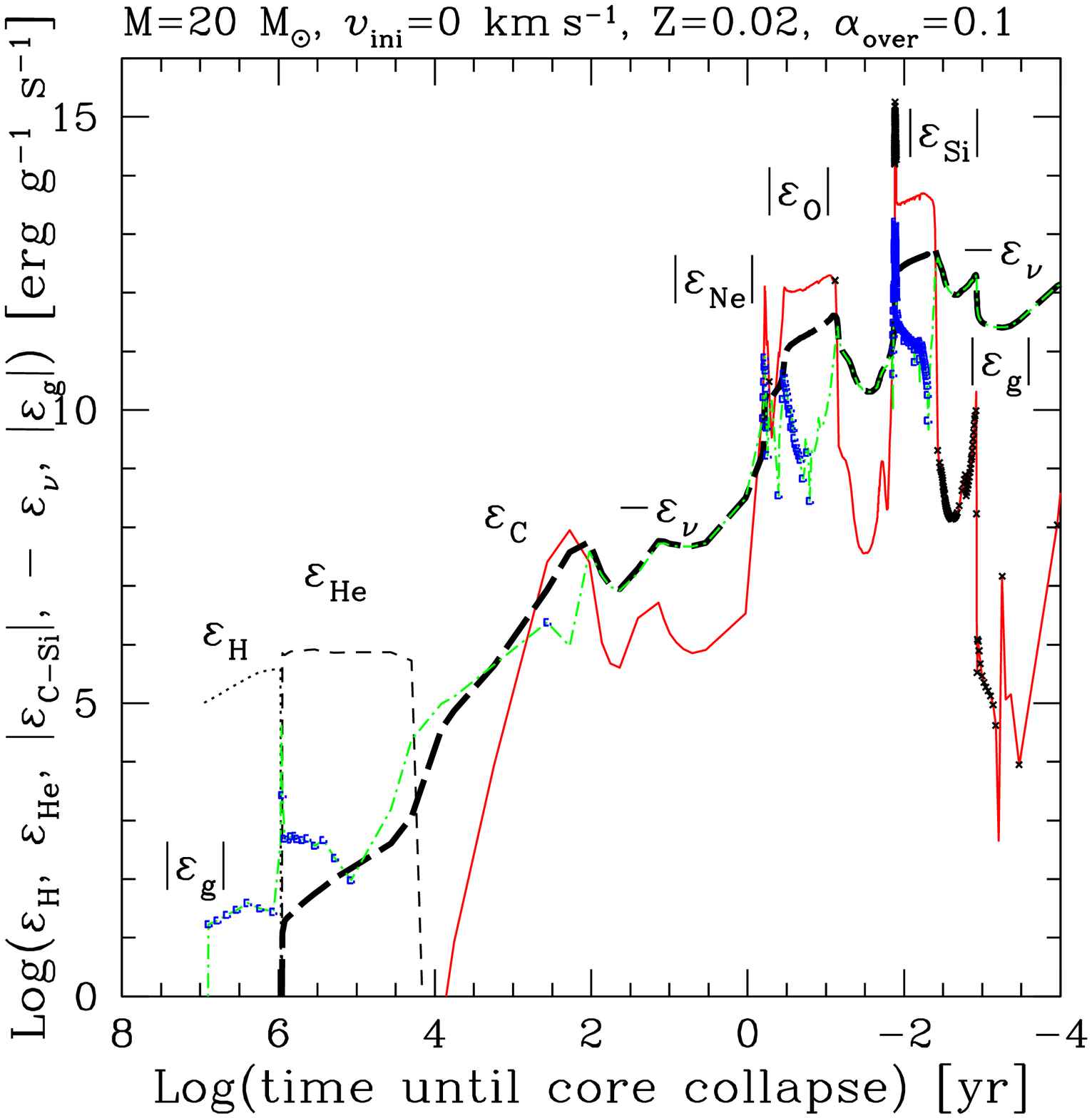}\includegraphics[width=8.8cm]{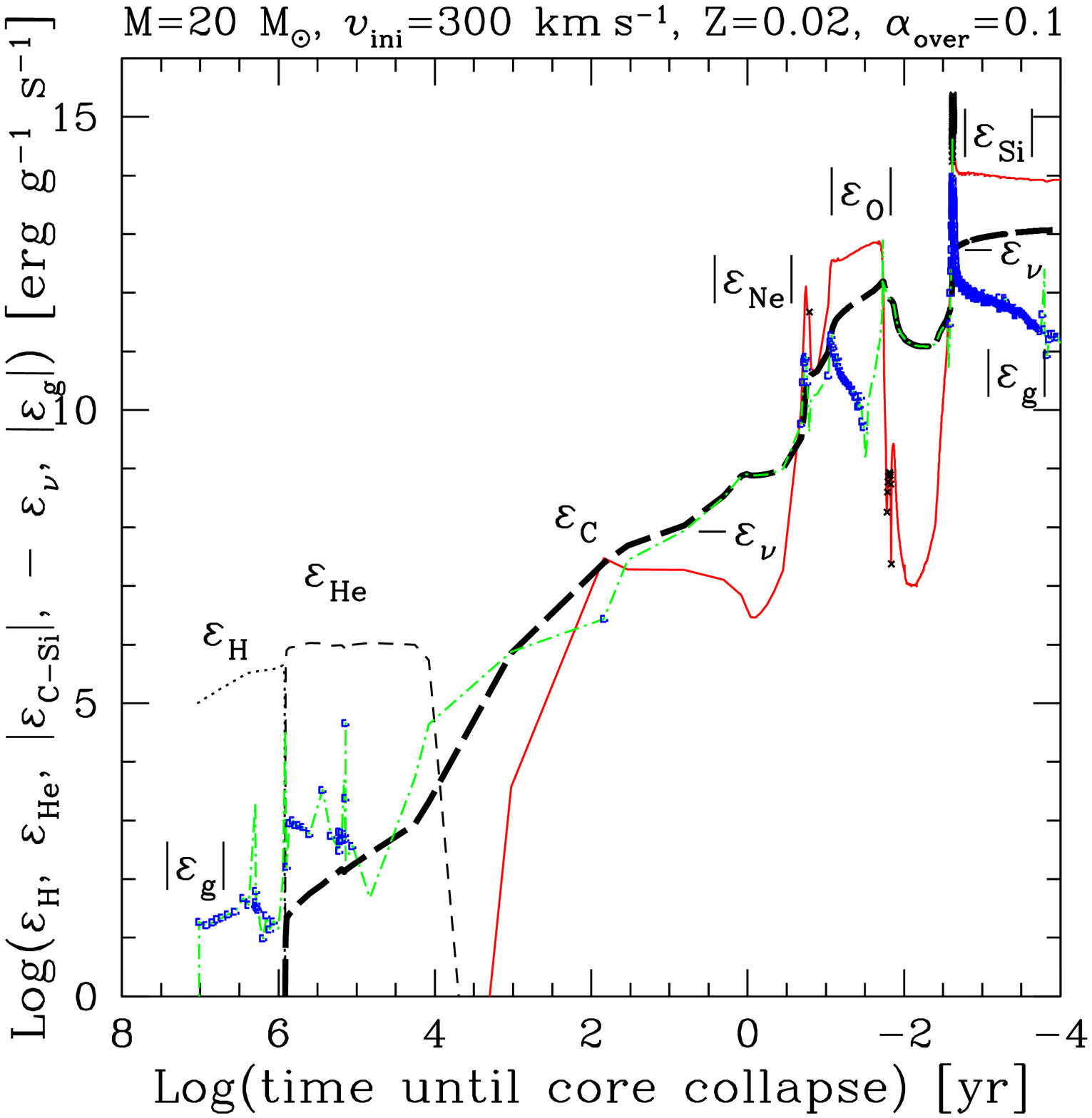}
\caption{Log of the energy production rate per unit mass at the star
center as a function of the time left until core
collapse for the non--rotating (left) and rotating (right) 20 $M_{\sun}$ models.
Nuclear energy production rates during H-- and He--burnings 
 are shown in dotted ($\varepsilon_H$) 
 and dashed ($\varepsilon_{He}$) 
lines respectively.
The solid line corresponds to the nuclear energy production rate in absolute value
 during the advanced  stages ($\varepsilon_C$--$\varepsilon_{Si}$). 
 Black crosses are drawn on top of the line
 whenever the energy production rate is negative. The thick long dashed
line is the energy loss rates due to neutrinos multiplied by -1 
($- \varepsilon_{\nu}$). Finally the
gravitational energy production rate in absolute value is plotted in the 
dotted--dashed line ($\varepsilon_g$). Blue squares are plotted on top 
when this energy is negative.
Note that negative gravitational energy production corresponds to an expansion.
}
\label{ent}
\end{figure*}

Figure \ref{trsl} (left) shows the tracks of the 
15 and 60 $M_{\sun}$ models
throughout their evolution 
in
the central temperature versus central
density plane (Log\,$T_{\rm{c}}$--Log\,$\rho_{\rm{c}}$ diagram).
Figure \ref{trsl} (right) zooms in 
the advanced 
stages of the 12, 20 and 40 $M_{\sun}$ models. It is also very 
instructive to look at Kippenhahn diagrams 
(Figs. \ref{dhr5m121520} and \ref{dhr5m2546}) in order 
to follow the evolution of the structure. Figure \ref{ent} helps understand 
the cause of the movements in the
Log\,$T_{\rm{c}}$--Log\,$\rho_{\rm{c}}$ diagram. 
\begin{figure*}[!tbp]
\centering
   \includegraphics[width=8.8cm]{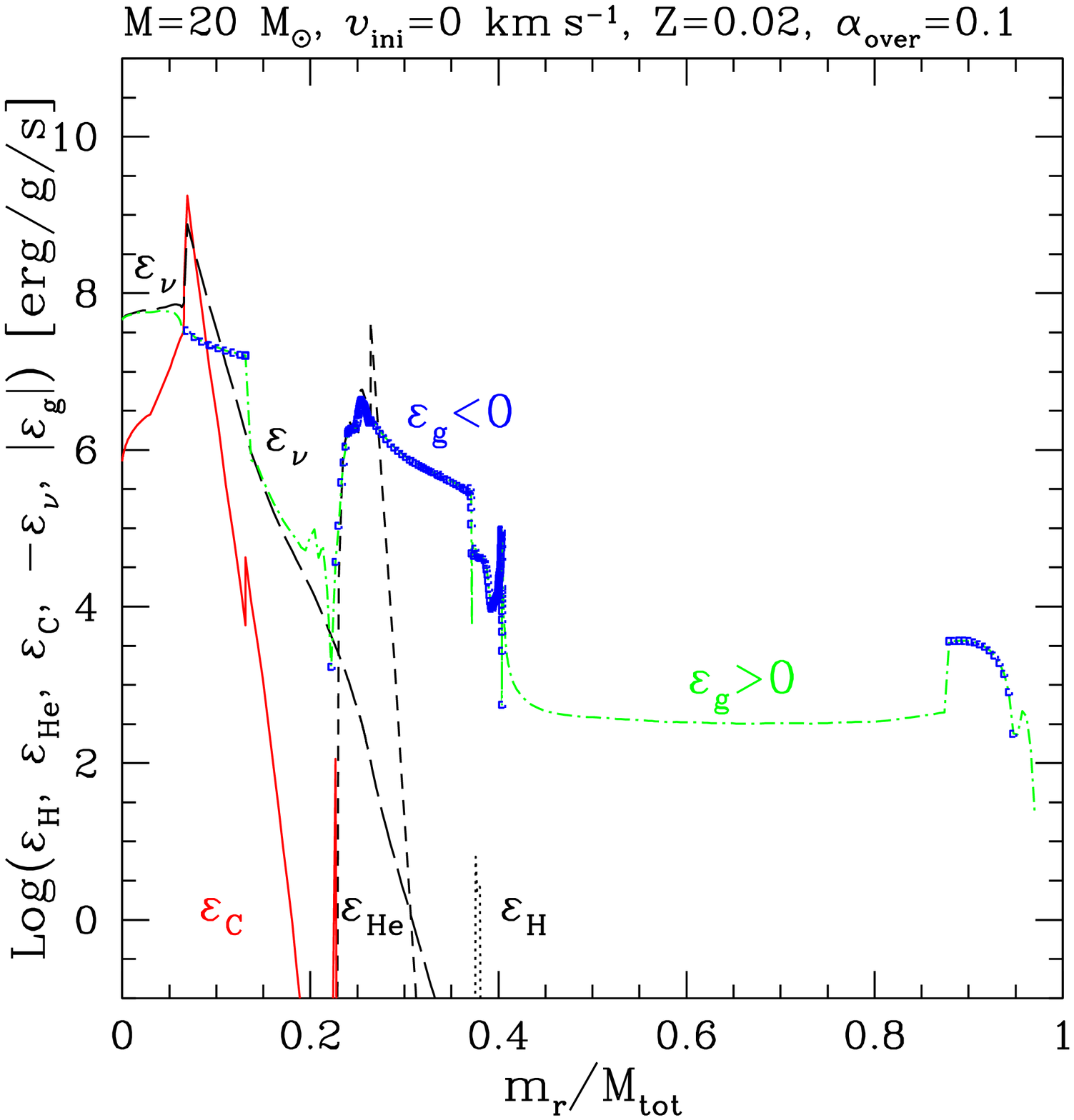}\includegraphics[width=8.8cm]{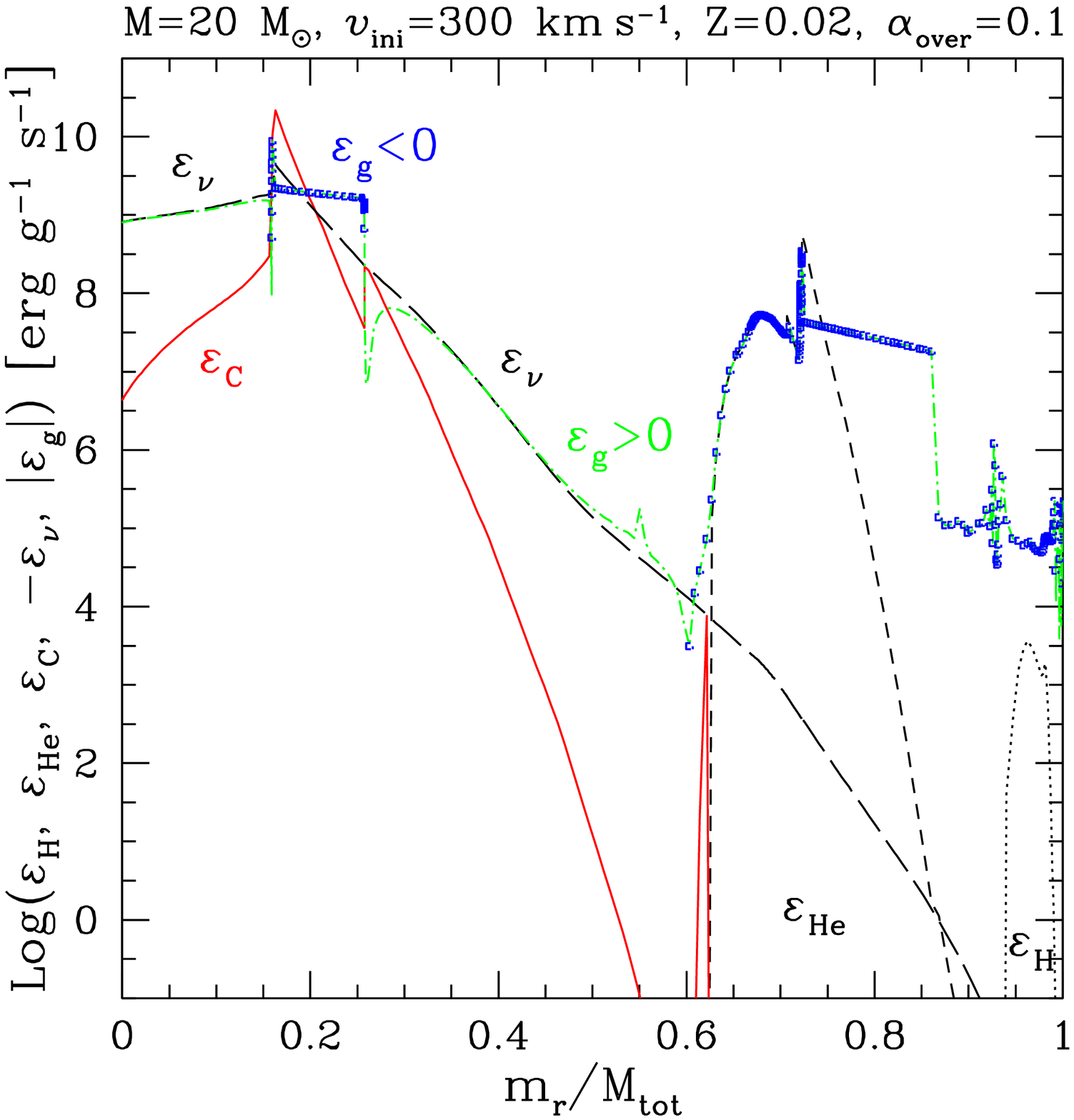}
\caption{Log of the energy production rate per unit mass as a function of 
$m_r/M_{tot}$ during shell C--burning for the non--rotating
 (left) and rotating (right) 20 $M_{\sun}$ models.
The solid line corresponds to the nuclear energy in absolute value
 during C--burning ($\varepsilon_C$). 
 Nuclear energy during H-- and He--burnings 
 are shown in dashed ($\varepsilon_{He}$) and
dotted ($\varepsilon_H$) lines respectively. The long dashed
line is the energy loss rates due to neutrinos multiplied by -1 
($- \varepsilon_{\nu}$). Finally the
gravitational energy production rate in absolute value is plotted in the 
dotted--dashed line ($\varepsilon_g$). Blue squares are plotted on top when this energy is negative.
Note that negative gravitational energy  corresponds to an expansion.
}
\label{enm}
\end{figure*}
We clearly identify two categories of stellar models: those whose evolution is mainly affected by
mass loss (with an inferior mass limit of about 30 $M_{\sun}$), and those whose evolution
is mainly affected by rotational mixing (see also Sect. \ref{ltau}).
We can see that for the 12, 15, 20 $M_{\sun}$ models, the rotating 
tracks have a higher temperature and lower density due to bigger cores.
The  bigger cores  are due to the effect of mixing, which largely
dominates the structural effects of the centrifugal force.
On the other hand, for the 40 and 60 $M_{\sun}$ models, mass loss dominates
mixing effects and the rotating model tracks in the 
Log\,$T_{\rm{c}}$--Log\,$\rho_{\rm{c}}$ plane
are at the same level or
below the non--rotating ones. 

In order to understand the evolutionary tracks in the
Log\,$T_{\rm{c}}$--Log\,$\rho_{\rm{c}}$ plane, we need to look at the different sources
of energy at play. These are the nuclear energy, 
the neutrino and photon 
 energy losses and the
gravitational energy (linked to contraction and expansion). The different energy
production rates at the star center are plotted in Fig. \ref{ent} as a function of the time 
left until core collapse. Going from the left to the right of
Fig. \ref{ent}, the evolution starts with H--burning where  $\varepsilon_H$ dominates. In
response, a small expansion occurs ($\varepsilon_g$ negative and very small
movement to lower densities in the 15 $M_{\sun}$ model during H--burning 
in Fig. \ref{trsl}). At the end of
H--burning, the star contracts non--adiabatically ($T \sim \rho^{1/3}$, every further
contraction is also non--adiabatic). The contraction increases the central temperature. 
This happens very quickly and is seen in the sharp
peak of $\varepsilon_g$ between H-- and He--burnings. When the temperature is
high enough, He--burning starts,
$\varepsilon_{He}$ dominates and contraction is stopped. Note that during the H-- and
He--burning phases, most of the energy is transferred by radiation on thermal timescale.
After He--burning, neutrino
losses ($\varepsilon_{\nu}<0$) overtake photon losses. This accelerates the
evolution because neutrinos escape freely.
During burning stages, the nuclear energy production 
stops the contraction if $\varepsilon_{nucl} \sim - \varepsilon_{\nu}$ (see
C--burning for the rotating model) or even provoke an expansion when
$\varepsilon_{nucl} > - \varepsilon_{\nu}$ (most spectacular during Si--burning).
Central density decreases when the central regions expand (see Fig.~\ref{trsl}).
Once the iron core is formed, there
is no more nuclear energy available while neutrino losses are still present 
and the core collapses. \\
Figure \ref{enm} shows the variation of the energy production
as a function of
the mass fraction inside a 20 $M_\odot$ stellar model at a stage
during the shell C--burning phase.
At the different burning
shells, expansion occurs due to positive nuclear energy production. In the
outer part, contraction and expansion are controlled by the
photon luminosity and therefore by the opacities.  In the inner regions, the
energy produced either by the nuclear reactions or by contraction is evacuated by the neutrinos.
In the non--rotating star,
partial ionisation of helium I in the outermost layers produces a peak in the opacity 
($\kappa \nearrow$). This induces an expansion of the star 
($\varepsilon_g<0$).
The situation is different for the rotating model
because it has lost most of its envelope and temperatures are higher than
the ionisation transition zone. \\
Numerically, it is important to note that the largest value for energy
production rates corresponds to the nuclear one. Its maximum value is 
therefore 
used in order to determine the evolutionary time steps in our code. 

\subsubsection{The fate of the 12 $M_{\sun}$ models}\label{12m}
By looking at the track of the 12 $M_{\sun}$
models in Fig.~\ref{trsl}, we can see that rotation has a noticeable effect on the
post C--burning phases.
Indeed, the non--rotating model starts Ne--burning off--centre and
the burning never reaches the centre. 
The unburnt Ne--O core, $M_{\rm{Ne-O}}$ is equal to 0.096 (see Table~\ref{table1b}). 
On the other hand, the rotating model 
starts Ne and O--burnings in the centre. This can be
seen in Fig. \ref{dhr5m121520} (top). 
The computation of the 12 $M_\odot$ models were stopped
during the Ne/O--burning phase.
To explore their further evolution, one can use the mass limits
for the Ne--cores, $M_{\rm{Ne}}$, given by
\citet{N84}:
\begin{itemize}
\item $M_{\rm{Ne}}=1.46$ $M_{\sun}$ is the lower limit for neon 
ignition in the centre.
\item $M_{\rm{Ne}}=1.42$ $M_{\sun}$ is the lower limit for off--centre
 neon ignition where the subsequent neon burning front 
reaches the centre.
\item $M_{\rm{Ne}}=1.37$ is the lower limit for off--centre
 neon ignition. In the mass range between 1.37 and 
1.42 $M_{\sun}$, neon
burning never reaches the centre.
\end{itemize} 
Our rotating 12 $M_{\sun}$ model has $M_{\rm{Ne}} \gtrsim 1.6$ well above
the lower mass limit for neon ignition in the centre.
In
the non--rotating model the Ne--core mass is around 1.4 $M_{\sun}$
(more quantitatively carbon mass fraction decreases from 0.01 to 0.001
between 1.45 and 1.34 $M_{\sun}$). As expected from the mass limits above,
in this model, neon burning, which starts off--centre, will probably 
reach the centre. Then \citep[see][ Sect. 3.2: fate of stars with 
$10\,M_{\sun}< M_{\rm{ms}}<13\,M_{\sun}$ and references therein]{NH88}, 
electron capture will
help the star to collapse making the neon/oxygen burning explosive and
possibly ejecting the H and He--rich layers.
Note that
in our models we only follow multiple--$\alpha$ elements.
We
did not follow the evolution of the electron mole number, $Y_e$,
or of neutron excess, $\eta$, neither include Coulomb corrections.
Let us recall that the electron mole number, $Y_e=\sum _i Z_i\,Y_i$,
and the neutron excess, $\eta= \sum_i (N_i-Z_i)Y_i$,
are linked by the following relation:
$Y_e=(1-\eta)/2$  ($N_i$, $Z_i$ and $Y_i$ are respectively the number of neutron
s,
protons and the number abundance of element $i$; $Y_i=X_i/A_i$, where
$X_i$ and $A_i$ are the mass fraction and the mass number of element
$i$).
Therefore the electron mole number, $Y_e$, is always equal to 
0.5\footnote{The mass limits given by \citet{N84} 
were also obtained from calculations with $Y_e=0.5$.}. 
Lower values of 
$Y_e$ (due to electron captures)
 and the inclusion of
Coulomb corrections in the equation of state have an impact 
in this context.
Electron captures remove electrons. This decreases the electron
pressure and facilitates the collapse. 
Coulomb corrections generally act
to decrease the iron core mass by about 0.1\,$M_{\sun}$ 
\citep[ and references therein]{WHW02}.
These omissions can be the cause of the failure of our models to follow
the evolution of the 12\,$M_{\sun}$ models further. 
These two effects however do not affect significantly the evolution of more massive
stars before the shell Si--burning phase.

\subsection{Kippenhahn diagrams}
\begin{figure*}[!tbp]
\centering
\includegraphics[width=7.5cm]{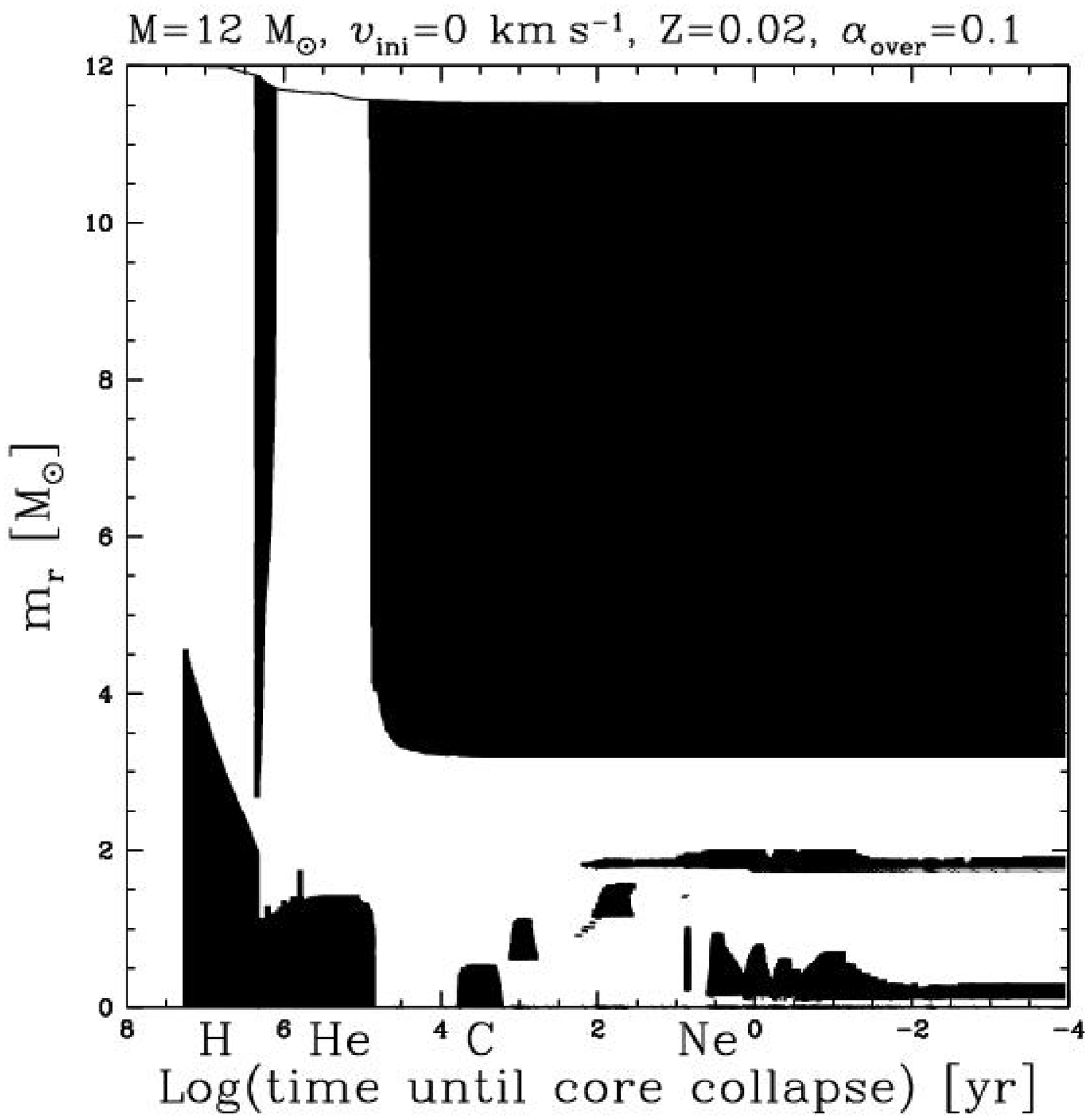}\includegraphics[width=7.5cm]{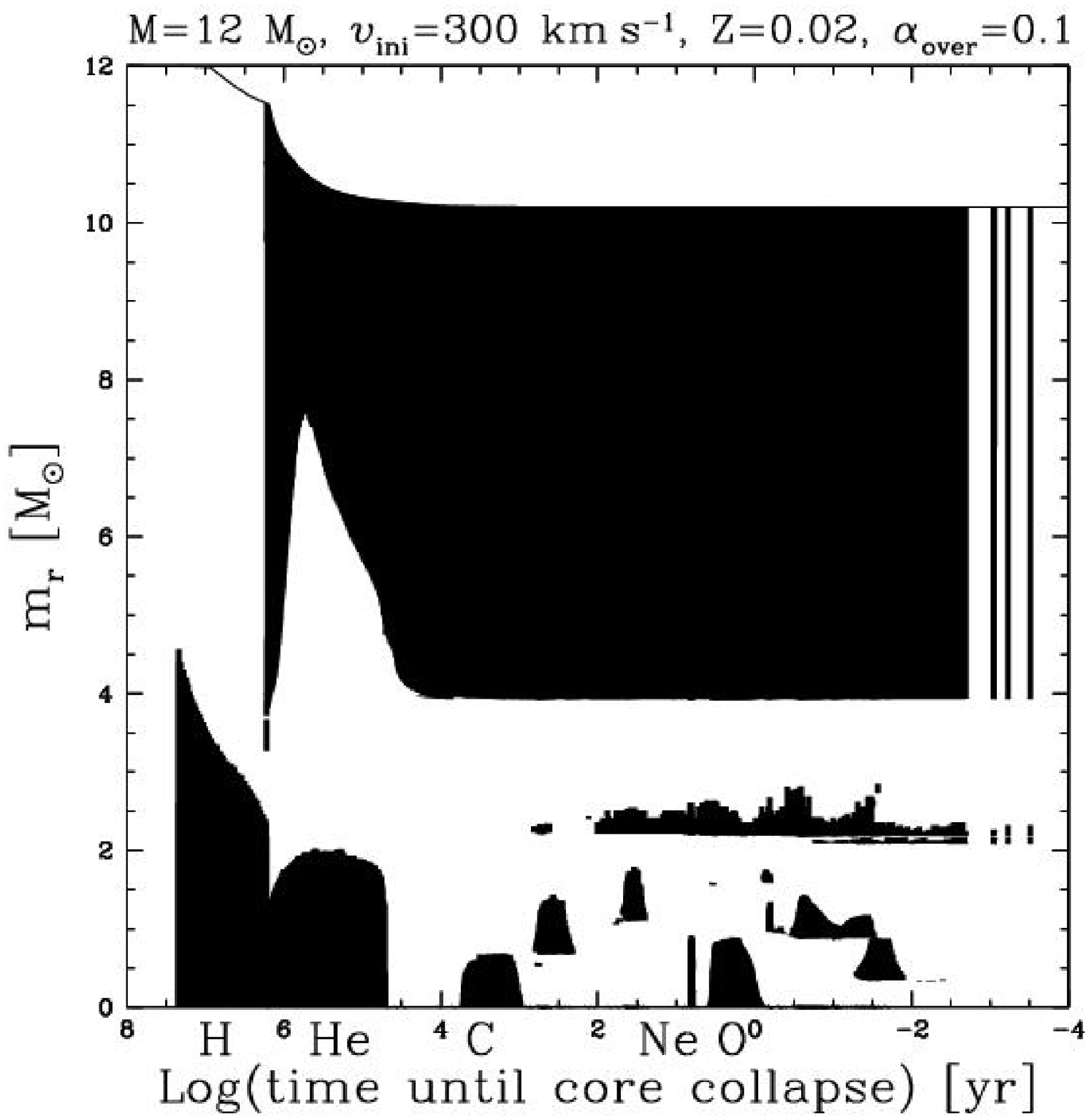}
\includegraphics[width=7.5cm]{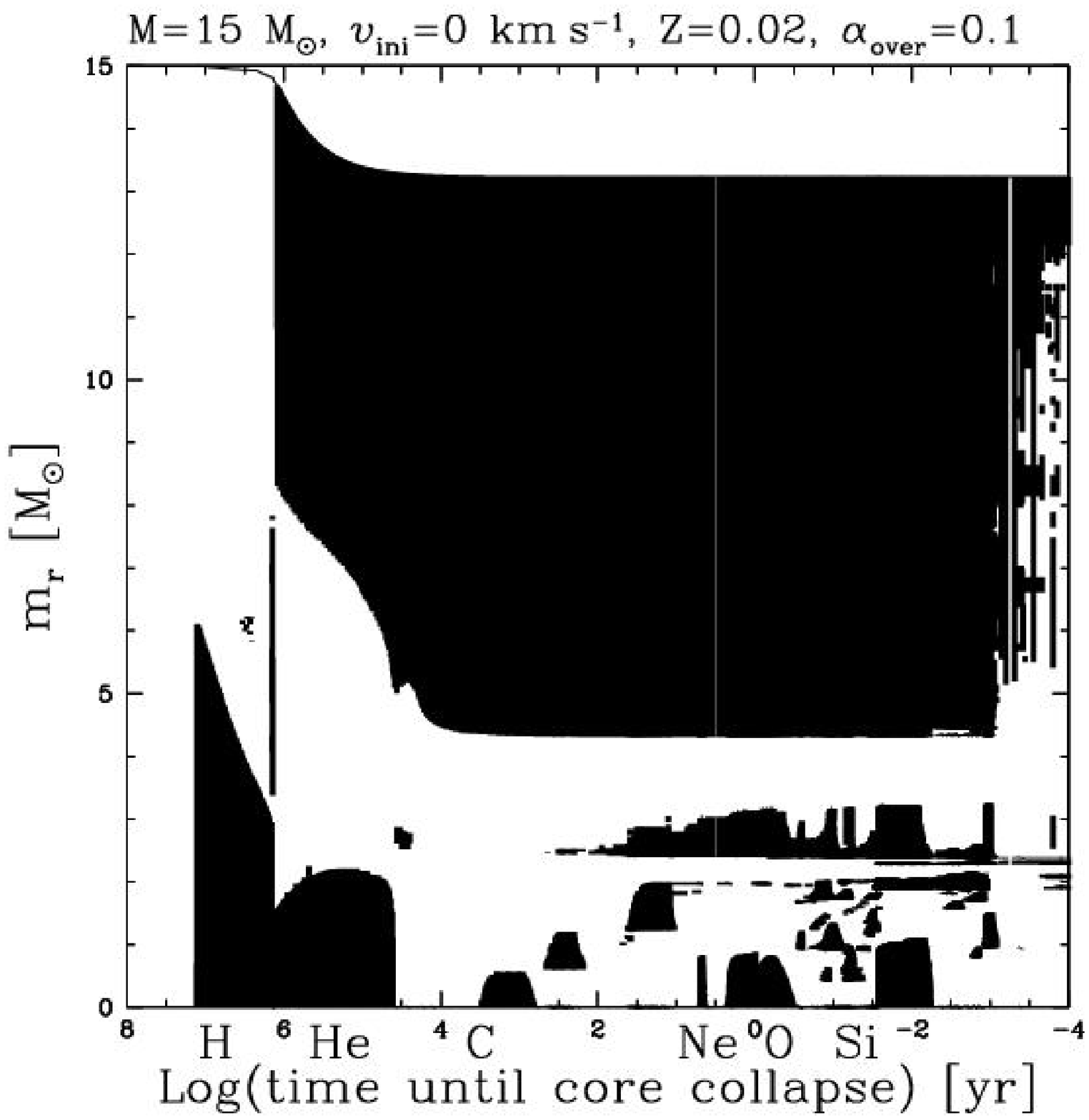}\includegraphics[width=7.5cm]{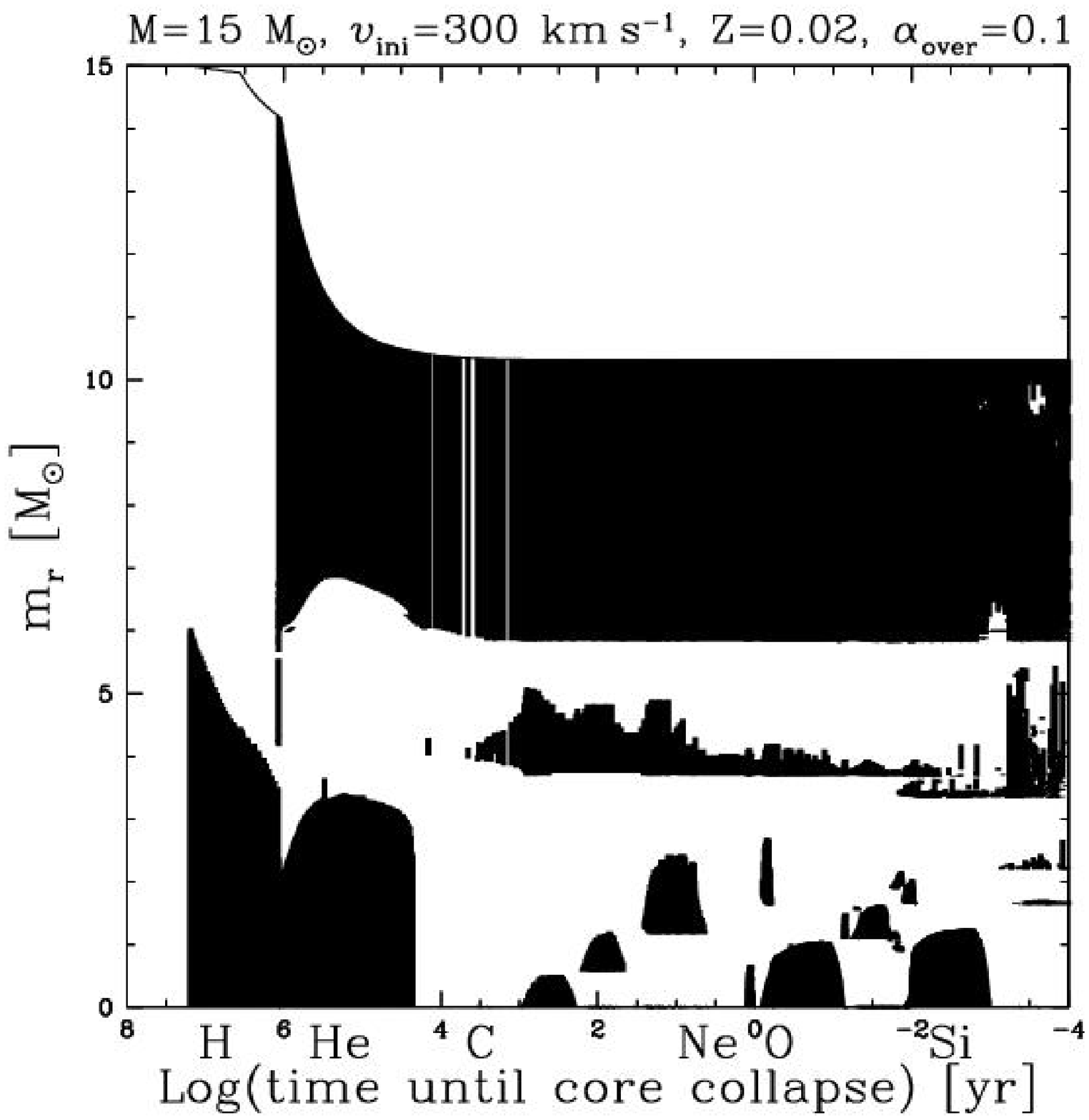}
\includegraphics[width=7.5cm]{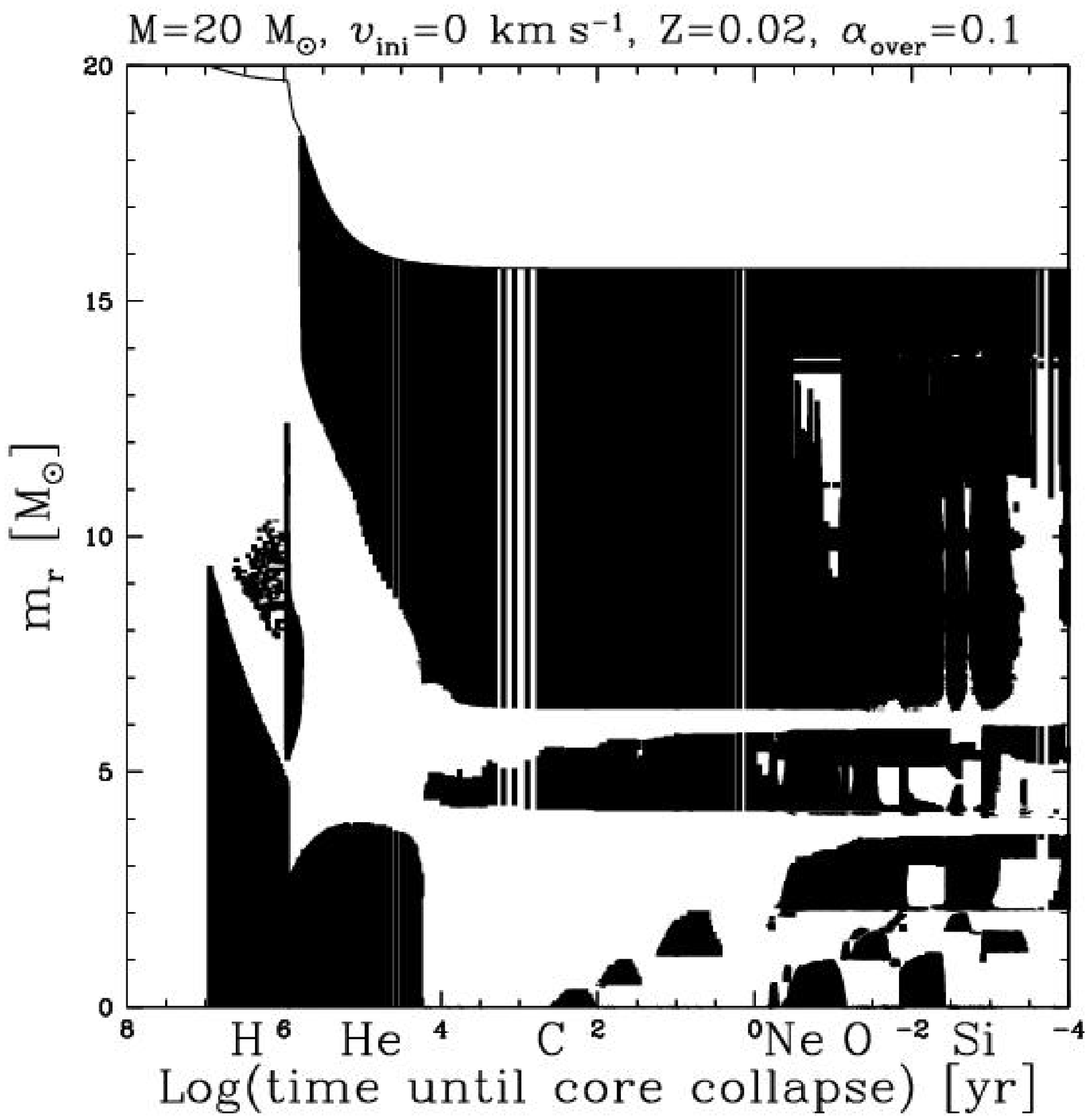}\includegraphics[width=7.5cm]{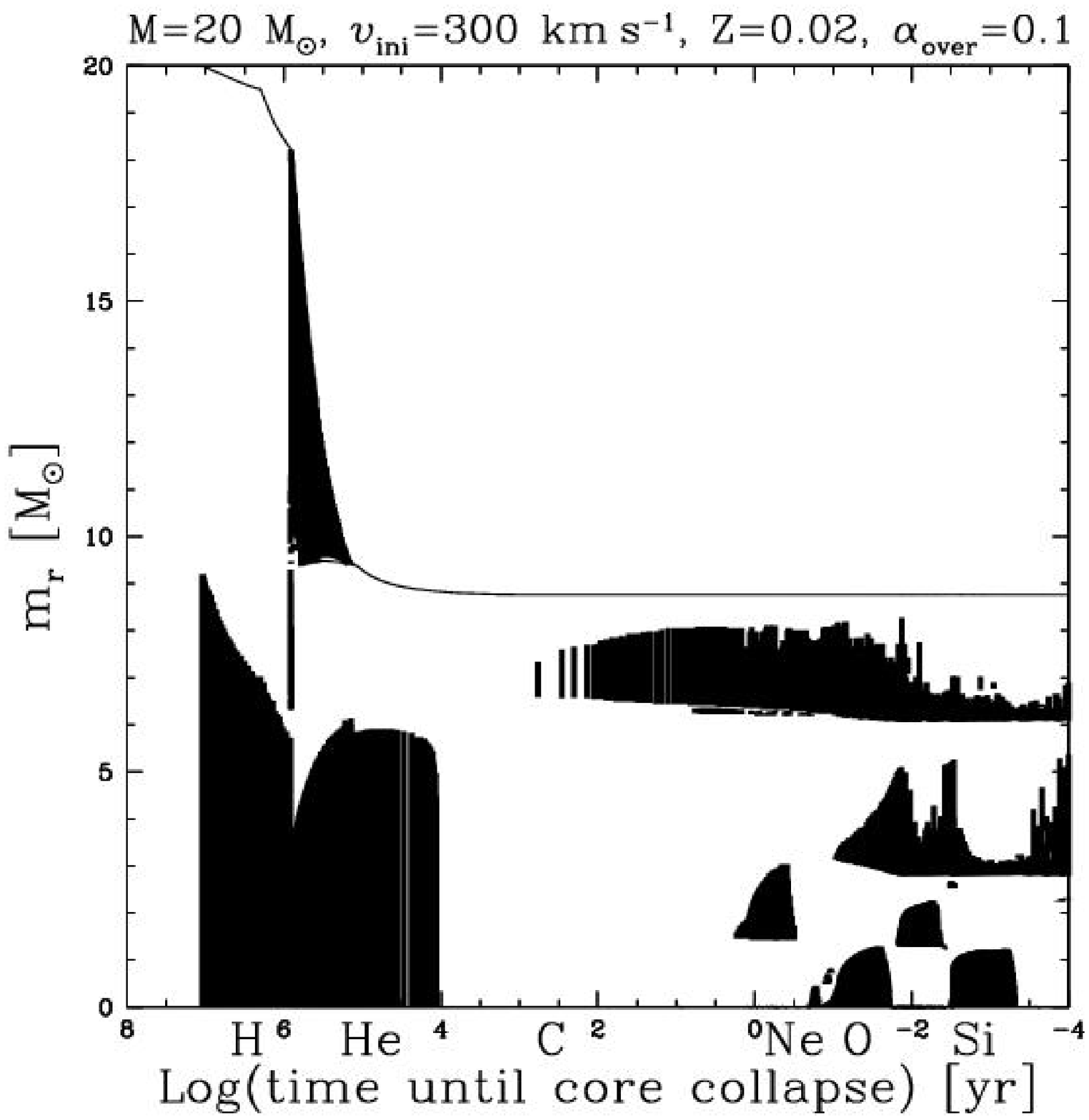}
\caption{Kippenhahn diagrams for the non--rotating (left) and 
$v_{\rm{ini}}$= 300 km\,s$^{-1}$ (right) 12 (top), 
15 (middle) and 20 (bottom) $M_{\sun}$ models. The black zones correspond to convective regions
(see text).
}
\label{dhr5m121520}
\end{figure*}

\begin{figure*}[!tbp]
\centering
\includegraphics[width=7.5cm]{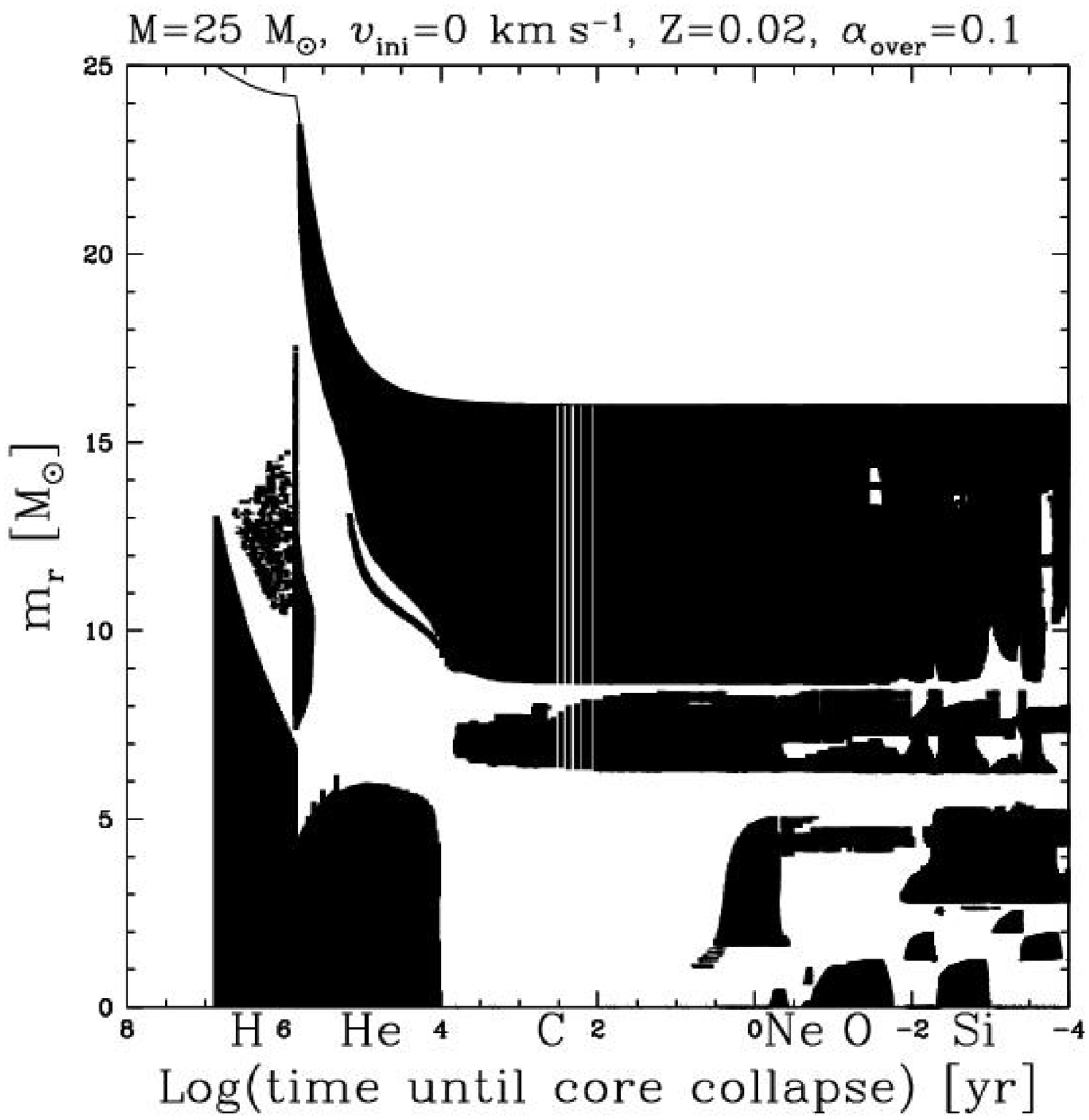}\includegraphics[width=7.5cm]{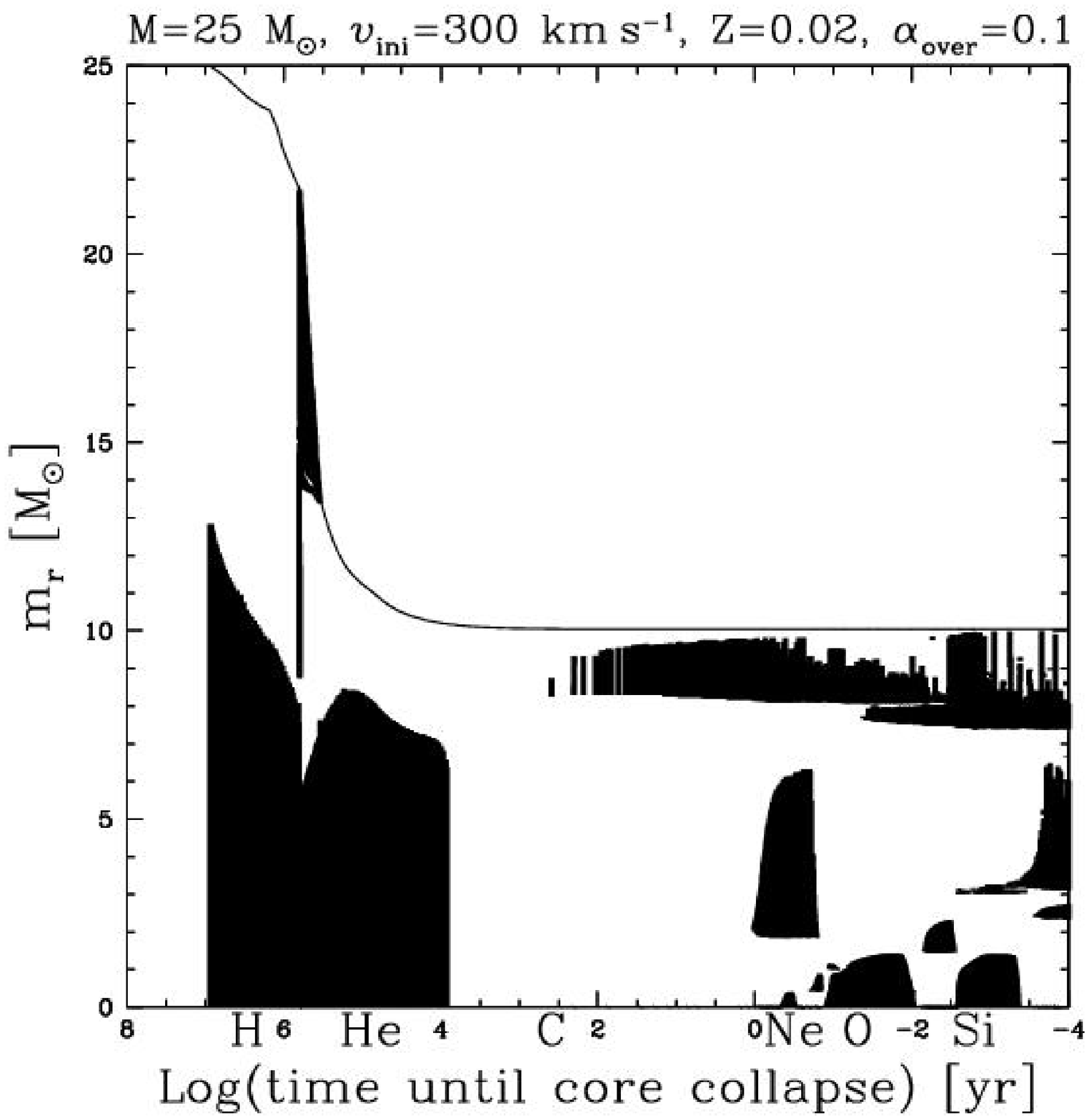}
\includegraphics[width=7.5cm]{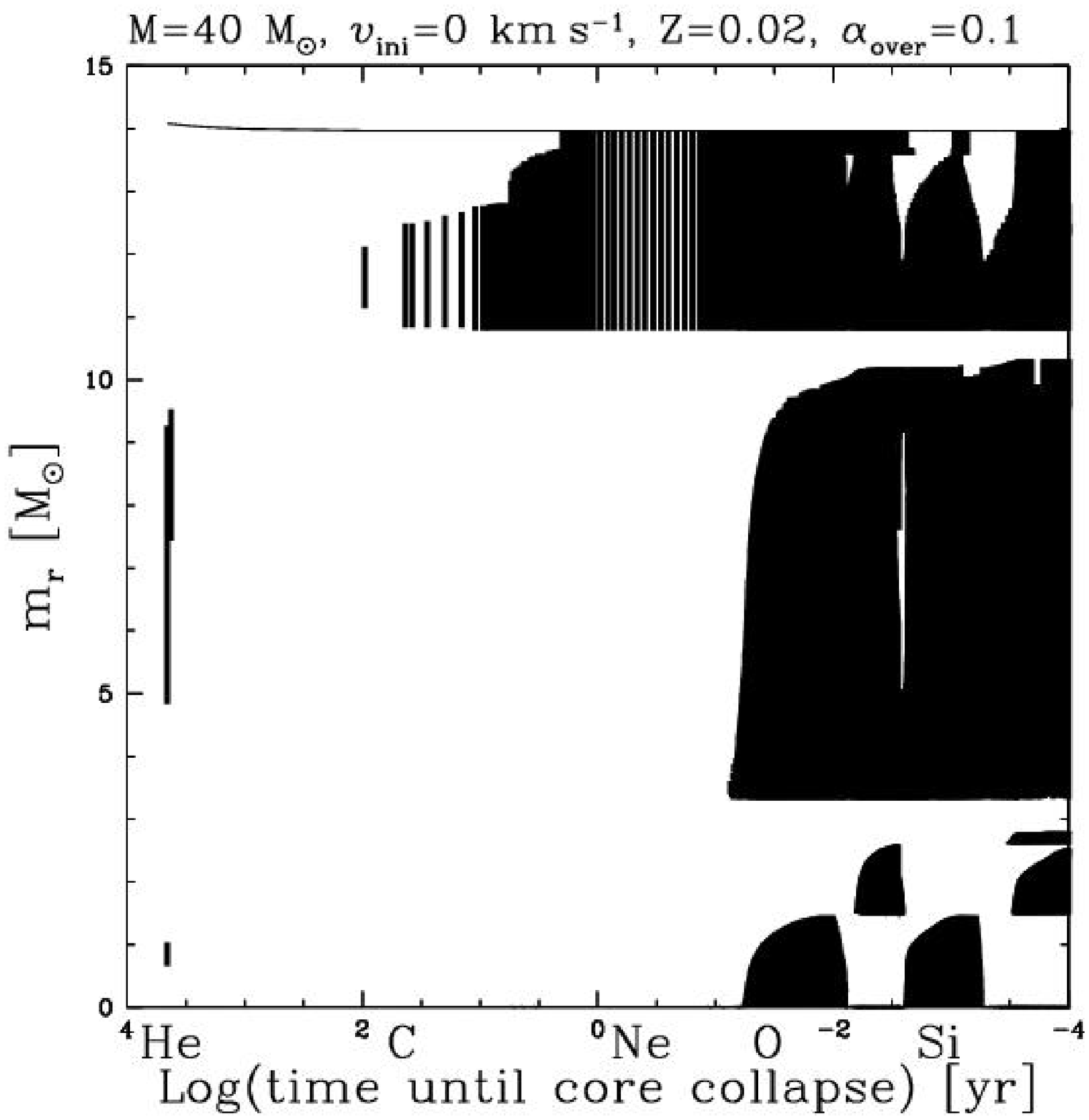}\includegraphics[width=7.5cm]{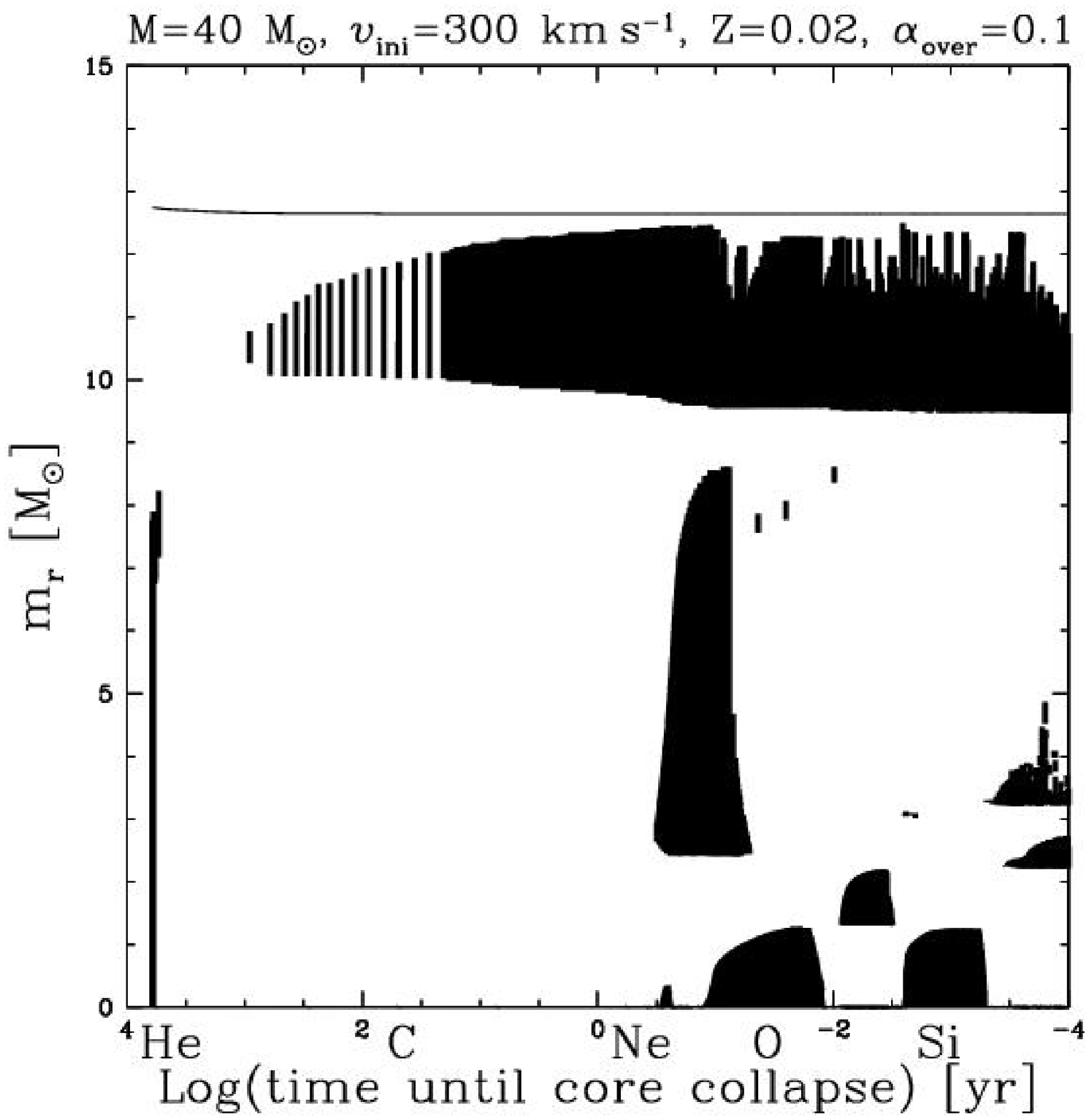}
\includegraphics[width=7.5cm]{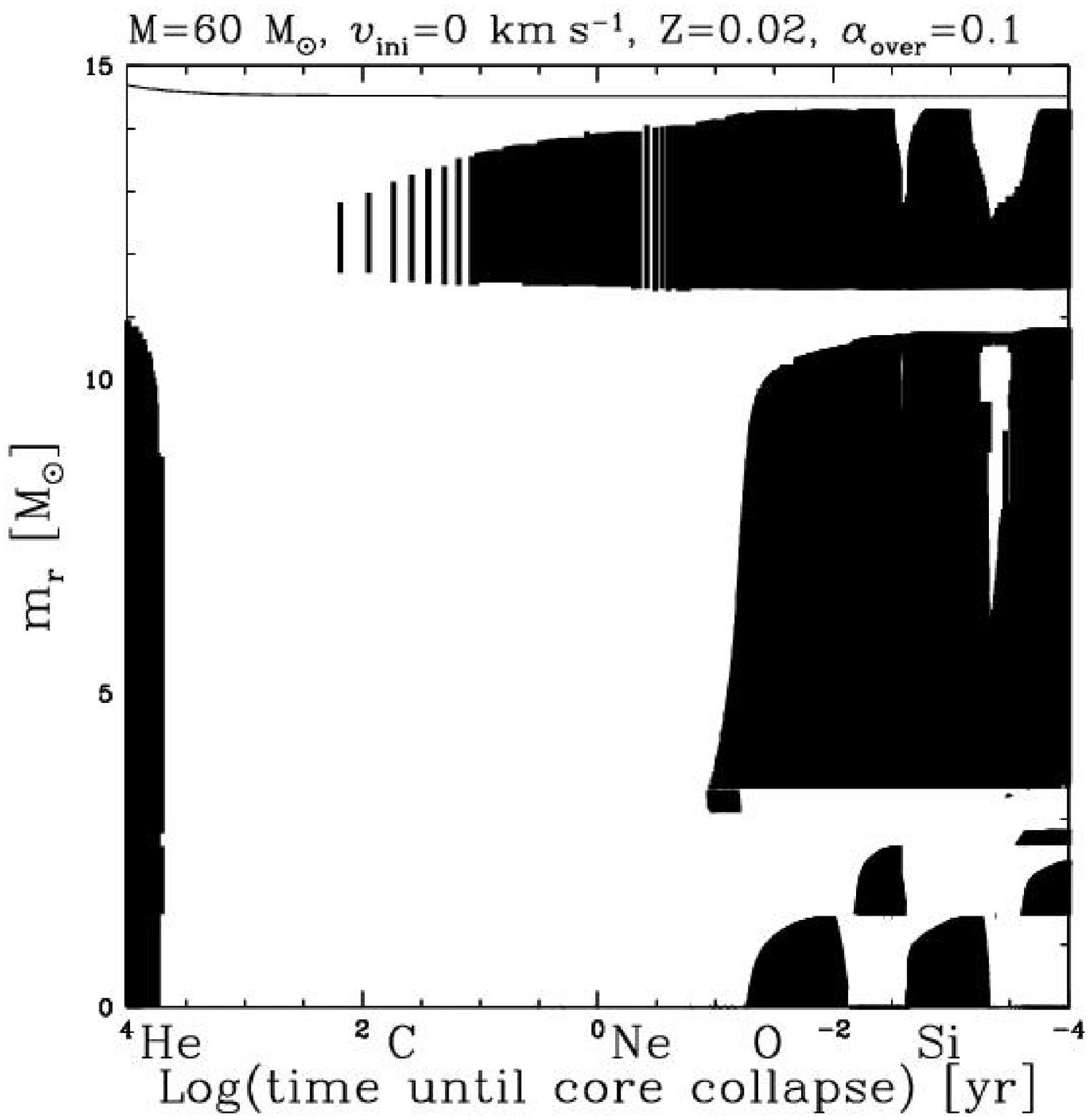}\includegraphics[width=7.5cm]{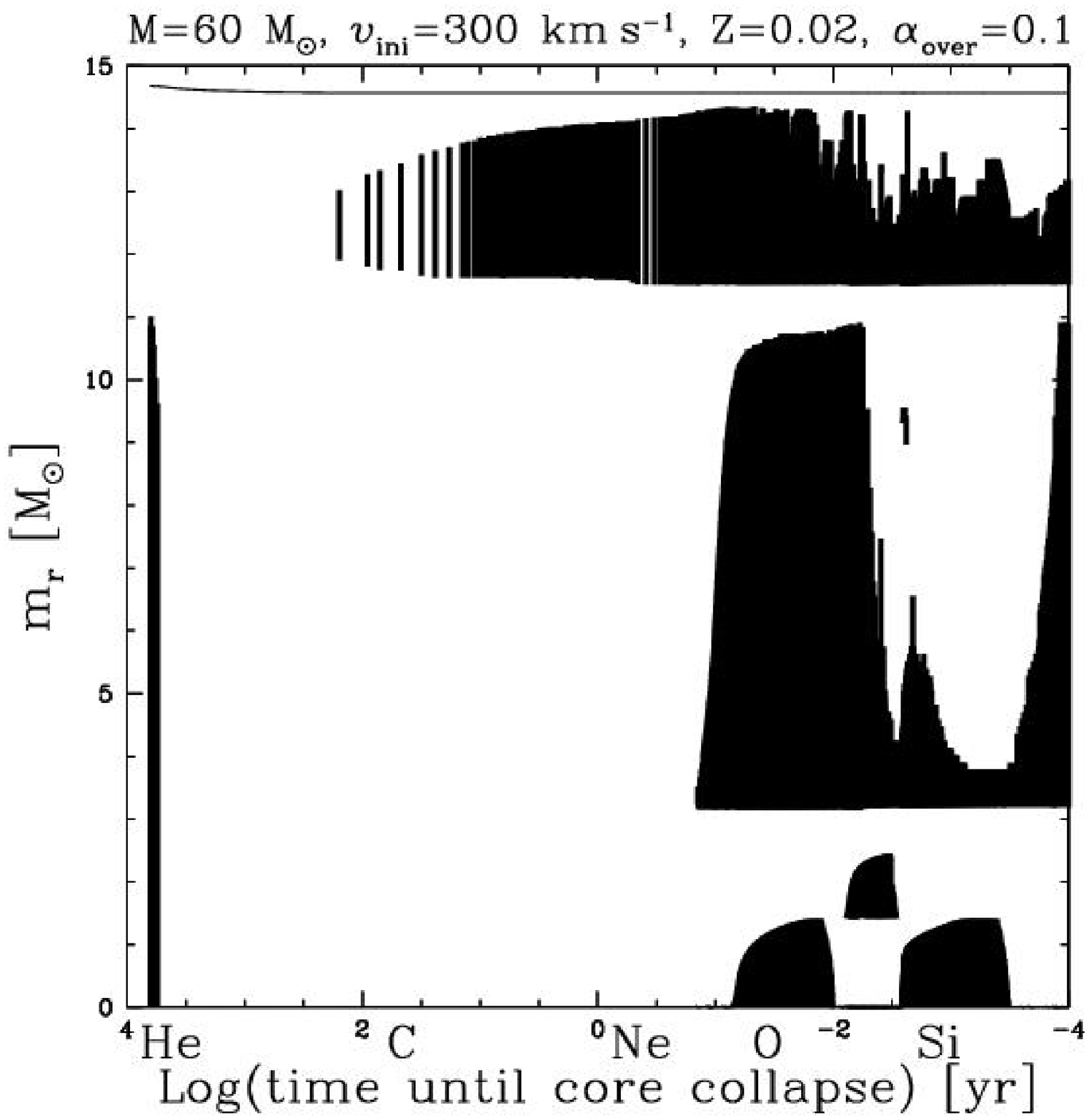}
\caption{Kippenhahn diagram for the non--rotating (left) and 
$v_{\rm{ini}}$= 300 km\,s$^{-1}$ (right) 25 (top), 
40 (middle) and 60 (bottom) $M_{\sun}$ models.
The black zones correspond to convective regions
(see text).}
\label{dhr5m2546}
\end{figure*}

\begin{figure}[!tbp]
\centering
   \includegraphics[width=8.8cm]{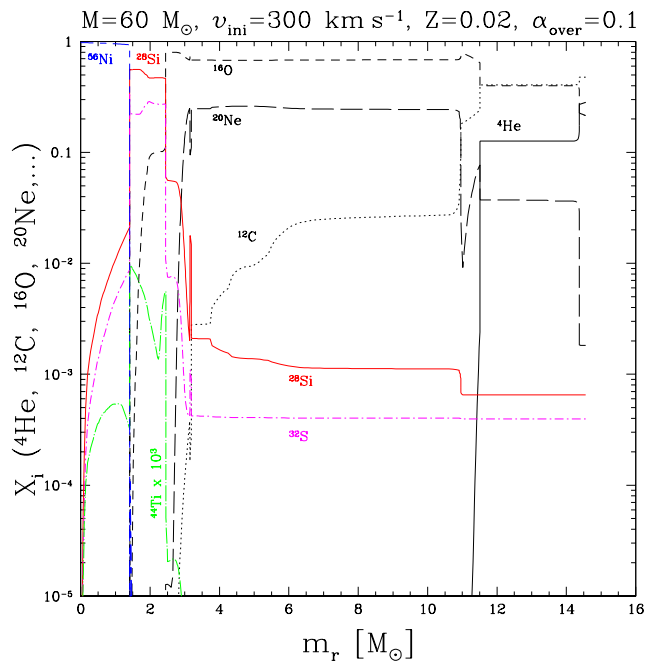}
\caption{Variations of the abundance (in mass fraction)  as a function of the 
lagrangian mass coordinate, $m_r$, at the end of central Si--burning 
for the rotating 60 $M_{\sun}$. Note that the $^{44}$Ti abundance
 (dotted--long dashed line) is enhanced by a factor 1\,000 for display 
 purposes.}
\label{a60s3endsi}
\end{figure}

Figures \ref{dhr5m121520}--\ref{dhr5m2546} show the Kippenhahn diagrams for the
different models. The y--axis represents the mass coordinate and the x--axis 
the time left until core collapse. The black zones represent convective zones. 
Since our calculations have not reached
core collapse yet, we estimate that there is $10^{-5}\,$yr between the last model and
the collapse. This value has no significant influence since it is only a small
additive constant. The graph is built by drawing vertical lines at each time step where
the star is convective. this discrete construction shows its weakness at the
right edge of each diagram and during shell He--burning 
where time steps are too distant from each other to cover the
surface properly.
The abbreviations of the various burning stages are written below the
graph at the time corresponding to the central burning stages.

We can see the effect of the blue loops \citep{ROTX} 
in the 12 $M_{\sun}$
models on the external convective zone during the core He--burning phase. The blueward motion reduces the
external convective zone or even suppresses it. 
We also note the complex succession of the different convective zones between
central O and Si--burnings (for instance in the non--rotating 15 $M_{\sun}$ model).
The difference between
non--rotating and rotating models is striking in the 20 and 25 $M_{\sun}$
models. We can see that small convective zones above the central H--burning
core disappear in rotating models. Also visible is the loss of the
hydrogen rich envelope in the rotating models. On the other hand non--rotating and rotating 40 and 60 $M_{\sun}$
models all have very similar convective zones history after He--burning. 

\subsubsection{Convection during core C--burning?}

Recent calculations \citep{HLW00} show that non--rotating 
stars with masses less than about 22 $M_{\sun}$
have a convective central C--burning core while heavier stars have a radiative
one. Our non--rotating models agree with this. What about models of rotating stars? 
Figure \ref{dhr5m121520} (bottom) shows
the Kippenhahn diagrams for the non--rotating and rotating 20 $M_{\sun}$
models. We can see that the rotating model has a radiative core
during central C--burning. It is due to the fact that the nuclear energy 
production rate $\varepsilon_C$ does not overtake 
$- \varepsilon_{\nu}$ (see Fig. \ref{ent} right) 
and therefore the central entropy does not increase
enough to create a convective zone. This behaviour results from
the bigger He--cores formed in rotating models. Bigger cores imply higher
central temperatures during the core He--burning phase and higher central temperatures 
imply lower carbon content at the end of the He--burning phase.
Thus less fuel is available for the core C--burning phase which does not succeed
to develop a convective core.
The same
explanation works for more massive (rotating or non--rotating) stars.
Thus the upper mass limit for a convective core during the C--burning phase
is lowered by rotation, passing from about 22 $M_\odot$ to a value inferior to
20 $M_\odot$ when the initial velocity increases from 0 to 300 km s$^{-1}$.

\subsection{Abundances evolution}
\begin{figure*}[!tbp]
\centering
\includegraphics[width=7.5cm]{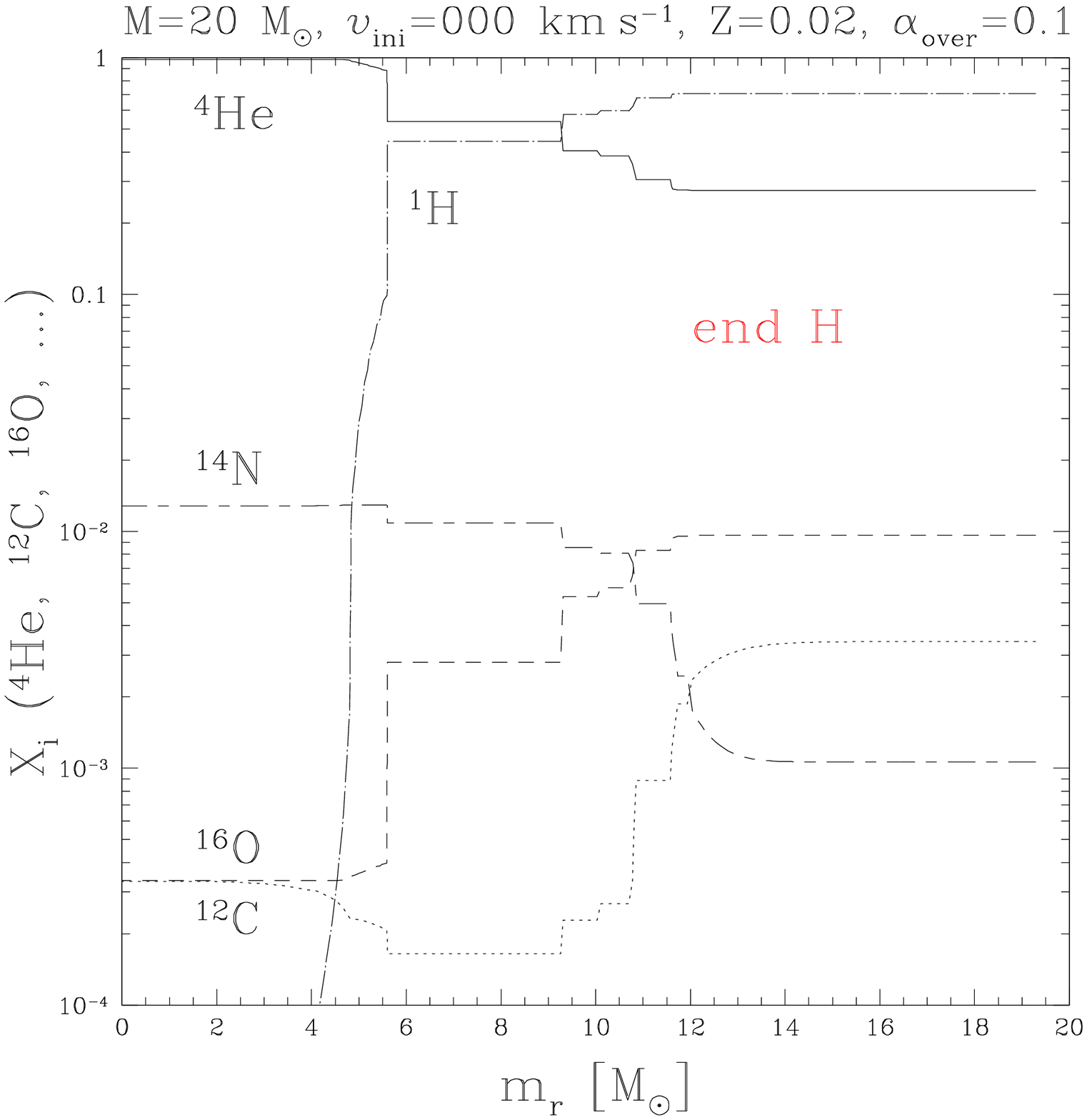}\includegraphics[width=7.5cm]{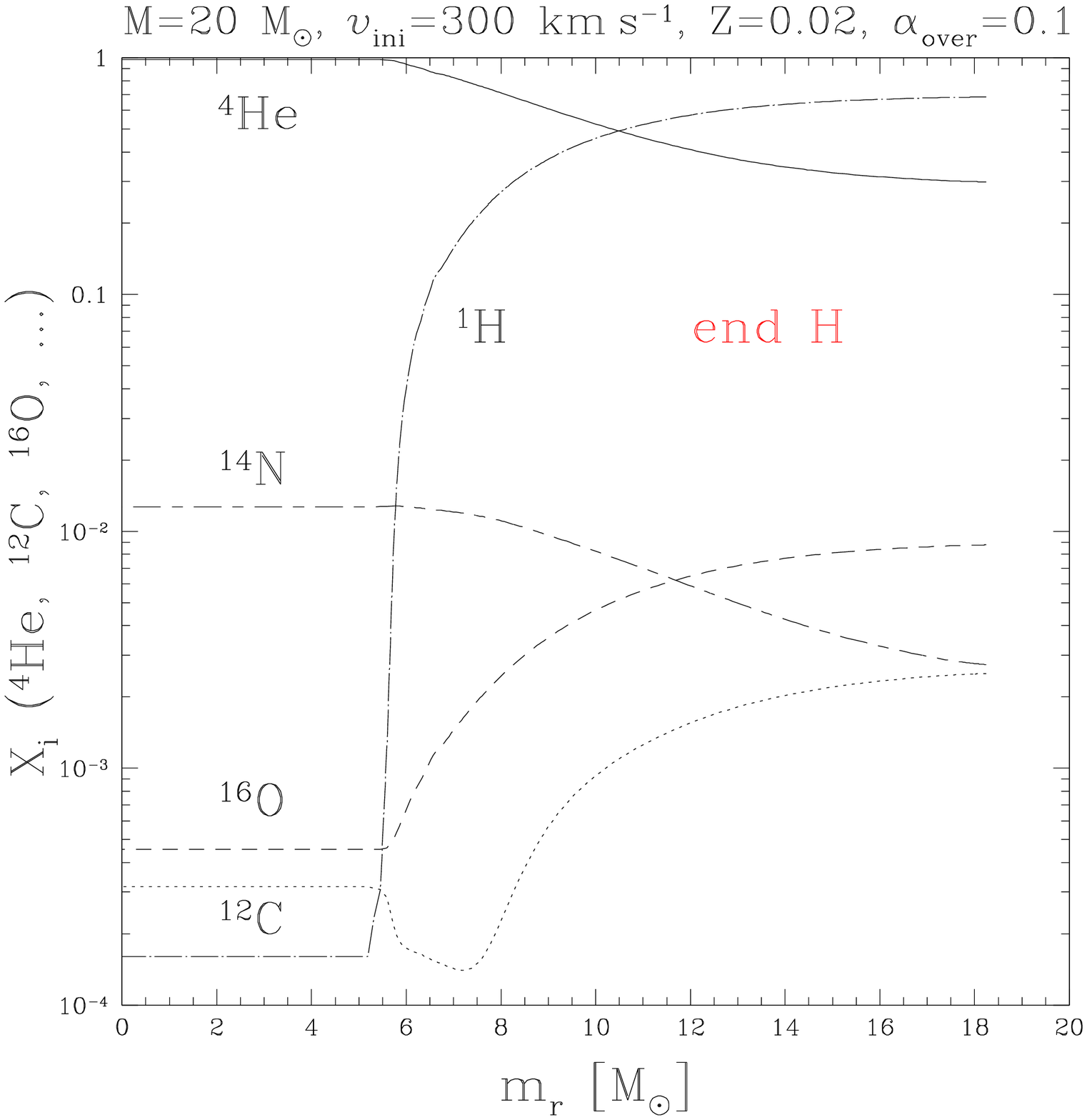}
\includegraphics[width=7.5cm]{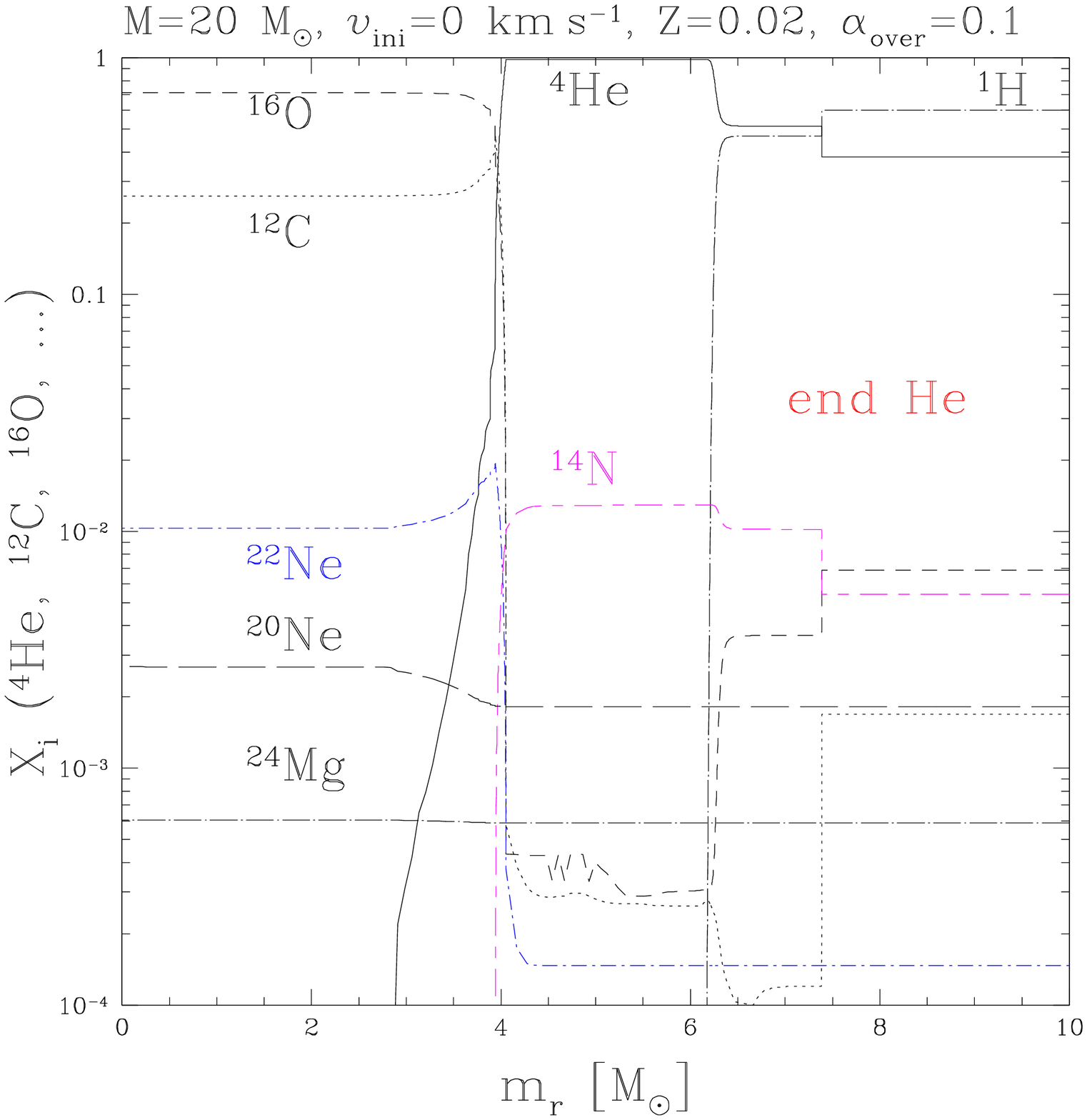}\includegraphics[width=7.5cm]{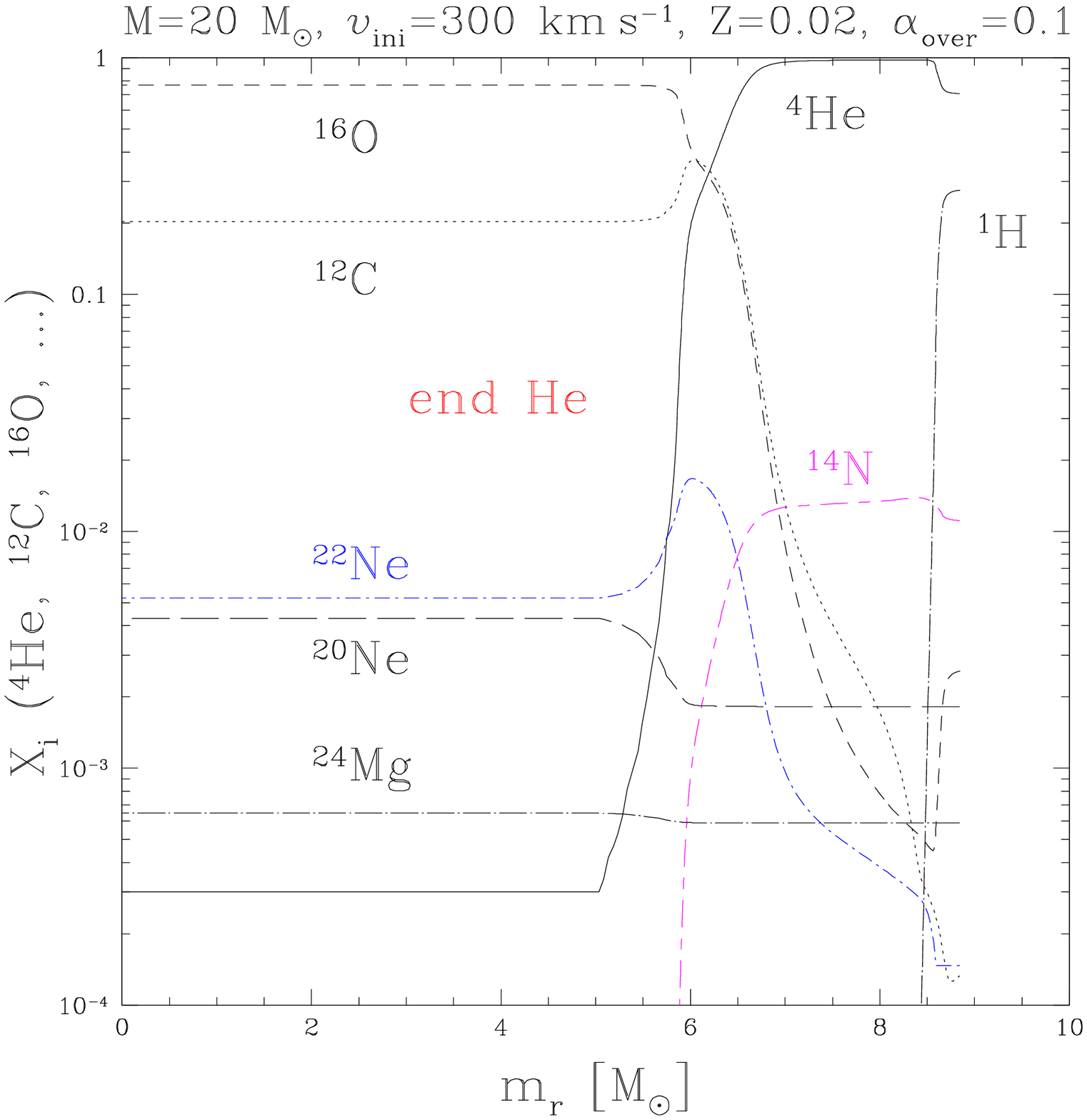}
\includegraphics[width=7.5cm]{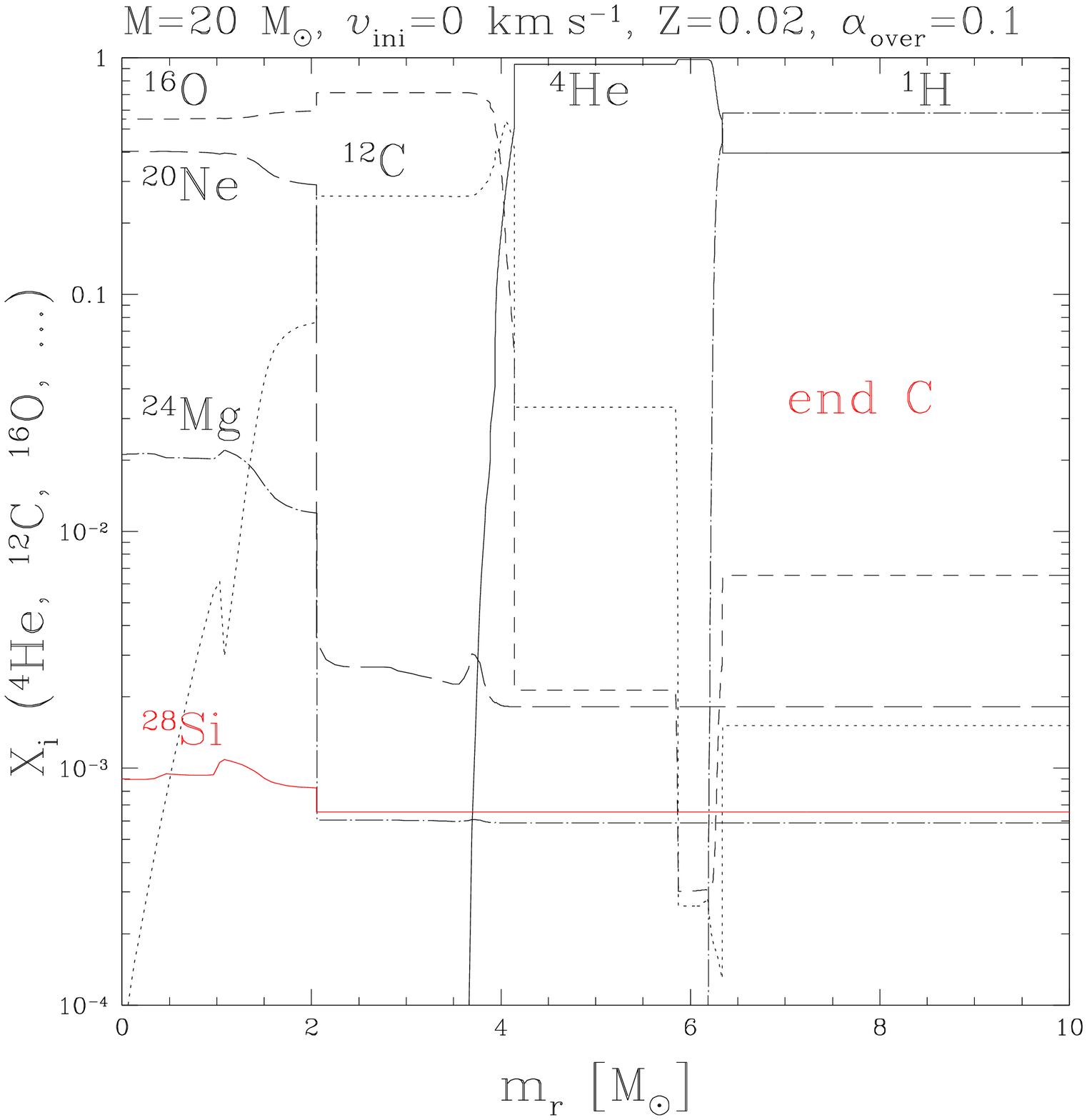}\includegraphics[width=7.5cm]{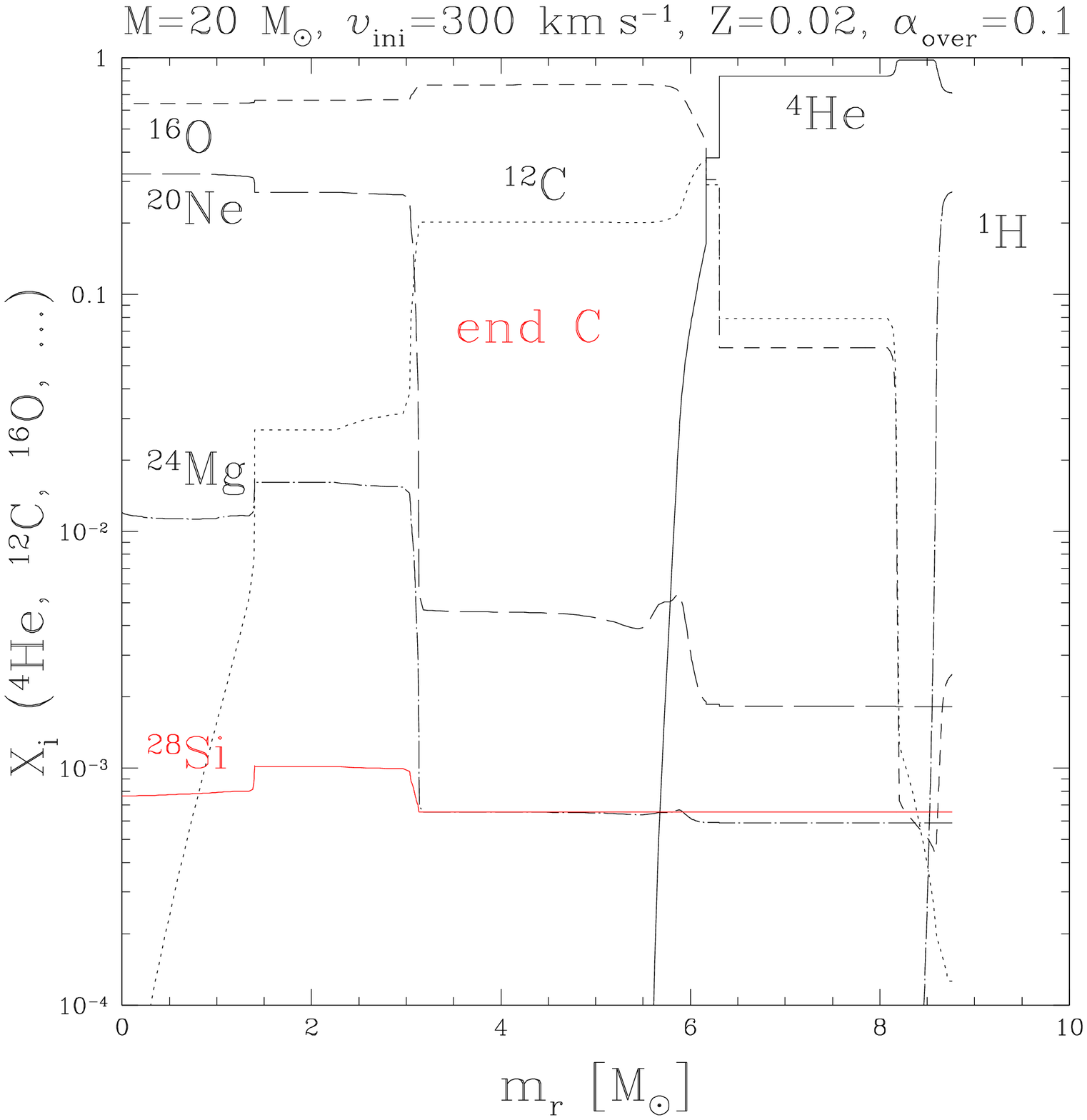}
\caption{Variation of the abundances in mass fraction as a function of the lagrangian mass
at the end of central hydrogen (top), 
helium (middle) and
carbon (bottom) burnings for the non--rotating (left) and rotating
 (right) 20$M_{\sun}$ models.}
\label{a20hhec}
\end{figure*}

\begin{figure*}[!tbp]
\centering
\includegraphics[width=7.5cm]{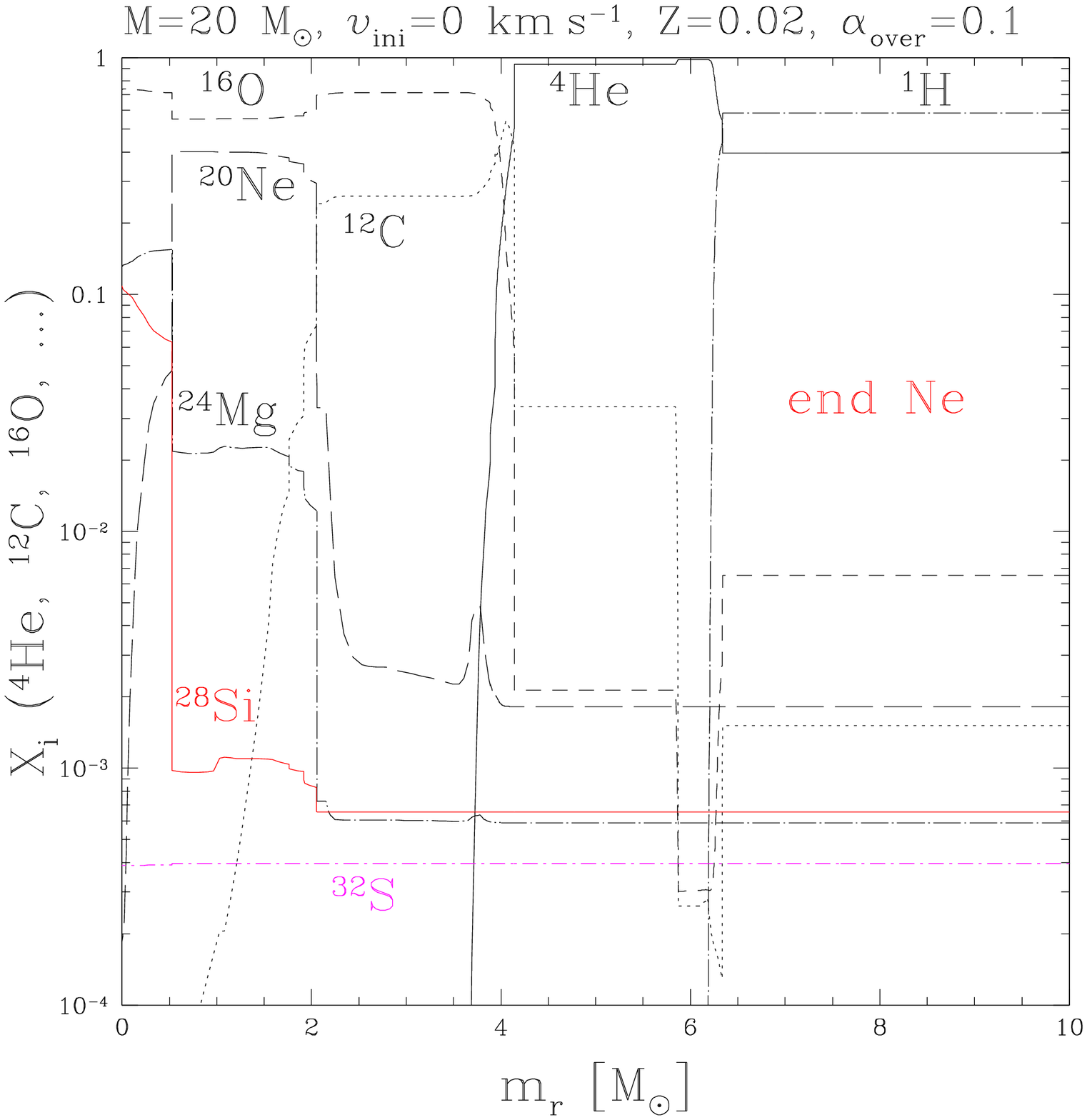}\includegraphics[width=7.5cm]{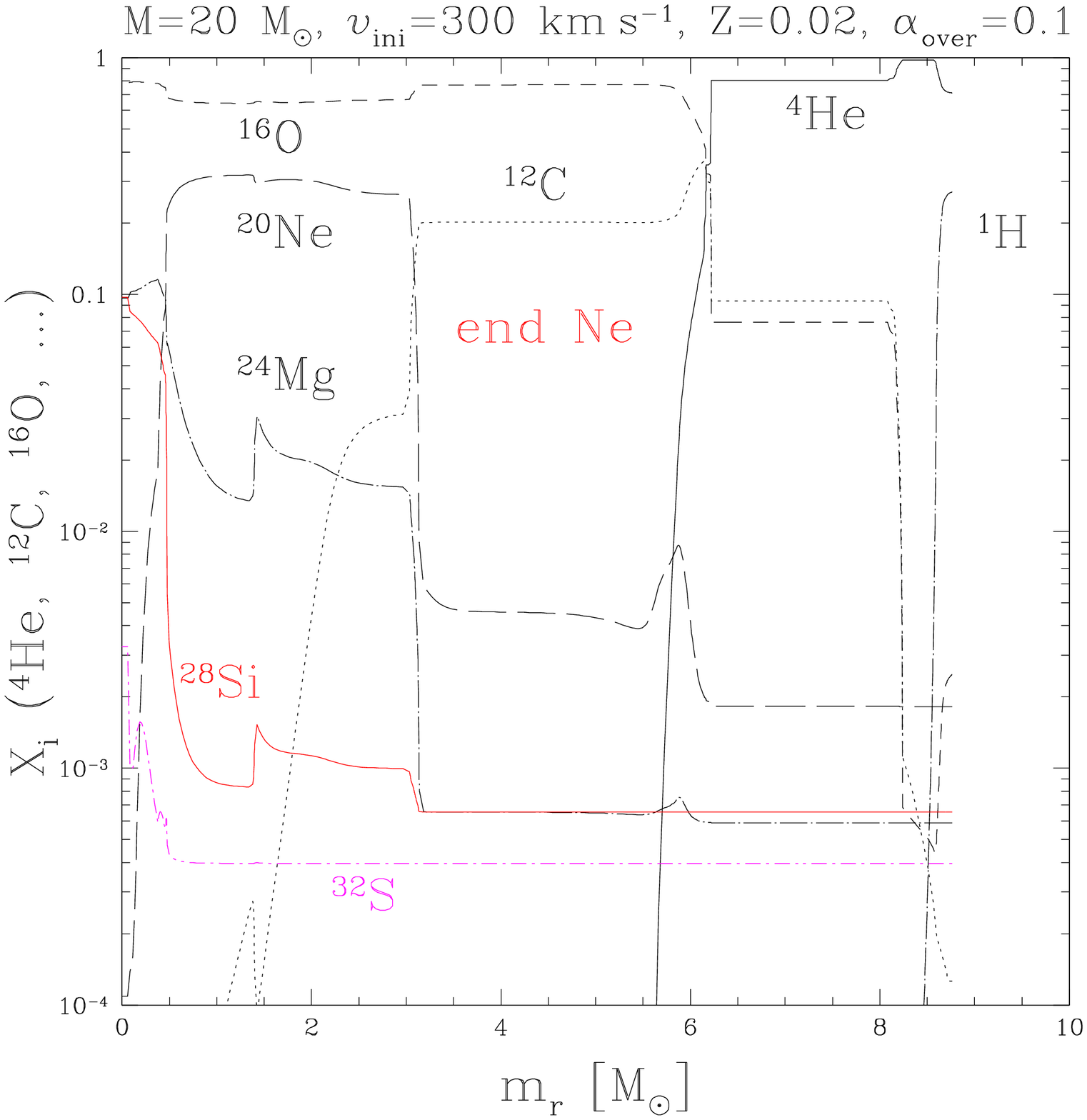}
\includegraphics[width=7.5cm]{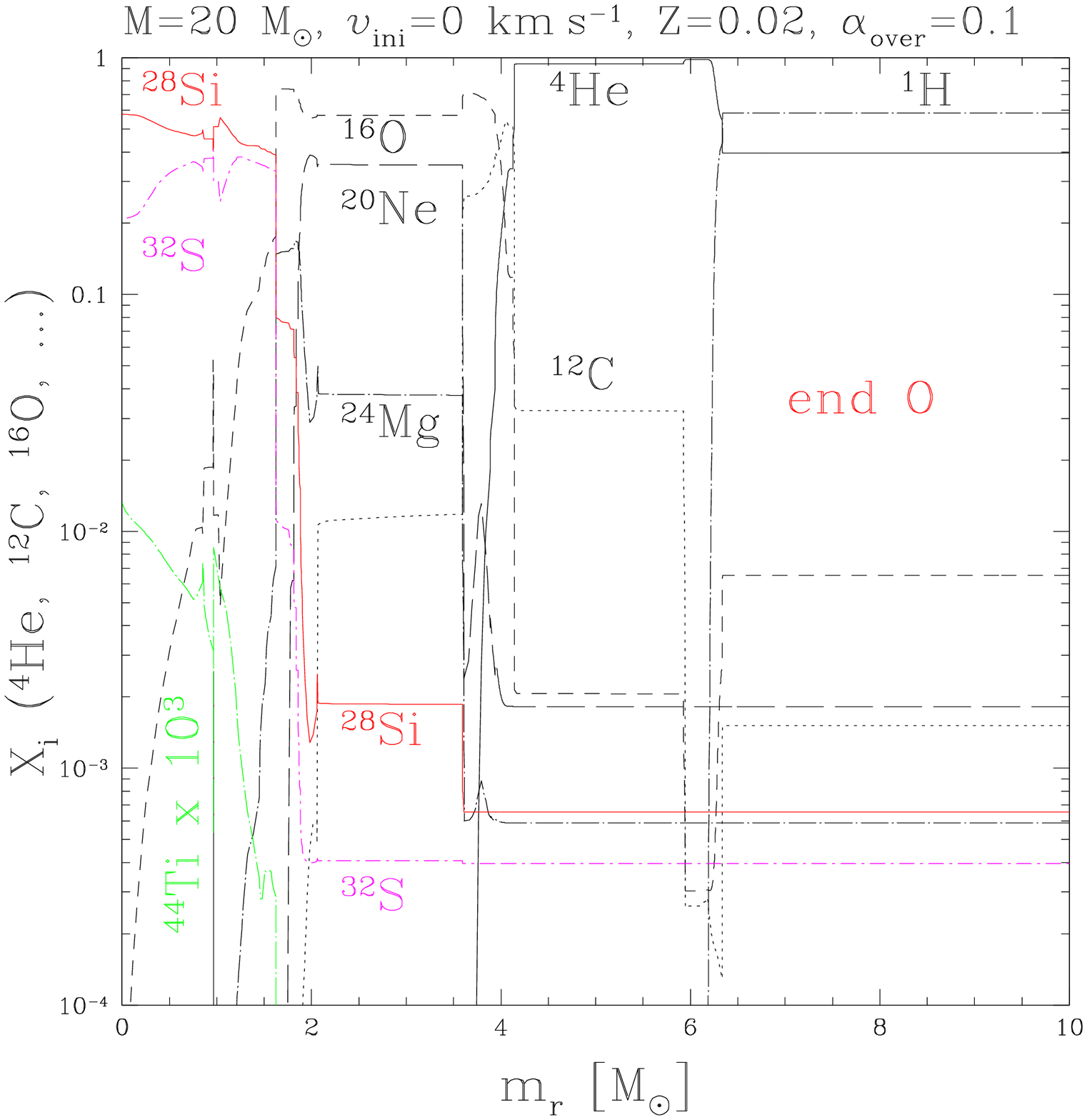}\includegraphics[width=7.5cm]{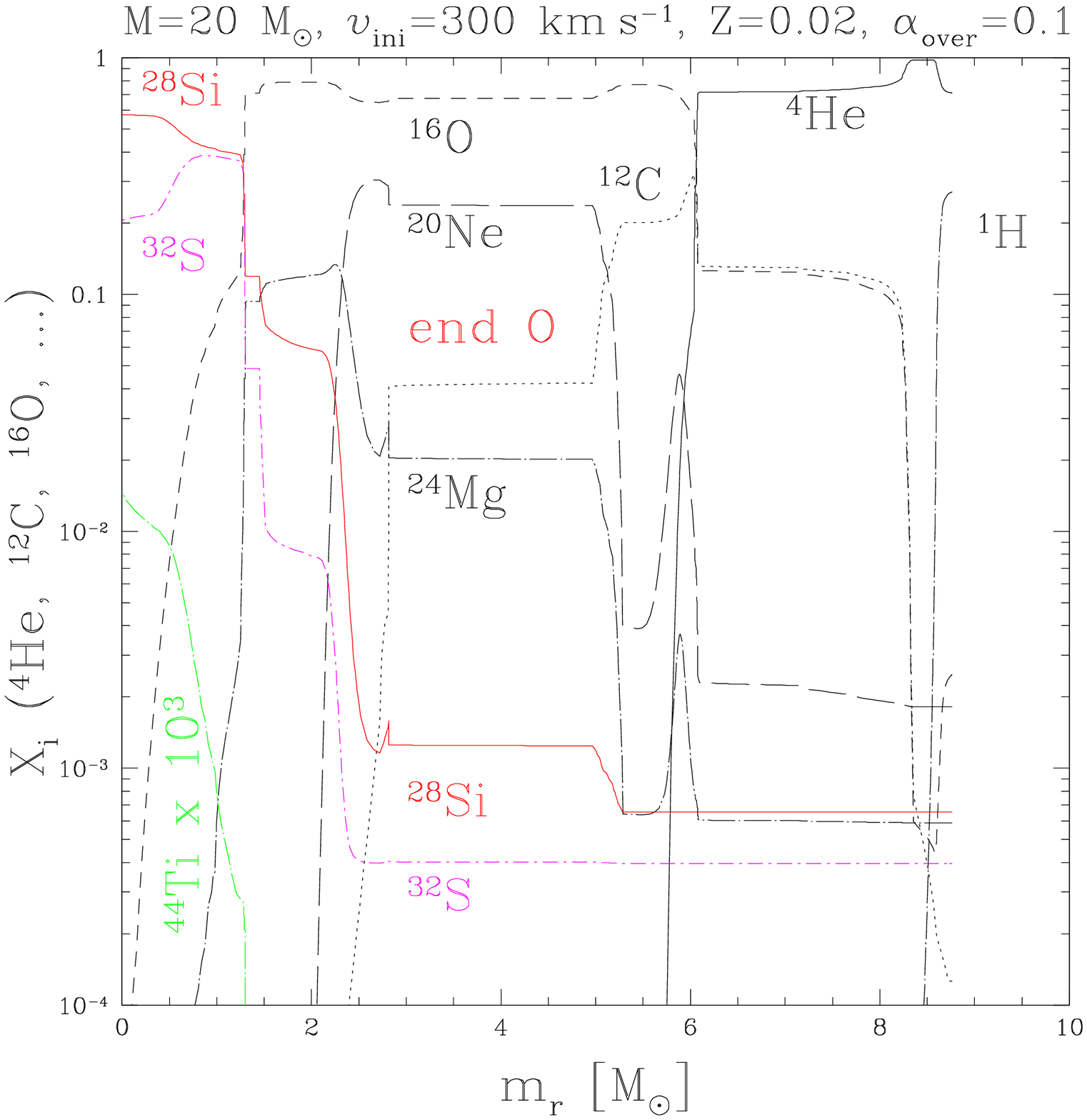}
\includegraphics[width=7.5cm]{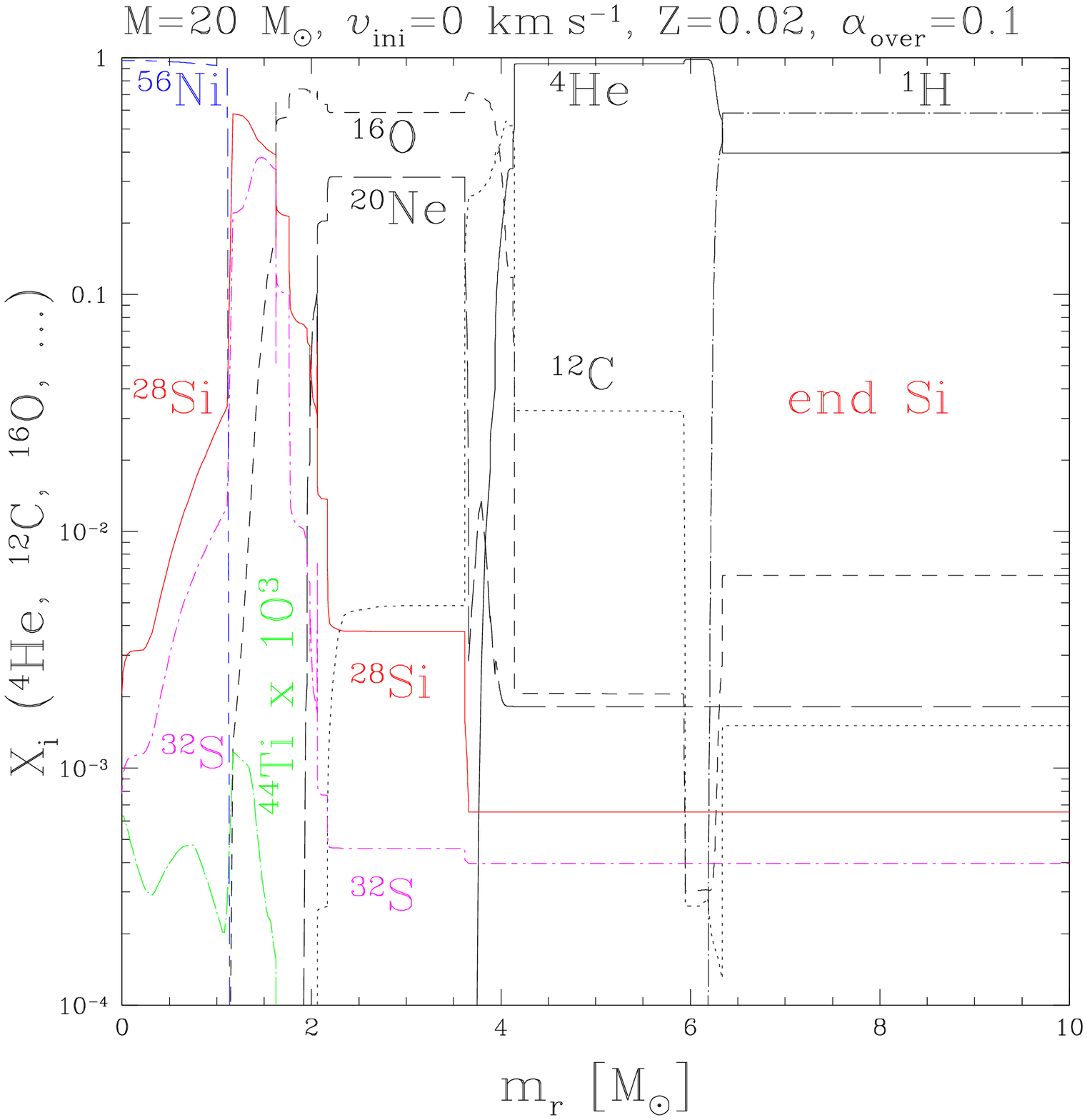}\includegraphics[width=7.5cm]{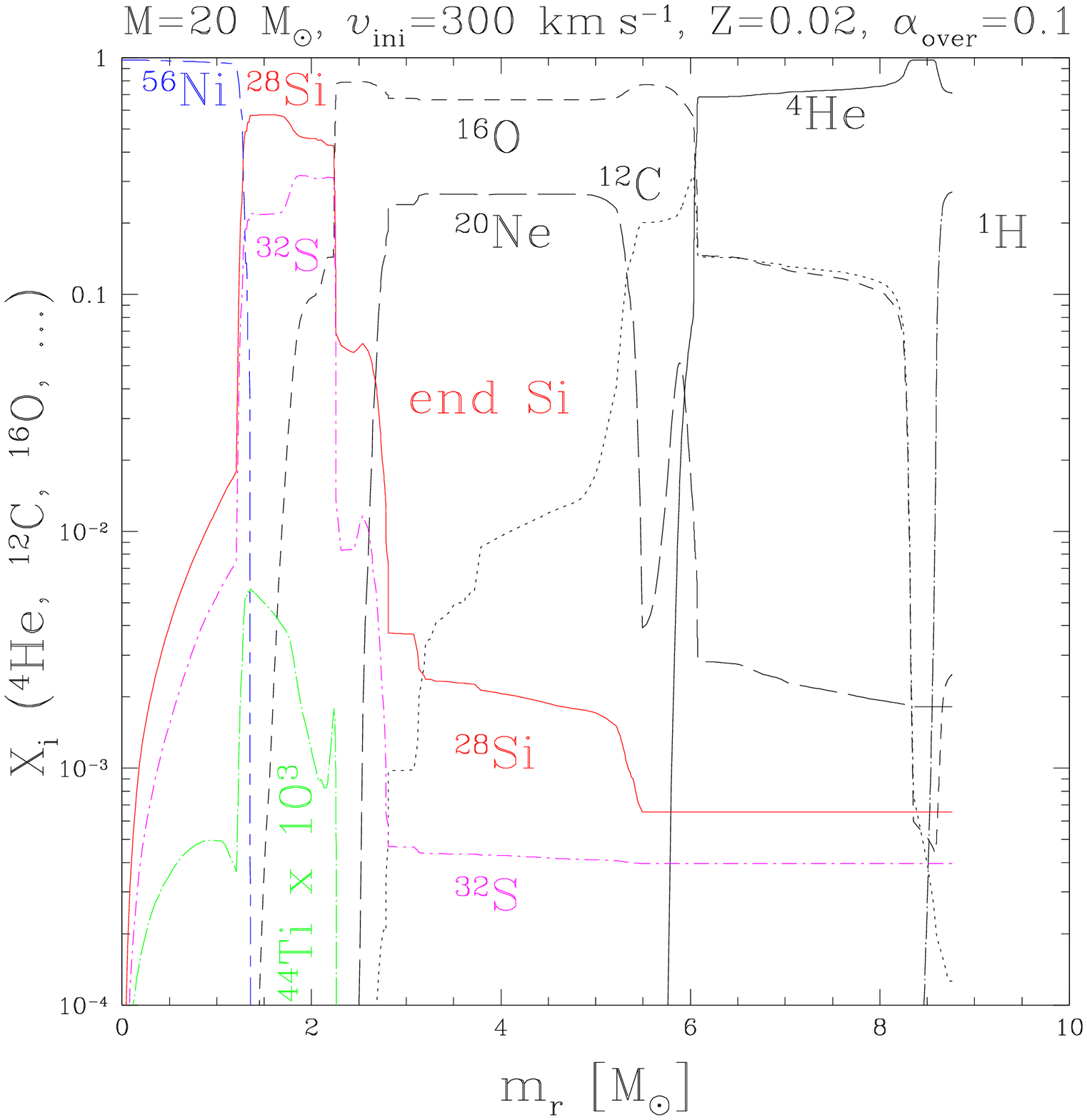}
\caption{Variation of the abundances in mass fraction as a function of
the lagrangian mass at the end of central neon (top), 
oxygen (middle) and silicon (bottom) 
 burnings for the non--rotating (left) and rotating
 (right) 20$M_{\sun}$ models.
 Note that the abundance of
$^{44}$Ti (dotted--long dashed line) is enhanced by a factor 1\,000 
for display purposes.}
\label{a20neosi}
\end{figure*}

Figures \ref{a20hhec} and \ref{a20neosi} show the evolution of the
abundances inside the non--rotating (left) and rotating (right)
20 $M_{\sun}$ models 
at the end of each central burning episode. A the end of
H--burning, we notice the smoother profiles in the rotating model, consequence of
the rotational mixing. At the end of He--burning, we can
already see the difference in core sizes and total mass. We also notice
the lower C/O ratio for rotating models. At the end of
O--burning, we can see that the rotating model produces much more oxygen
compared to the non--rotating model (about a factor two). At the end of
Si--burning, the iron and Si--cores are slightly bigger in the rotating
model (see also Table~\ref{table1}). The yields of
oxygen are therefore expected to increase significantly with rotation.
This will be discussed in an forthcoming article. 

\section{Pre--supernova models}
\subsection{Core masses}
\begin{figure}[!tbp]
\centering
\includegraphics[width=8.8cm]{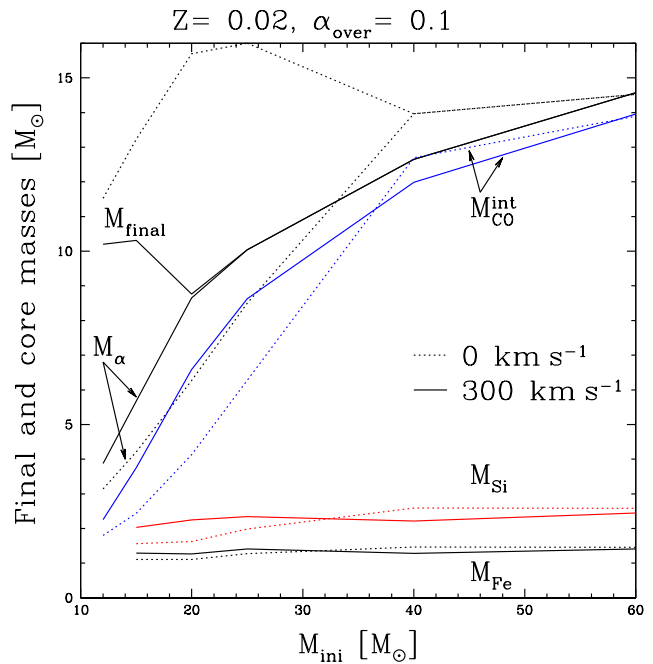}
\caption{Core masses as a function of the initial mass and velocity at
the end of core Si--burning.}
\label{mcores}
\end{figure}
Figure \ref{mcores} shows the core masses (Tables \ref{table1} and \ref{table1b})
as a function of initial mass
for non--rotating (dotted lines) and rotating  (solid lines) models. 
Since rotation increases mass loss, the final
mass, $M_{\rm{final}}$, of rotating models is always smaller than that of non--rotating
ones. Note that for very massive stars ($M \gtrsim 60 M_{\sun}$) mass
loss during the WR phase is proportional to the actual mass
of the star. This produces a convergence of the final masses (see for instance
Meynet \& Maeder 2004).
We can also see a general difference between the effects of rotation
below and above $30\,M_{\sun}$.
For $M \lesssim 30\,M_{\sun}$, rotation significantly increases 
the core masses due to mixing.  
For $M \gtrsim 30\,M_{\sun}$, rotation makes the star enter at an earlier
stage into the WR phase. The rotating star spends therefore a longer time in this
phase characterised by heavy mass loss rates. This results in smaller
cores at the pre--supernova stage.
We can see on Fig.~\ref{mcores} that the difference between rotating and
non--rotating models is the largest between 15 and 25 $M_{\sun}$.

Concerning the initial mass dependence, one can make the following remarks:
\begin{description}
\item[$M_{\rm{final}}$:] There is no
simple relation between the final mass and the initial one. The important 
point is that 
a final mass between 10 and 15 $M_{\sun}$ can correspond to any star with
an initial mass between 15 and 60 $M_{\sun}$.
\item[$M_{\alpha}$ and $M_{\rm{CO}}$:] The core masses increase
significantly with the initial mass. For very massive stars, these core
masses are limited by the very important mass loss rates undergone by these
stars: typically $M_{\alpha}$ is equal to the final mass for
$M \gtrsim 20\,M_{\sun}$ for rotating models and for $M \gtrsim
40 M_{\sun}$ for the non--rotating ones. The mass of the
carbon--oxygen core is also limited by the mass loss rates for 
$M \gtrsim 40\,M_{\sun}$ for both rotating and non--rotating models.
\item[$M_{\rm{Si}}$ (at the end of central Si--burning):]
 For rotating models, $M_{\rm{Si}}$ oscillates between 2 and
2.5 $M_{\sun}$. For non--rotating models, the mass increases
regularly between 15 $M_{\sun}$ ($M_{\rm{Si}} \simeq 1.56 M_{\sun}$)
and 40 $M_{\sun}$ 
($M_{\rm{Si}} \simeq 2.6 M_{\sun}$) and stays constant
for higher masses (due to mass loss). 
\item[$M_{\rm{Fe}}$ (at the end of central Si--burning):] 
 Follows the same trend as $M_{\rm{Si}}$.
\end{description}

\subsubsection{Final iron core masses}\label{finfe}
For non--rotating models, the masses of the iron core $M_{\rm{Fe}}$ in
the last computed model (end of shell
Si--burning) are very close (within 8\%)
to the silicon core masses, $M_{\rm{Si}}$, at the end of central
Si--burning. This occurs because
the extent of shell Si--burning is limited by the entropy increase
produced by the second episode of shell O--burning.
Therefore even though our rotating models have not reached core collapse,
we can have an estimate of the final iron core mass by taking 
the
value of $M_{\rm{Si}}$ at the end of central Si--burning. In this way, 
we obtain iron core masses for rotating models between 2 and
2.5 $M_{\sun}$. For non--rotating models, the mass is between
1.56 $M_{\sun}$ and  2.6 $M_{\sun}$. 
Rotating models have therefore more
massive iron cores and we expect the lower mass limit for black
hole (BH) formation to decrease with rotation.

As said in Sect. \ref{12m} about the fate of the 12 $M_{\sun}$, we
did not follow the evolution of the electron mole number, $Y_e$,
 neither include Coulomb corrections.
Coulomb corrections generally act
to decrease the iron core mass by about 0.1\,$M_{\sun}$ 
\citep[ and references therein]{WHW02}.
Electron captures during Si--burning
increases neutron excess and also reduces the electron pressure and this
(with photodisintegration) will allow the core to collapse \citep{WHW02}.
It is therefore possible that some of our 
models should collapse before shell Si--burning occurs. Taking this
argument into account and the fact that we used Schwarzschild criterion 
for convection, we have to consider the value of $M_{\rm{Si}}$ 
at the end of core Si--burning as an upper limit for the final iron core
mass.


\subsection{Internal structure}

\begin{figure*}[!tbp]
\centering
   \includegraphics[width=8.8cm]{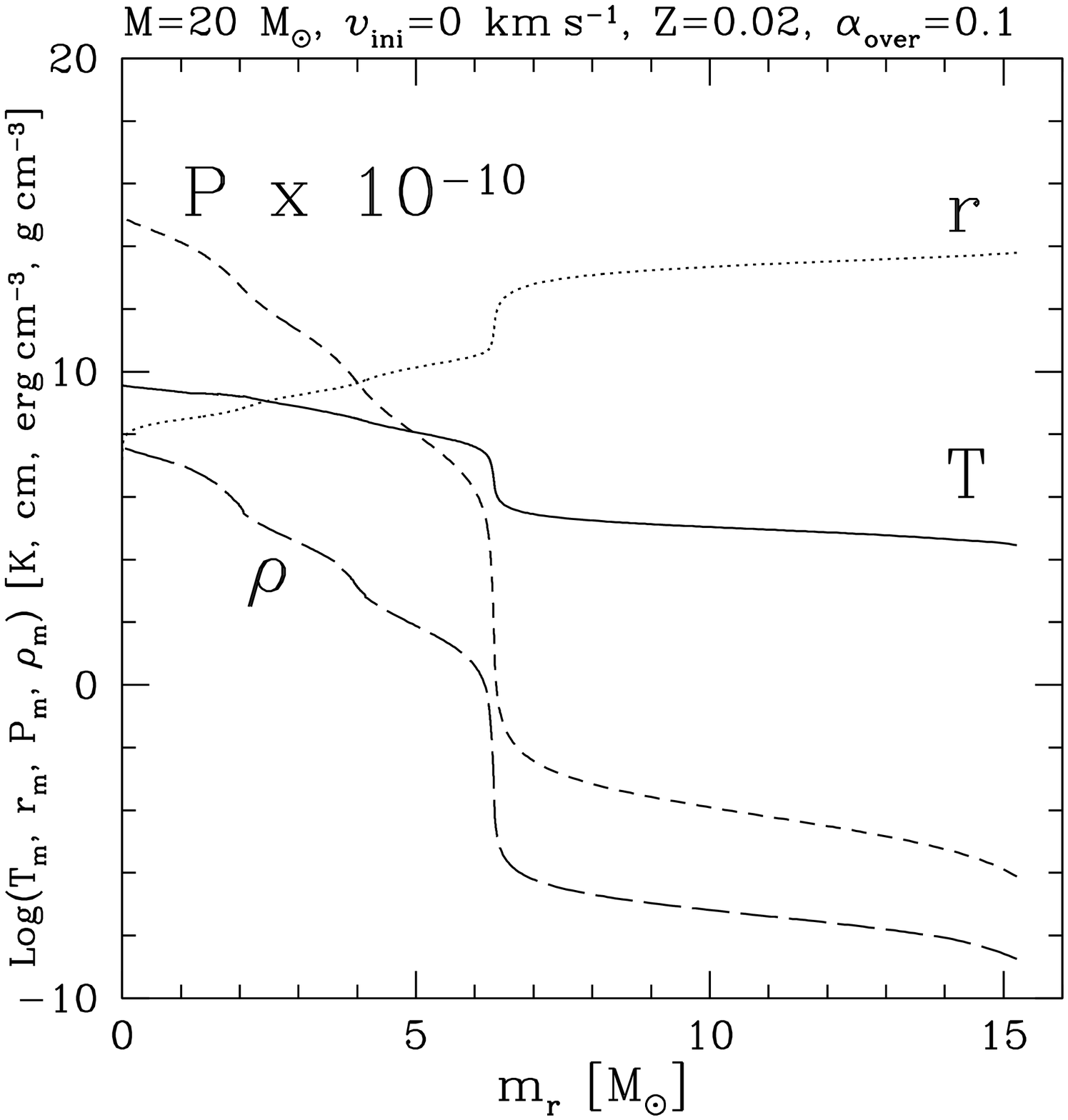}\includegraphics[width=8.8cm]{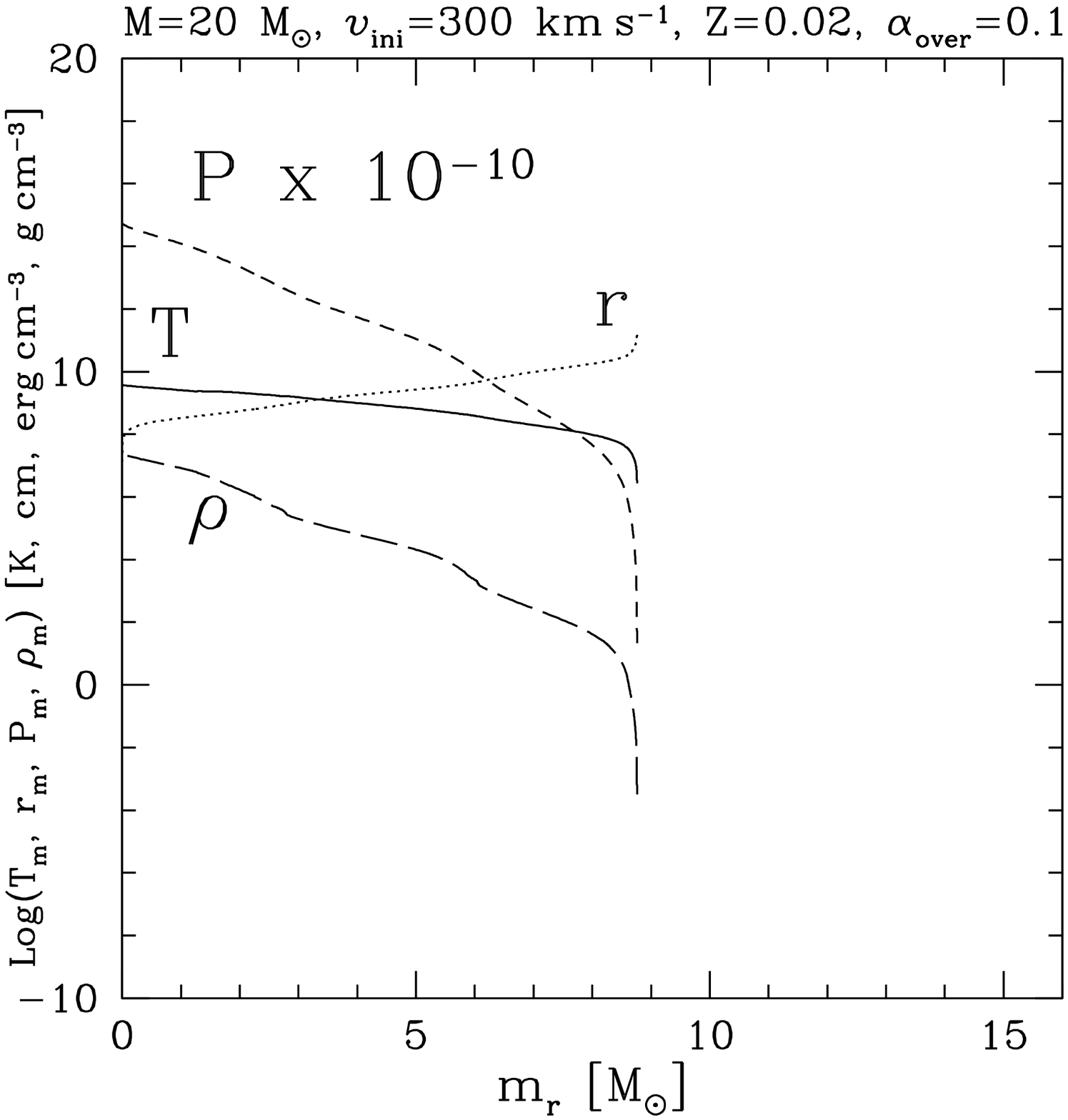}
\caption{Profiles of the radius, $r$, density, $\rho$, temperature, $T$
and pressure $P$ at the end of 
core Si--burning for the
non--rotating (left) and rotating (right) 20 $M_{\sun}$ models. The
pressure has been divided by $10^{10}$ to fit it in the diagram.
}
\label{steq}
\end{figure*}

As well as the chemical composition (abundance profiles and core masses) 
of the pre--supernova star, other parameters, like the density profile, 
the neutron excess (not followed in our calculations), the entropy and 
the total radius of the star, play an important role in the supernova
explosion. Figure \ref{steq} shows the density, temperature, radius and pressure
variations as a function of the lagrangian mass coordinate at the end
of the core Si--burning phase. Since the
rotating star has lost its envelope, this truly affects the parameters
towards the surface of the star. The radius of the star (BSG) is about
one percent that of the non--rotating star (RSG). As said above this modifies
strongly the supernova explosion. We also see
that temperature, density and pressure profiles are flatter 
in the
interior of rotating models due to the bigger core sizes. 

\section{Comparison with the literature}
In this section, we compare our results (HMM hereinafter) with four 
other recent papers: 
\citet{LSC00} (LSC hereinafter), 
\citet{WHW02} (WHW), 
\citet{RHHW02} (RHW) 
and \citet{HLW00} (HLW).
Before we start the comparison, we need to mention which physical
ingredients (treatment of
convection, $^{12}$C$(\alpha,\gamma)^{16}$O reaction rate, \ldots) 
they use:
\begin{itemize}
\item LSC use Schwarzschild criterion for convection without 
overshooting (except for core He--burning for which semiconvection and
an induced overshooting are taken into account). 
For $^{12}$C$(\alpha,\gamma)^{16}$O, they use the rate
of \citet{CF85} (CF85). Mass loss is not included.
\item WHW use Ledoux criterion for convection with 
semiconvection. They use a relatively large diffusion coefficient
for modeling semiconvection. Moreover non--convective zones
immediately adjacent to convective regions are slowly mixed
on the order of a radiation diffusion time scale to 
approximately allow for the effects of convective overshoot.
For $^{12}$C$(\alpha,\gamma)^{16}$O, they use the rate
of \citet{CF88} (CF88) multiplied by 1.7.
\item RHW use Ledoux criterion for convection with 
semiconvection. They use the same 
method as WHW for semiconvection. 
For $^{12}$C$(\alpha,\gamma)^{16}$O, they use the rate
of \citet{BU96} (BU96) multiplied by 1.2.
\item HLW use Ledoux criterion for convection with 
semiconvection using a small diffusion coefficient (about one percent 
of WHW's coefficient) 
and without overshooting. For 
$^{12}$C$(\alpha,\gamma)^{16}$O, they use a rate close to
\citet{CF85} (CF85). They present models with and without
rotation.
\item In this paper (HMM), we used Schwarzschild criterion 
for convection with overshooting for core H and He--burnings. For 
$^{12}$C$(\alpha,\gamma)^{16}$O, we used the rate
of \citet{NACRE} (NACRE).
\end{itemize}

\subsection{HR diagram} 

We remark that the present evolutionary tracks (as well
as those from LSC) do not decrease in luminosity when they cross the Hertzsprung gap.
This is in contrast with the tracks
from HLW which present a significant decrease in luminosity when
they evolve from the MS phase to the RSG phase. Models
computed with the present code but using the Ledoux criterion
for convection (without semiconvection) present a very
similar behaviour to those of HLW. Thus the difference 
between the two sets of models mainly results from the different
criterion used for convection. 
 
\subsection{Lifetimes} 
\begin{table*}
\caption{Lifetimes of central burning stages of
solar metallicity models. Lifetimes are in years with 
exponent in brackets
(2.14 (-2)=$2.14\ 10^{-2}$). }
\begin{tabular}{l | r r r | r r r | r r r}
\hline \hline
$M_{\rm{ZAMS}}$    &   15 (HMM) &   15 (WHW) &   15 (LSC) &    20 (HMM)&   20 (WHW) &   20 (LSC) &  25 (HMM)  &  25 (WHW)  &  25 (LSC)  \\
\hline
$\tau_{\rm{H}}$    &   1.13 (7) &   1.11 (7) &   1.07 (7) &  7.95 (6)  &  8.13 (6)  &  7.48 (6)  &   6.55 (6) &   6.70 (6) &   5.93 (6) \\
$\tau_{\rm{He}}$   &   1.34 (6) &   1.97 (6) &   1.4  (6) &  8.75 (5)  &  1.17 (6)  &  9.3  (5)  &   6.85 (5) &   8.39 (5) &   6.8  (5) \\
$\tau_{\rm{C}}$    &   3.92 (3) &   2.03 (3) &   2.6  (3) &  9.56 (2)  &  9.76 (2)  &  1.45 (3)  &   3.17 (2) &   5.22 (2) &   9.7  (2) \\
$\tau_{\rm{Ne}}$   &   3.08     &   0.732    &   2.00     &  0.193     &  0.599     &  1.46      &  0.882     &  0.891     &  0.77      \\
$\tau_{\rm{O}}$    &   2.43     &   2.58     &   2.43     &  0.476     &  1.25      &  0.72      &  0.318     &  0.402     &  0.33      \\
$\tau_{\rm{Si}}$   &   2.14 (-2)&   5.01 (-2)&   2.14 (-2)&   9.52 (-3)&   3.15 (-2)&   3.50 (-3)&   3.34 (-3)&   2.01 (-3)&   3.41 (-3)\\
\hline
\end{tabular}
\label{table2}
\end{table*}
We can compare the lifetimes of the non--rotating 15, 20, 25 
$M_{\sun}$ models with recent calculations presented in WHW and 
LSC. The comparison is shown in Table \ref{table2}. As said earlier, LSC
use Schwarzschild criterion with overshooting only for He--burning. 
WHW use Ledoux criterion with a very efficient semiconvection 
and allow for some overshoot.
Despite important differences in the treatment of convection,
all the models give very similar H--burning lifetimes which differ by less than 10\,\%.
For the He--burning lifetimes, during which the convective core grows in mass,
one can expect that the results will be significantly different depending on which
convection criterion is used. This is indeed the case.
Inspecting Table~3, one sees that
our results are shorter by 30--50\% with respect to those of WHW.
In contrast when the Schwarzschild criterion is used with some overshooting
as in LSC, the results are very similar (differences inferior to six percents).
In the advanced stages one sees that the lifetimes obtained by the different
groups are of the same order of magnitude.
Let us note that the definition of the duration of the nuclear burning stages may
differ between the various authors and this tends to enhance the scatter
of the results. Keeping in mind this source of difference and the fact that
the lifetimes vary by eight or nine orders of magnitude between the H--burning 
and the Si--burning phases,
the agreement between the various authors appears remarkable. 

\subsection{Kippenhahn diagrams and convection during central C--burning}
Our Kippenhahn diagrams for the non--rotating models  are in good agreement 
with those of \citet{RHHW02} except that in our model the carbon and 
oxygen shells do not merge for the 20 $M_{\sun}$.
The only noticeable
difference between the structures in the advanced phase obtained in the present work and those obtained by LSC
is that their 20 $M_{\sun}$ model does not have a central convective core
during C--burning. This can be explained by the fact that they use the
$^{12}$C$(\alpha,\gamma)^{16}$O rate from \citet{CF85}. This
rate is larger than the NACRE rate we use in our models
(see Fig.~\ref{c12ag}) and therefore more $^{12}$C is burnt
during He--burning.

\subsection{Core masses} 
\subsubsection{Non--rotating models} 
\begin{table}
\caption{Final core masses at the pre--supernova stage for different 
models of non--rotating stars at solar metallicity.}
\begin{tabular}{ l  r r r r}
\hline \hline 
$M_{\rm{ZAMS}}$    &15 (HMM)&15 (RHW)&15 (HLW)&15 (LSC)         \\
\hline 
$M_{\rm{total}}$   & 13.232 & 12.612 & 13.55  &  15             \\
$M_{\alpha}^{01}$  &  4.168 &  4.163 &  3.82  &  4.10           \\
$M_{\rm{CO}}^{01}$ &  2.302 &  2.819 &  1.77  &  2.39           \\
$M_{\rm{Si}}^{50}$ &  1.842 &  1.808 &    -   &    -            \\
$M_{\rm{Fe}}^{50}$ &  1.514 &  1.452 &  1.33  &  1.429          \\
\hline
$M_{\rm{ZAMS}}$    &20 (HMM)&20 (RHW)&20 (HLW)& 20 (LSC)        \\
\hline 
$M_{\rm{total}}$   & 15.694 & 14.740 & 16.31  &  20             \\
$M_{\alpha}^{01}$  &  6.208 &  6.131 &  5.68  &  5.94          \\
$M_{\rm{CO}}^{01}$ &  3.840 &  4.508 &  2.31  &  3.44          \\
$M_{\rm{Si}}^{50}$ &  2.002 &  1.601 &    -   &    -           \\
$M_{\rm{Fe}}^{50}$ &  1.752 &  1.461 &  1.64  &  1.552         \\
\hline
$M_{\rm{ZAMS}}$    &25 (HMM)&25 (RHW)&25 (HLW)& 25 (LSC)      \\
\hline 
$M_{\rm{total}}$   & 16.002 & 13.079 & 18.72  &  25            \\
$M_{\alpha}^{01}$  &  8.434 &  8.317 &  7.86  &  8.01         \\
$M_{\rm{CO}}^{01}$ &  5.834 &  6.498 &  3.11  &  4.90         \\
$M_{\rm{Si}}^{50}$ &  2.577 &  2.121 &    -   &    -          \\
$M_{\rm{Fe}}^{50}$ &  1.985 &  1.619 &  1.36  &  1.527        \\
\hline
\end{tabular}
\label{table3}
\end{table}
%
\begin{figure}[!tbp]
\centering
\includegraphics[width=8.8cm]{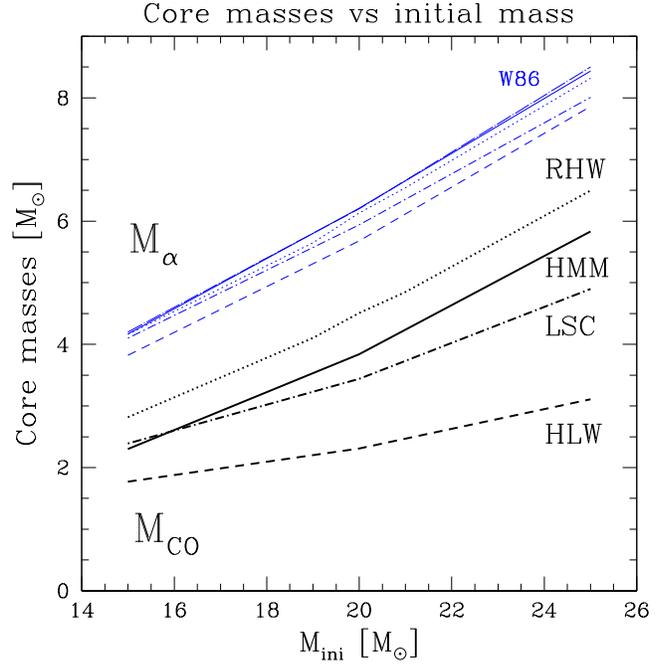}
\caption{Variation of the He core masses, $M_\alpha$ (light lines), and of the CO core masses, $M_{CO}$
(heavy lines),
at the pre--supernova stage in different initial mass models.
Only non--rotating models are shown.
The different types of line correspond to results obtained by
different groups: HMM labels our results,
W86 those of \citet{W86},
RHW those of \citet{RHHW02},
LSC those of \citet{LSC00} and
HLW those of \citet{HLW00}.}
\label{cc2}
\end{figure}
\begin{figure}[!tbp]
\centering
\includegraphics[width=8.8cm]{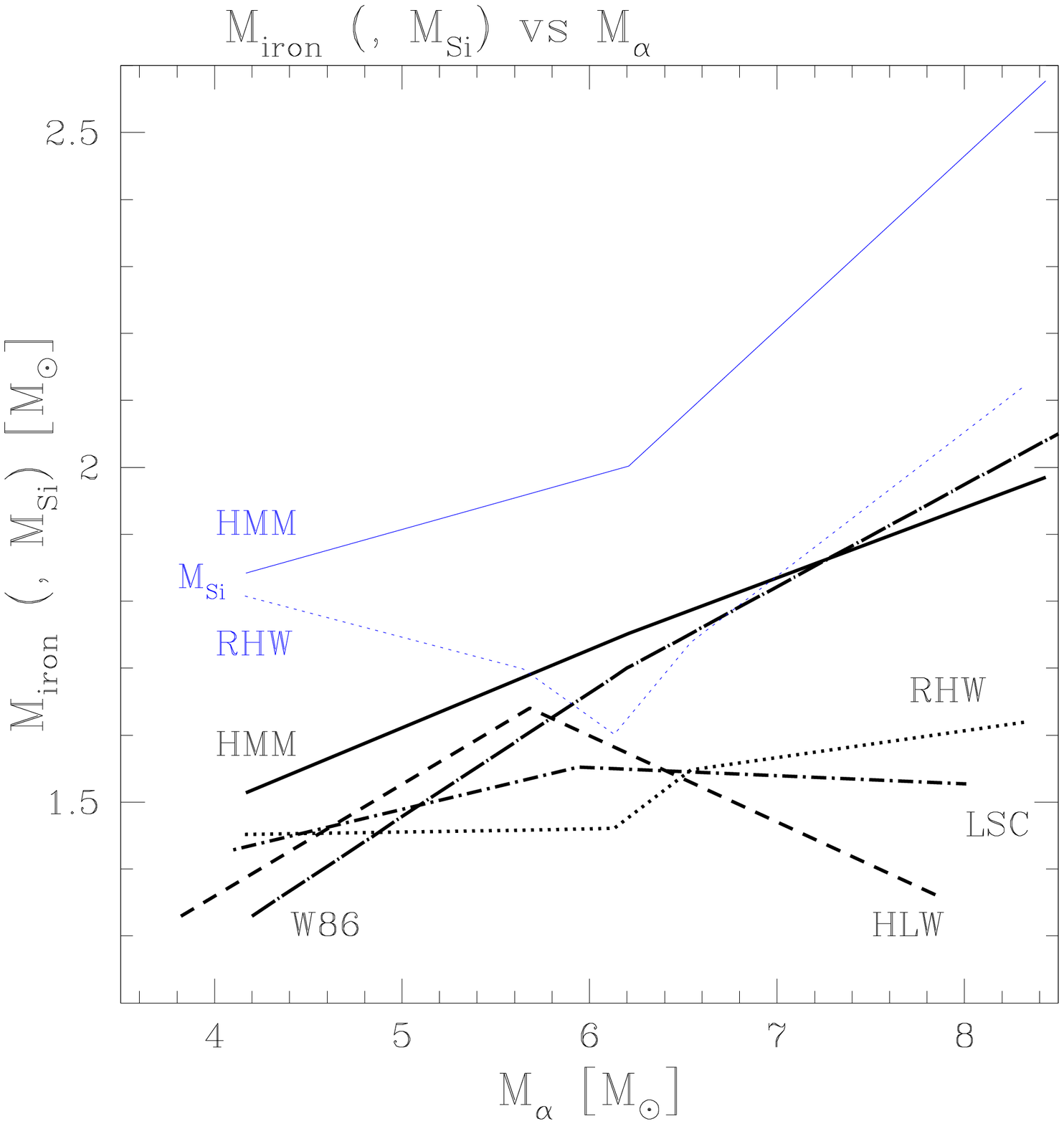}
\caption{$M_{\rm{iron}}$ (and $M_{\rm{Si}}$) as a function of 
$M_{\alpha}$ for non-rotating models from
different authors (see Table \ref{table3}). 
The labels are the same as in Fig.~\ref{cc2}.
The light lines show the variation of $M_{\rm{Si}}$,
the heavy lines those of $M_{\rm{iron}}$.
}
\label{cc}
\end{figure}
\begin{figure}[!tbp]
\centering
\includegraphics[width=8.8cm]{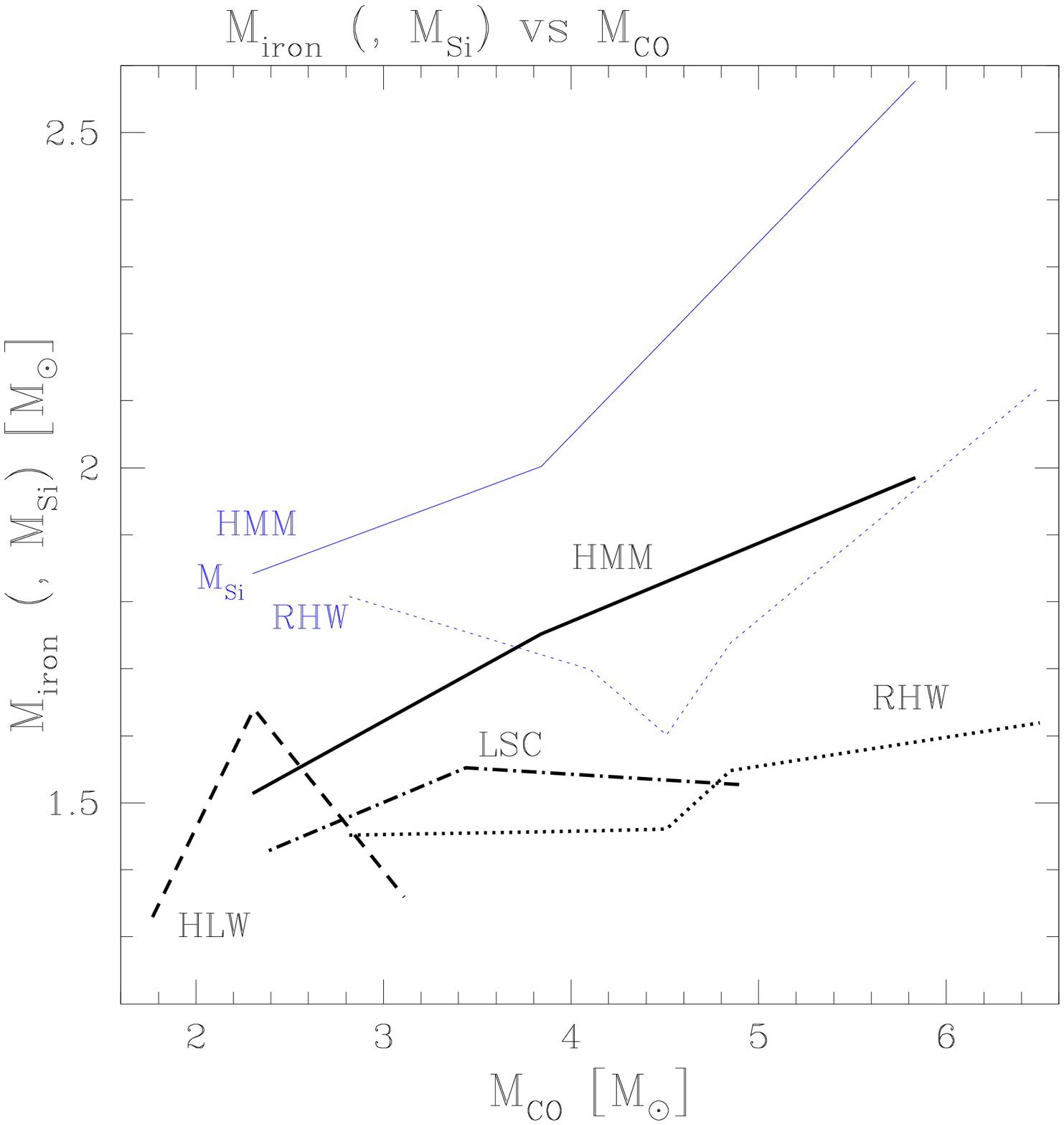}
\caption{$M_{\rm{iron}}$ (and $M_{\rm{Si}}$) as a function of 
$M_{\rm{CO}}$ for non-rotating models from
different authors (see Table \ref{table3}). 
The labels are the same as in Fig.~\ref{cc2}.
The light lines show the variation of $M_{\rm{Si}}$,
the heavy lines those of $M_{\rm{iron}}$.
}
\label{cc1}
\end{figure}

In Table \ref{table3} the final masses and the core masses at the pre--supernova stage
are given for different 15, 20 and 25 $M_\odot$ stellar models. The second column
corresponds to the present non--rotating models, the third shows the results
of \citet{RHHW02}, the fourth those of \citet{HLW00} and the fifth those
of \citet{LSC00}.
In Fig.~\ref{cc2}, we see that for $M_{\alpha}$, the results are very similar 
(within 5\%) between our models and those of LSC and RHW. This can be
understood by the similar outcome of the convection treatment. 
HLW use a small diffusion coefficient for
semiconvection and logically obtain slightly smaller helium cores.

\begin{figure}[!tbp]
\centering
\includegraphics[width=7.5cm]{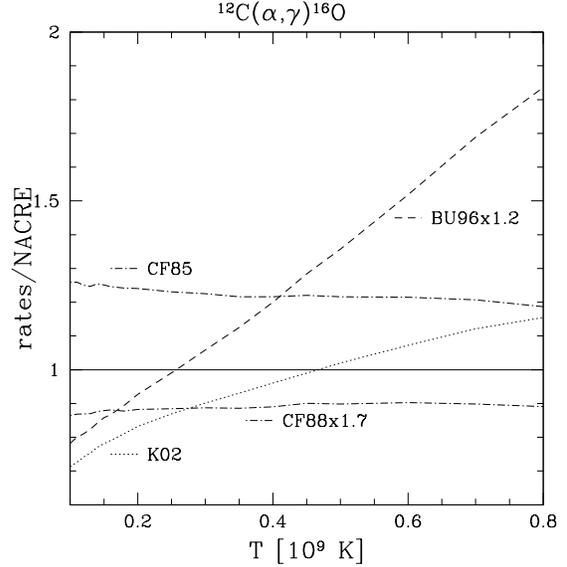}
\caption{Comparison of different $^{12}$C$(\alpha,\gamma)^{16}$O
reaction rates according to various authors:
CF85 labels the rate given by \citet{CF85},
CF88, the rate of \citet{CF88}, BU96 the rate of
\citet{BU96} and K02 the one of \citet{K02}.
 All the rates are normalised
to the rate given by NACRE (\citet{NACRE}).}
\label{c12ag}
\end{figure}

The differences between the mass of the CO cores are much greater.
Let us recall here that
the size of this core depends a lot on the convective criterion
and also on the rate of the $^{12}$C($\alpha,\gamma$)$^{16}$O reaction.
This reaction becomes one of the main source of energy at the end
of the core He--burning phase. 
A faster rate implies smaller central temperatures and thus
increases the He--burning lifetime. This
in turn will produce larger CO cores \citep{L91} with a smaller fraction
of $^{12}$C.
Figure \ref{c12ag} shows the rates 
used by various authors divided by the NACRE rate 
for the temperature range of interest.

Since HLW and LSC use the same rate for this reaction, most of the difference
between the mass of the CO cores must have its origin in the different treatment
of convection. 
One notes also that
LSC still have
slightly smaller cores than us even though they added some semiconvection and use
the CF85 rate for $^{12}$C($\alpha,\gamma$)$^{16}$O which is greater than the one adopted
in our models.
RHW, although they
had slightly smaller $M_{\alpha}$, have larger $M_{\rm{CO}}$. This can be
explained in part by the use of the rate BU96x1.2 for 
$^{12}$C($\alpha,\gamma$)$^{16}$O which is larger
than the NACRE rate at the end of
He--burning (see Fig. \ref{c12ag}).

In Figs~\ref{cc} and \ref{cc1} the Si and iron core masses
obtained at the end of shell Si--burning
are plotted as a function of $M_\alpha$ and $M_{CO}$ respectively.
The present results are well in the range of values obtained by different authors
for $M_\alpha \lesssim 6\,M_\odot$ ($M_{\rm ini} \lesssim 20\,M_\odot$). Above this mass range,
our results are in agreement with those of \citet{W86} and significantly
above the results obtained more recently by the other groups.
As discussed in Sect. \ref{finfe}, we did not follow the evolution of
the neutron excess or include Coulomb corrections. This does not affect
our results until the end of core Si--burning but may affect the results
plotted in Figs.~\ref{cc} and \ref{cc1} obtained at the end of the shell Si--burning.
In this last case, the present results 
have to be considered as upper limits. This might be part of the explanation
why our iron core masses appear to be systematically greater than those
obtained in recent calculations. 
However 
one notes that
\citet{WHW02} give a Chandrasekhar mass (lower mass limit for collapse) of 1.79 $M_{\sun}$ for
the 25 $M_{\sun}$ model
which is large compared to the iron core we obtain at the end of core Si--burning,
implying that our 25 $M_{\sun}$ ($M_\alpha = 8.4$ $M_\odot$) model may experience shell Si--burning
before collapsing.
Thus if we cannot discard that the final iron core masses are overestimated due to the above
reason, they may also be greater than the masses obtained by other groups for other reasons.
In this context it is interesting to compare the masses of the Si--burning core.
The Si--cores are created by
O--burning before Si--burning (except possibly a small fraction due to 
an additional
shell O--burning during Si--burning). Their sizes are thus not dependent on the
neutron excess or the Chandrasekhar mass. Looking at Fig.~\ref{cc} and \ref{cc1}
where our Si--core masses are compared to those obtained by RHW, we see that
our core masses are systematically larger.
In that case the difference cannot be attributed to the neglect in our models
of the electron capture reactions and of the Coulomb corrections. 
Our bigger cores
result from the different prescription we used for convection in our models.
Thus it is possible that the bigger iron cores we have obtained are due, at least in part, 
to the prescriptions we used for convection.

\subsubsection{Rotating models}\label{corot} 
\begin{table*}
\caption{Final core masses at the pre--supernova stage for different 
models of rotating stars at solar metallicity. 
Note that we use $M_{\rm{Si}}$ at the end of central Si--burning 
for the value of $M_{\rm{Fe}}$
as discussed in the text.}
\begin{tabular}{l | r r r | r r r | r r}
\hline \hline 
$M_{\rm{ZAMS}}$    &   15   &F15B (HLW)&E15 (HLW)&20    &F20B (HLW)&E20 (HLW)&  25   &E25 (HLW) \\
\hline
$M_{\rm{total}}$   & 10.316 & 12.89 & 10.86  & 8.763 & 14.76 & 11.00  & 10.042 &  5.45         \\
$M_{\alpha}^{01}$  &  5.604 &  3.88 &  5.10  & 8.567 &  5.99 &  7.71  & 10.042 &  5.45         \\
$M_{\rm{CO}}^{01}$ &  3.325 &  2.01 &  3.40  & 5.864 &  2.75 &  5.01  &  7.339 &  4.07         \\
$M_{\rm{Fe}}^{50}$ &  2.036 &  1.38 &  1.46  & 2.245 &  1.36 &  1.73  &  2.345 &  1.69         \\
\hline
\end{tabular}
\label{table4}
\end{table*}
We can also compare core masses of the rotating 15, 20, 25 
$M_{\sun}$ models with recent calculations by \citet{HLW00} (HLW).
For $M_{\rm{Fe}}$, we use $M_{\rm{Si}}^{50}$. As discussed in section
\ref{finfe}, this assumes that shell  Si--burning occurs before the
collapse and our value has to be considered as an upper limit. 

The comparison is shown in Table \ref{table4}. 
``F..B'' models are the models with the same initial rotational velocity and 
inclusion of the $\mu$--gradients inhibiting effects on rotational
mixing. These are the models which should give approximately the 
same results as us if uncertainties concerning the treatment of convection and
particular reaction rates were small. We also show the  ``E..'' models 
with a lower initial rotational velocity but without the $\mu$--gradients 
inhibiting effects. 

One can see by comparing the results from the two 
models of HLW, the great dependence of the core masses on the treatment
of the $\mu$--gradient inhibiting effect. 
The more efficient the rotational mixing
(or less strong are the inhibiting effects of the $\mu$--gradients), the
greater the core masses.
Compared to the results obtained by HLW, one sees that our core masses are
significantly greater. This essentially results from two facts:
first the effects of rotation are included in our models in a different way
than in the models by HLW. In particular in our models, the treatment of rotational mixing
includes the inhibiting effect of $\mu$--gradients without any ad hoc parameters, 
and the transport of the angular momentum by the meridional circulation is
properly accounted for by an advective term \citep{ROTIII}. Secondly, the present stellar models
were computed with the Schwarzschild criterion with a moderate overshooting while the models of HLW
were computed with the Ledoux criterion with 
semiconvection using a small diffusion coefficient 
and without overshooting.

\subsection{Final angular momentum}\label{compam}
\begin{figure}[!tbp]
\centering

\includegraphics[width=8.8cm]{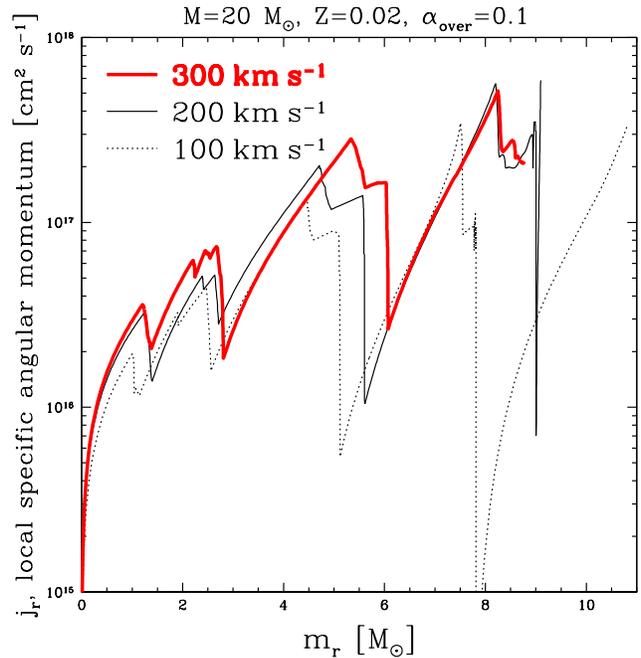}
\caption{Comparison of the final local specific angular momentum 
profiles for different 20 $M_{\sun}$ models. 
Models with different initial velocities, $\upsilon_{\rm{ini}}$=
100, 200 and 300\,km\,s$^{-1}$ are drawn with dotted, solid
and thick solid lines respectively. We can see the convergence of the
final of the final angular momentum of the core above
$\upsilon_{\rm{ini}}$= 200\,km\,s$^{-1}$.
}
\label{jevoc123}
\end{figure}
\begin{figure}[!tbp]
\centering

\includegraphics[width=8.8cm]{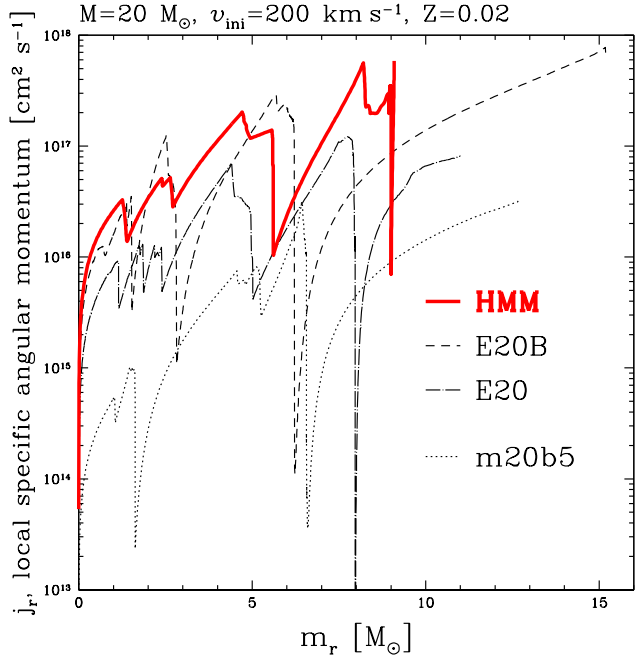}
\caption{Comparison of the final local specific angular momentum 
profiles for different 20 $M_{\sun}$ models all with the same initial 
rotational velocity, $\upsilon_{\rm{ini}}$= 200 km\,s$^{-1}$.
The thick solid line (HMM) corresponds to our model.
The models from \citet{HLW00} are drawn with a dashed line for model E20
(no $\mu$--barrier) and with a dotted--dashed line for model E20B
($\mu$--barrier with $f_{\mu}=0.05$). Finally, model m20b5 from
\citet{HWLS03} including the effect of the magnetic fields according to
\citet{Sp02} is drawn with the dotted line.
}
\label{jevoc2}
\end{figure}
Long soft gamma--ray bursts (GRBs) were recently connected with SNe
\citep[see][ for example]{Ma03}.
One scenario for GRB production is the collapsar mechanism devised by
\citet{W93}. In this mechanism, a star
collapses into a black hole and an accretion disk due to the high
angular momentum of the core. Accretion from the disk onto the central
black hole produces bi--polar jets. These jets can only reach the
surface of the star (and be detected) if the star loses 
its hydrogen rich envelope before the collapse. 
WR stars are therefore good candidates for collapsar
progenitors since they lose their hydrogen rich envelope during the
pre--SN evolution. The question
to answer is whether the core of WR stars contains 
enough angular momentum
at the pre--SN stage
(the specific angular momentum, $j$, of the material just outside the
core must be larger than
 $10^{16}\,cm^2\,s^{-1}$). 
So far only Heger and co--workers have obtained
values for the angular momentum of the cores of massive stars at the
pre--SN stage \citep{HLW00,HWLS03}. The physical
ingredients of their model have been given in Sect.
\ref{corot}. The comparison between our models and theirs 
shows that the size of the various cores
depends significantly on the treatment of both convection and rotation.
The evolution of angular velocity and angular momentum in the
models of \citet{HLW00,HWLS03}, is
presented in \citet[ HLW00
hereinafter; in Figs. 8 and 9]{HLW00}.
The evolution of angular velocity and momentum in our models is 
described in
\citet{ROTV} and in Sect. \ref{omjevo}. 

HLW00 show with their Fig. 9 the convergence of the final angular 
momentum of the core for a wide range of initial angular momentum. 
The dependence of the final angular momentum
on the initial one for our models is displayed in Fig. \ref{jevoc123}.
Models with $\upsilon_{\rm{ini}}$=
100 and 200 km\,s$^{-1}$ have been computed until the end of O--burning.
This should not affect the comparison since the angular momentum profile
does not change during Si--burning.
One can see in our case that convergence only occurs above 
$\upsilon_{\rm{ini}}$=200\,km\,s$^{-1}$. 
Indeed, the average specific angular momentum of
the core (assuming a 1.7 $M_\odot$ core) at the end of the calculation
is $1.326\,10^{16}$, $1.801\,10^{16}$ and 
$2.106\,10^{16}\,cm^2\,s^{-1}$ for $\upsilon_{\rm{ini}}$=
100, 200 and 300 km\,s$^{-1}$ respectively.

Concerning the evolution of the angular momentum, the general picture is
the following. Mass loss removes angular momentum from the surface and
transport processes (convection and rotational mixings) redistribute
angular momentum inside the star (see Sect. \ref{omjevo} and the 
references given above for details). 
Here we are only concerned about the evolution of the
angular momentum of the core of the star.  
During H--burning, both our models and models
without the inhibiting effect of the $\mu$--gradient on mixing 
(models without ``B'') from  HLW00 show
a large decrease of the angular momentum of the core. On the other hand,
in HLW00 models including the inhibiting effect of the $\mu$--gradient 
(models with ``B''), the core
does not lose much angular momentum during H--burning. 
In our models, thermal
turbulence is taken into account and is able to overcome the 
inhibiting effect of the $\mu$--gradient. HLW00 do not include the 
thermal effects and in their situation, the inhibiting effect of the
$\mu$--gradient is almost complete even with a reduction parameter 
$f_{\mu}$ equal to 0.05.
The different treatment of rotation (and especially the different way 
the inhibiting effect of the 
$\mu$--gradient is included) has therefore a strong impact on the
evolution of the angular momentum of the core during the MS 
and explains the
difference between the various models. 

At the end of H--burning, the core contracts and the envelope expands. 
This restructuring phase is accompanied by a formation of a very deep
external convective zone. At the same time, shell H--burning
creates a short--lived intermediate convective zone. 
These changes may affect
the angular momentum profile. The largest change in our models is the
creation of a large drop of the angular momentum at the bottom of the
external convective zone (see Fig. \ref{jevo}). This is due to the fact 
that convection
enforces solid body rotation and therefore angular momentum is
transported at the outer edge of the convective zone. No significant
change is seen in the core.

During He--burning, the trend is the opposite from H--burning. 
In both our models and models
without the inhibiting effect of the $\mu$--gradient on mixing from 
HLW00, the angular momentum in the core decreases slightly. 
On the other hand,
in HLW00 models with the inhibiting effect of the $\mu$--gradient, 
the core
loses a significant amount of angular momentum after H--burning. 
The reason is the following. During H--burning,
in HLW00 models with the $\mu$--gradient effects on mixing, 
even though the core does not lose much angular momentum, the
layers just above it lose angular momentum (due to various
transport processes). This creates a large angular velocity gradient at
the edge of the core which increases rotational mixing during 
He--burning. 
Furthermore the successive convective and
semiconvective zones (due to the restructuring phase and 
shell H--burning) mix as well the angular momentum of the outer parts 
of the core with 
layers above the core and a large
amount of angular momentum is transfered out of the core at this time. 
In our models (as well as those from HLW00
without $\mu$--gradient), angular momentum is transfered to the
layers above the core during H--burning. 
This creates a smaller gradient of angular velocity at the edge of
the core at the end of H--burning 
and thus rotational mixing is weaker during He--burning.
Therefore, the angular momentum of the core does not change as much
during He--burning. Note also that He--burning is ten times shorter than
H--burning and that there is less time to mix. 
 As said in Sect. 
\ref{omjevo}, during the advanced stages, the angular momentum profile
does not change substantially. Only convective zones create spikes 
along the profile. 

The comparison of the final
angular momentum profile of the different models, all with the same
initial mass and surface angular velocity, is shown in 
Fig. \ref{jevoc2}. 
The thick solid line (HMM) corresponds to our model.
The models from \citet{HLW00} are drawn with a dotted--dashed line for 
model E20 (no $\mu$--barrier) and with a dashed line for model E20B
($\mu$--barrier with $f_{\mu}=0.05$). Finally, model m20b5 from
\citet{HWLS03} including the effect of the magnetic fields according to
\citet{Sp02} is drawn with the dotted line. 
Even though the evolution of angular momentum differs between our model
and model E20B (with the $\mu$--gradient effects on mixing) from
\citet{HLW00}, the final
value of the angular momentum of the core is very similar for these two
models. This confirms the possibility of the formation of GRBs via 
collapsars
from rotating massive stars \citep{WH03,HWLS03} if the effects of
magnetic field (not included in our work) are small. 
Indeed, for example, the 25 $M_\odot$ model ends up
as a WR star with a core having enough angular momentum 
($j\gtrsim 10^{16}\,cm^2\,s^{-1}$ for the material just outside the
core, see Fig. \ref{jevo}) to create a
collapsar.
The difference between our model and 
model E20 (without the $\mu$--gradient effects on mixing) from
\citet{HLW00} is probably due to the combination of the non--inclusion
of the $\mu$--gradient effects on mixing and of the different treatment 
of meridional circulation (see Sect. \ref{corot}).
From model m20b5 (see Fig. \ref{jevoc2}), 
one sees that the
inclusion of the effects of magnetic fields according to \citet{Sp02}
decreases significantly the final angular momentum of the core. In
this situation, the core rotates too slowly and cannot produce a 
collapsar. 
We can also compare our models with the observed rotation period of
young pulsars. Rotating models without the effects of magnetic fields 
have about 100 times more angular momentum at the pre--SN stage
 than the observed young pulsars \citep{HLW00}. 
 Models including the effects of magnetic fields according
to \citet{Sp02} have about 5--10 times more angular momentum at the
pre--SN stage than the observed young pulsars \citep{HWLS03}. 
This means that in any case, additional slow down is necessary during
the core collapse \citep{WH03,F04} in order to reproduce the observed 
rotation periods of young pulsars. The question that needs to be
answered is when and how this slow down occurs.
Further developments will therefore be of great importance for the
formation of both NSs and GRBs. 
The topic of the final angular momentum of our models 
and its implications for further
evolution will be developed in a future article. 

\subsection{Lower mass limit for models to reach iron core collapse}
As said above, our 12 $M_{\sun}$ models have not been pursued beyond
the O and Ne--burning phases for the rotating and non--rotating models
respectively.
Nevertheless, we think that the rotating model has the potential to
reach an iron core while the non--rotating model does not.
Recent calculations done by \citet{HLW00} are similar to ours on that
point: Their non--rotating models as well as the rotating models E12B,
F12B and G12B neither reach core--collapse. Only the model E12 reaches
core collapse but the physics used in that last model does not include
$\mu$--gradient inhibiting effects on rotationally induced mixing. At
the same time, recent non--rotating 13 $M_{\sun}$ models in \citet{WHW02} and 
\citet{LSC00}  reach core--collapse. Therefore we expect the lower mass 
 limit for non--rotating models to reach the standard iron core collapse 
 to be around 12--13
 $M_{\sun}$ in agreement with \citet{NH88}. Our rotating models
 tend to show that this limit should be lower for rotating stars. 
A finer grid of models around 12 $M_{\sun}$ would help constraining 
this limit.

\section{Conclusion}

The Geneva evolution code has been improved in order to model the pre--supernova
evolution of rotating massive stars. We extended the nuclear reaction
network with a multiple--$\alpha$ elements chain between carbon and
nickel for the advanced burning stages. We also stabilized the internal
structure equations using Sugimoto's prescription \citep{SU70}. Finally,
we added dynamical shear to the other rotationally induced mixing processes
(secular shear and meridional circulation). 

We calculated a grid of stellar models at solar
metallicity with and without rotation and with masses equal
to 12, 15, 20, 25, 40 and 60 $M_{\sun}$. 

Concerning the evolution of
rotation itself during the advanced stages, the angular velocity
increases regularly with the successive contraction of the core while
the angular momentum does not change significantly (only convection
creates spikes along its profile). This means that we can have a good
estimate of the pre--collapse angular momentum at the end of
He--burning. Comparing our pre--SN models with the criteria for
collapsar progenitors \citep{WH03}, we find that WR stars are possible
progenitors of collapsars. However, in this work we neglected the
effects of magnetic fields. Further developments will
be very interesting for the formation of both GRBs and neutron stars. 
Dynamical shear, although very efficient, only smoothens
sharp angular velocity gradients but does not transport angular momentum
over great distances.

We find that rotation significantly
affects the pre--supernova models by the impact it has during
H and He--burnings. We clearly see the two mass groups where either
rotationally induced mixing dominates for $M<30 M_{\sun}$ or rotationally
increased mass loss dominates for $M>30 M_{\sun}$ as already discussed in
\citet{ROTX}.

We show that rotation
affects the lower mass limits for the presence of 
convection during
central carbon burning,
for iron core
collapse supernovae and for black hole formation. The effects of rotation on pre--supernova models
are most spectacular for stars between 15 and 25 $M_{\sun}$. Indeed,
rotation changes the supernova type (IIb or Ib instead of II), 
the total size of progenitors (Blue instead of Red SuperGiant) and the core
sizes by a factor $\sim 1.5$ (bigger in rotating models). For Wolf-Rayet
stars ($M>30 M_{\sun}$) even if the pre--supernova models are not 
different between rotating and
non--rotating models, their previous evolution is different
\citep{ROTX}. 
We also compare our results with the literature. The biggest
differences are the final mass and the various core masses.
We obtain bigger core masses and this should have a strong impact on
yields.

Future developments are planned to be able to follow the evolution until
core collapse as well as follow neutron excess and 
detailed nucleosynthesis during the entire evolution.

\begin{acknowledgements} We wish to thank S. Goriely for his help in 
the development of the numerical method for the nuclear
reaction network, F. K. Thielemann and W. R. Hix for their help
concerning silicon burning, N. Langer for the discussions about angular
momentum and A. Heger for the information concerning
his models. R. Hirschi also wants to thank the Universiti Malaya in
Kuala Lumpur and in particular Dr. Hasan Abu Kassim 
for their hospitality during his visits to the physics department.
\end{acknowledgements}


\bibliographystyle{aa}

\end{document}